\documentclass[twocolumn]{aastex631}
\usepackage{natbib}
\usepackage{amssymb}
\usepackage{amsmath}
\def\be{\begin{eqnarray}}
\def\ee{\end{eqnarray}}
\usepackage{comment}
\usepackage{graphicx,subfigure}
\usepackage{enumerate}
\usepackage{rotating}
\usepackage{longtable}
\usepackage{booktabs}
\usepackage{float}
\usepackage{hyperref}
\hypersetup{colorlinks,urlcolor={blue},linkcolor={blue},citecolor={blue}}

\shorttitle{Kilonova radio flares in \textit{Swift}/BAT short GRBs}

\shortauthors{A. Eddins et al.}

\begin{document}

\title{A search for kilonova radio flares in a sample of \textit{Swift}/BAT short GRBs}

\author[0000-0001-7937-2007]{Avery~Eddins}
\affiliation{Department of Physics and Astronomy, Texas Tech University, Box 1051, Lubbock, TX 79409-1051, USA.}
\email{avery.eddins@ttu.edu}

\author[0000-0002-4832-0420]{Kyung-Hwan Lee}
\affiliation{Department of Physics, University of Florida, PO Box 118440, Gainesville, FL 32611-8440, USA}

\author[0000-0001-8104-3536]{Alessandra Corsi}
\affiliation{Department of Physics and Astronomy, Texas Tech University, Box 1051, Lubbock, TX 79409-1051, USA.}

\author[0000-0001-5607-3637]{Imre Bartos}
\affiliation{Department of Physics, University of Florida, PO Box 118440, Gainesville, FL 32611-8440, USA}

\author[0000-0003-1306-5260]{Zsuzsanna M\'arka}
\affiliation{Columbia Astrophysics Laboratory, Columbia University in the City of New York, New York, NY 10027, USA}

\author[0000-0002-3957-1324]{Szabolcs M\'arka}
\affiliation{Department of Physics, Columbia University in the City of New York, New York, NY 10027, USA}

\begin{abstract}
\label{abstract}
The multi-messenger detection of GW170817 showed that binary neutron star (BNS) mergers are progenitors of (at least some) short gamma-ray bursts (GRBs), and that short GRB jets (and their afterglows) can have  structures (and observational properties) more complex than predicted by the standard top-hat jet scenario. Indeed, the emission from the structured jet launched in GW170817 peaked in the radio band (cm wavelengths) at $\approx 100$\,d since merger ---a timescale much longer than the typical time span of radio follow-up observations of short GRBs. Moreover, radio searches for a potential late-time radio flare from the fast tail of the neutron-rich debris that powered the kilonova associated with GW170817 (AT\,2017gfo) have extended to even longer timescales (years after the merger). In light of this, here we present the results of an observational campaign targeting a sample of seven, years-old GRBs in the \textit{Swift}/BAT sample with no redshift measurements and no promptly-identified X-ray counterpart. Our goal is to  assess whether this sample of short GRBs could harbor nearby BNS mergers,  searching for the late-time radio emission expected from their ejecta. We found one radio candidate  counterpart for one of the GRBs in our sample, GRB\,111126A, though an origin related to emission from star formation or from an AGN in its host galaxy cannot be excluded without further observations.
\end{abstract}


\section{Introduction}
\label{intro}
                 
Since the discovery of the binary neutron star (BNS) merger GW170817 in gravitational waves (GWs) and light at all wavelengths \citep{Abbott2017a, Abbott2017b, Abbott2017c}, the question of whether there may be a population of nearby, GW170817-like events in the known sample of short gamma-ray bursts (GRBs) has received particular attention \citep[e.g.,][]{Horesh2016,Bartos2019, Dichiara2020, Schroeder2020, Matsumoto2020, Ricci2021}. Notably, GW170817 had a delayed electromagnetic afterglow \citep{Abbott2017c}, first detected in X-rays about 9 days after the merger and the prompt detection of $\gamma$-rays \citep{Haggard2017, Margutti2017,Troja2017}. A radio afterglow detection followed, about 15 days after the merger \citep{Alexander2017,Hallinan2017}. This delayed, non-thermal, radio-to-X-ray emission was related to a structured jet launched after the merger and observed off-axis \citep[e.g.,][]{Granot2018, Lazzati2017, Lazzati2018, Mooley2018a, Mooley2018b, Nakar2018,Ghirlanda2019, Makhathini2021}. X-ray emission in excess to that expected from the structured jet was tentatively detected over 900 days after the GW170817 merger \citep{Hajela2022,Troja2022}. Possible explanations for this emission include a flat radio-to-X-ray spectrum afterglow from the slower kilonova ejecta that powered AT\,2017gfo via the $r$-process \citep[e.g.,][]{Arcavi2017,Chornock2017,Coulter2017,Cowperthwaite2017,Dorut2017,Kasliwal2017,Kasen2017,Kilpatrick2017,Pian2017,Shappee2017,Smartt2017,Tanvir2017,Valenti2017,Metzger2019,Radice2018,Radice2020}, or radiation from accretion processes \citep{Hajela2022, Metzger2021, Nedora2021,Troja2021}.  \citet{Balasubramanian2021} and \citet{Balasubramanian2022} found no evidence for radio emission in excess to that expected from a structured jet at late times, though efforts to detect a kilonova radio afterglow are still ongoing.  \citet{Oconnor2022} also did not find further evidence of an X-ray excess at 4.8\,yrs since the GW170817 merger. \cite{Kilpatrick2021} observed the field of GW170817 in the optical band at late times, finding no remnant of the kilonova emission in their observations. 

As evident from the above discussion, continued radio (and X-ray) observations are needed to confirm the presence of late-time emission from GW170817 and disentangle its origin. However, these results have spurred new interest in the search for so-called late-time radio flares from short GRBs \citep[e.g.][]{Nakar2011,Hotokezaka2015,Hotokezaka2018,Kathirgamaraju2019,Nedora2021}, especially considering that the kilonova afterglow could be visible in radio for years after the initial event. Several recent efforts have targeted both well-localized short GRBs with known redshifts and short GRBs lacking accurate X-ray localizations and redshift measurements \citep[e.g.][]{Schroeder2020,Bruni2021,Grandorf2021,Ricci2021, Ghosh2022}. 

Searches targeting well-localized, cosmological short GRBs have been motivated by the expectation that radio flares can be brighter in the presence of a long-lived magnetar formed after the BNS merger and thus are potentially detectable also at cosmological distances \citep{Ricci2021}. For example, \citet{Schroeder2020} conducted 6\,GHz observations of nine, low-redshift ($z<0.5$) short GRBs with the Karl G. Jansky Very Large Array (VLA) on rest-frame timescales of $\approx 2-14$\,yr following the bursts. The lack of detections constrained the energy deposited into the ejecta to $\lesssim {\cal O}(10^{52})$\,ergs \citep[see also][]{Ricci2021, Ghosh2022}. A re-analysis of 27 short GRBs with GHz radio observations also enabled them to conclude that $\lesssim 50\%$ of events could have formed a stable magnetar after the mergers.

\citet{Grandorf2021} used the VLA to observe four short GRBs in the \textit{Swift}/BAT sample without X-ray localizations. They found a previously uncatalogued radio source within the error region of GRB\,130626 with a $3-6$\,GHz flux density consistent with a radio flare associated with a BNS at a distance of $\approx 100$\,Mpc. However, an origin related to a persistent radio source, not the GRB, could not be excluded. 

Here, we present the results of continued efforts aimed at uncovering potentially nearby, GW170817-like short GRBs via a search for their late-time radio emission. Specifically, we carried out observations of 7 short GRBs in the \textit{Swift}/BAT sample that lack accurate X-ray localizations and have unknown distances. Our short GRB selection criteria, as well as observations and analysis procedures, are very similar to those described in \citet{Grandorf2021}. 
Our paper is organized as follows. In Section \ref{GRBsample}, we describe how we select our GRB sample.
In Section \ref{VLA_Obs}, we describe our observations. In Section \ref{VLA_sources} we describe the counterpart radio candidates identification. In Sections \ref{sec:results} and \ref{Results}, we present and discuss our results. In Section \ref{conclusion}, we summarize and conclude. 

Hereafter we assume cosmological parameters  $H_0 = 69.6$\,km\,s$^{-1}$\,Mpc$^{-1}$, $\Omega_{\rm M }= 0.286$, $\Omega_{\rm vac} = 0.714$ \citep{Bennett2014}. All times are given in UT unless otherwise stated. 

\section{GRB sample selection}
\label{GRBsample}
For this study, we consider only short GRBs in the \textit{Swift}/BAT sample. We focus on short GRBs detected by the \textit{Swift}/BAT because the VLA can cover the typical BAT localization error region in a single-field observation at $3-6$\,GHz. We consider events with $T_{90}\le 2$\,s as those have been traditionally linked to BNS merger progenitors \citep{Nakar2007}, though see e.g. \citet{Rastinejad2022} for a recent discovery of a kilonova associated with the nearby, minute-long duration GRB\,211211A. 

We further select short GRBs with no (promptly) identified X-ray counterparts. Short GRBs in the {\it Swift} sample that have identified afterglows reside at larger distances at which the detection of long-term radio flares is unlikely with short VLA observations. Instead, we select GRBs with no redshift measurement because those could hide a population of nearby events. In fact, we note that without a GW detection, GW170817 would have remained one of the several short GRBs with no follow-up observations and no redshift measurement. This motivates our search.

We exclude  from our short GRB sample events in regions of the sky inaccessible to the VLA, particularly those with declination below $-40^{\circ}$. In addition, we avoided the Clarke belt of satellites, ranging from a declination of $-5^{\circ}$ to 15$^{\circ}$, because VLA observations of sources in the belt are subject to strong radio frequency interference (RFI) from satellites in C-band ($4–8$\,GHz), X-band ($8–12$\,GHz), and Ku-band ($12–18$\,GHz). 

Finally, we remove short GRBs for which existing observations would have already found evidence for kilonova emission had the GRB originated from a nearby BNS merger \citep[e.g.][]{Xu2014, Yurkov2012}. See \cite{Bartos2019} for further details on the sample selection criteria.

\section{VLA Observations and data reduction}
\label{VLA_Obs}

\begin{table*}
\begin{center}\begin{footnotesize}
\caption{VLA observations of the GRBs in our sample. All observations were carried out with the VLA in its C configuration (nominal FWHM of the synthesized beam of 3.5\arcsec \ at 6\,GHz and 7.0\arcsec \ at 3\,GHz). For each GRB field we report the observed central frequency, the image central RMS noise, the date and epoch of the VLA observation, the time spent on source with the VLA, the center of the BAT error region, the BAT position error radius, and the reference for the BAT position and position error.  }
\label{Grandorf_Table}
\setlength{\tabcolsep}{0.36\tabcolsep}
\begin{tabular}{ccccccccc}
\hline
\hline
GRB & $\nu$\footnote{Observation central frequency.} & RMS\footnote{The RMS noise at the center of the image.} & Date Observed & $\Delta T$\footnote{Time between the GRB trigger and our VLA observations.} & Time On Source & BAT Center\footnote{\emph{Swift}/BAT refined localization center.} & BAT Radius\footnote{Error radius of the refined \emph{Swift}/BAT localization.} & Reference
\\
& (GHz) & ($\mu$Jy) & (UT) & (yr) & (mm:ss) & (R.A.~~~Dec.) & (\arcmin) &
\\
\hline
\hline
  080121A & 2.8 & $1.1\times 10^2$ & May 22, 2020 & 12.3 & 33:12 & 09h09m01.8s $+41$d50m21.3s & 2.5 & (\ref{GSFC GRB080121A})
  \\
  & 6.0 & 39 & April 17/18, 2020 & 12.2 & 34:00 &'' &'' &''
  \\
  \hline
  090417A & 2.7 & 16 & May 23, 2020 & 11.1 & 33:09 & 02h19m58.3s $-07$d08m28.9s & 2.8 & \cite{Baumgartner2009}
  \\
  & 6.1 & 9.1 & April 14, 2020 & 11.0 & 34:00 & '' & '' & ''
  \\
  \hline
  101129A & 2.9 & 21 & May 25, 2020 & 9.5 & 33:09 & 10h23m41.0s $-17$d38m42.0s & 3.1 & (\ref{GSFC GRB101129A})
  \\
  & 6.1 & 5.7 & April 16, 2020 & 9.4 & 34:05 & ''&'' & ''
  \\
  \hline
  111126A & 3.0 & 7.4 & May 23, 2020 & 8.5 & 33:09 & 18h24m07.1s $+51$d28m06.1s & 2.5 & (\ref{GSFC GRB111126A})
  \\
  & 6.0 & 4.8 & April 23, 2020 & 8.4 & 34:00 & '' & '' & ''
  \\
  \hline
  120403A & 3.0 & 15 & May 23, 2020 & 8.1 & 33:12 & 02h49m49.8s $+40$d29m21.8s & 2.3 & \cite{Sakamoto2012}
  \\
  & 6.1 & 5.2 & April 16, 2020 & 8.0 & 34:05 & '' & '' & ''
  \\
  \hline
  140606A & 3.0 & 9.7 & May 29, 2020 & 6.0 & 33:06 & 13h27m11.7s $+37$d35m56.5s & 2.4 & \cite{Cummings2014}
  \\
  & 6.1 & 5.5 & April 25, 2020 & 5.9 & 34:05 & '' & ''& ''
  \\
  \hline
  160726A & 2.8 & 13 & May 24, 2020 & 3.8 & 33:09 & 06h35m14.3s $-06$d37m1.4s & 1.29 & (\ref{GSFC GRB160726A})
  \\
  & 6.1 & 7.9 & April 18, 2020 & 3.7 & 34:00 & '' & ''& ''
  \\
  \hline
\end{tabular}
\end{footnotesize}\end{center}\end{table*}

We carried out VLA observations of the  7 GRBs in our sample via project VLA/20A-239 (PI: Bartos). Each GRB was observed at nominal central frequencies of 6\,GHz (C-band) and 3\,GHz (S-band) with the VLA in its C configuration. The choice of frequency was motivated by the need to match the VLA field of view to the typical size of the error region of the \textit{Swift}/BAT GRBs in our sample. Moreover, late-time radio flares from BNS mergers are expected to be powered by optically thin radio emission \citep{Nakar2011}, thus lower frequencies offer better chances for discovery. On the other hand, the lower the frequency, the larger the expected number of unrelated sources in the field and the slower the time variability (hence, the harder the identification of false positives). Based on the above considerations, S- and/or C-band observations are overall advantageous compared to both L-band (nominal central frequency of 1.4\,GHz) observations, or observations at frequencies above 6\,GHz. 

\begin{table*}
\begin{center}\begin{footnotesize}
\caption{Candidate radio counterparts found within the BAT error regions of the GRBs in our sample. We list only radio detections that passed all criteria discussed in Section \ref{VLA_sources}. For each candidate, we report the sky position, the class of the closest cataloged source, the epoch of the VLA detection, the central frequency of the observation, measured peak flux density, the offset between the measured VLA position and the position of the closest cataloged source, the estimated VLA position error, and the compactness parameter of the radio emission (see Section \ref{VLA_sources} for discussion).}
\label{GRB_Table}
\setlength{\tabcolsep}{0.36\tabcolsep}
\begin{tabular}{ccccccccccc}
\hline
\hline
Source Name &  R.A.~~Dec. &  Class\footnote{NED classification of the closest cataloged object} & Epoch\footnote{The mid-time of our VLA observation.} & $\Delta T$\footnote{The time between the BAT trigger and the mid-time of our VLA observation (see the ``Epoch'' column).} & $\nu$\footnote{The central frequency of our VLA observation. } & $F_{\nu}$\footnote{The {\fontfamily{qcr}\selectfont imstat} peak flux density.} 
& Offset\footnote{The angular distance between the radio candidate and the closest known object in NED} & Pos. Err.\footnote{VLA position error calculated as described in Section \ref{VLA_sources}.} & Compactness\footnote{Compactness parameter calculated as described in Section \ref{VLA_sources}.}
\\
&  & & (MJD) & (yr) & (GHz)& ($\mu$Jy)
& (\arcsec) & (\arcsec) & 
\\
\hline
\hline
 $101129$A-Candidate-1
 & 10h23m45.15s $-17$d35m51.20s & IrS & 58955.07 & 9.4 & 6.1 & $139 \pm 11$ & .84 & .25 & $1.12 \pm .12$
 \\
 &'' & '' & 58994.14 & 9.5 & 2.9 & $372 \pm 29$ & '' & .48 & $1.25 \pm .12$
 \\
 \hline
 $111126$A-Candidate-1
 & 18h24m20.56s $+51$d28m44.71s & IrS & 58962.53 & 8.4 & 6.0 & $105.2 \pm 7.5$ & 18 & .13 & $1.07 \pm .10$
 \\
 & '' & '' & 58992.38 & 8.5 & 3.0 & $194 \pm 12$ &'' & .20 & $1.22 \pm .11$
 \\
 \hline
 $111126$A-Candidate-2
 & 18h23m55.44s $+51$d28m08.04s & IrS & 58962.53 & 8.4 & 6.0 & $84.7 \pm 8.4$ & 12 & .18 & $1.24 \pm .17$
 \\
 & '' & '' & 58992.38 & 8.5 & 3.0 & $210 \pm 13$ & '' & .20 & $1.133 \pm .099$
 \\
 \hline
\end{tabular}
\end{footnotesize}\end{center}\end{table*}

The VLA data were calibrated using the automated VLA calibration pipeline in {\fontfamily{qcr}\selectfont CASA}\footnote{\url{https://casa.nrao.edu/index.shtml}}.
The calibrated data were then inspected manually for additional flagging, especially to mitigate RFI.  The {\fontfamily{qcr}\selectfont CASA} task {\fontfamily{qcr}\selectfont tclean} was used in interactive mode to image the observed fields, with a robust parameter of $0.5$ and Briggs weighting. Sources were identified in the cleaned images both by hand and by using Blobcat, a Python program designed to locate sources in an image, to ensure all sources were investigated \citep{Hales2012a,  Hales2012b}.

The central noise RMS of each image was determined using {\fontfamily{qcr}\selectfont imstat} by measuring the RMS of the residual image within a central circular region of radius 10 times the FWHM of the nominal synthesized beam. The efficiency-corrected RMS was then calculated by dividing this central RMS by the efficiency (primary beam correction) as measured at the location of each candidate. Our results are summarized in Table \ref{Grandorf_Table}.

\section{Candidate radio counterpart identification}
\label{VLA_sources}
We visually inspected all of our calibrated images to search for candidate radio sources. Most sources identified via visual inspection were also confirmed by running Blobcat. We used the {\fontfamily{qcr}\selectfont CASA} task {\fontfamily{qcr}\selectfont imfit} to estimate the VLA position errors. Specifically, errors were calculated by dividing the clean beam semi-major axis, as measured using  {\fontfamily{qcr}\selectfont imfit}, by the signal-to-noise ratio (SNR) of a given source. Following \citet{Mooley2013}, only radio sources found within the BAT error region of each GRB with ${\rm SNR}\gtrsim 7$ were considered reliable detections. Hereafter, the SNR is defined as the ratio between the source peak flux density, and the peak flux density error calculated as described in what follows.

The peak flux density of each radio source was measured using {\fontfamily{qcr}\selectfont imstat} and a circular region centered on the source, with radius equal to the FWHM of the nominal synthesized beam. The peak flux density error was determined by adding the efficiency-corrected RMS in quadrature with the absolute flux calibration error. The last is estimated to be 5\% of the peak flux density for observations using 3C\,286 as absolute flux calibrator, and 10\% of the peak flux density if the flux calibrator was 3C\,48 (due to a recent flare). 

For each of the radio sources with SNR$\gtrsim 7$, we further used the {\fontfamily{qcr}\selectfont CASA} task {\fontfamily{qcr}\selectfont imfit} to calculate the integrated flux density. The integrated flux density error was determined by adding the error of the integrated flux density as reported by {\fontfamily{qcr}\selectfont imfit} and 5\% (or 10\%) of the {\fontfamily{qcr}\selectfont imfit} integrated flux density in quadrature, similarly to what done for the peak flux density. 

Next, we derive the compactness parameter by dividing the integrated flux by the {\fontfamily{qcr}\selectfont imstat} peak flux density. 
We used the compactness parameter to determine whether each radio source is more likely to be a compact object, like a merger, or an extended object, like a galaxy \citep{Itoh2020}. We take $0.9 < C < 1.5$ as the range of compactness values for a point-like source \citep{Mooley2013}. To ensure point-like morphology and avoid contamination from sidelobes,  following \citet{Mooley2013}, we also require that the size of the detected radio source, as reported by {\fontfamily{qcr}\selectfont imfit}, is smaller than $1.5\times$ the size of the clean beam FWHM on corresponding axes and that the source is located at a distance $\gtrsim 20\times$ the geometric mean of the clean beam from sources with peak flux density $\gtrsim 500\,\mu$Jy and extended sources. We require all of the above conditions to be met at both 3\,GHz and 6\,GHz. This helps ensure that the 3\,GHz and 6\,GHz flux are both likely to be dominated by the same emission process, and that the spectral indices $\beta$ (where $\beta$ is defined so that $f_{\nu}\propto \nu^{\beta}$) can be reliably estimated.

Finally, we narrowed down our remaining candidate radio counterparts by checking for the existence of previously cataloged sources at their location \citep{Grandorf2021,Mooley2013}. We consulted 
%the NASA/IPAC Extragalactic Database \footnote{\url{https://ned.ipac.caltech.edu/}} \citep[NED;][]{Helou1995},
NED \citep{NED2019},
the VLA FIRST catalog \citep{Becker1994}, the VLASS quick look image repository \citep[via CIRADA;][]{Lacy2020}, and the \textit{Chandra} Source Catalog 2.0 \citep{Evans2010}. NED notably queries NVSS among other catalogs. If a source we observed with the VLA was found to be within 2\arcsec \ of a cataloged \textit{radio} source detected in catalog images taken before the GRB trigger time, it was discarded from further analysis (as this indicates that the radio source is likely persistent, not a transient).

The results of this selection process are reported in Table \ref{GRB_Table}. As evident from this Table, only 2 of the 7 GRBs in our sample are associated with candidate radio counterparts that pass all of the selection criteria described here. In the following Section, we discuss each GRB in more detail.

\section{Results}
\label{sec:results}
After observing the fields of the GRBs in our sample (Section \ref{GRBsample}) with the VLA, we calibrated our observations as described in Section \ref{VLA_Obs} and selected candidate radio counterparts as described in Section \ref{VLA_sources}. In what follows, we detail our analysis for each of the GRBs in our sample.
\subsection{GRB\,080121A}
\label{GRB080121A}
GRB\,080121A triggered the \textit{Swift}/BAT at 21:29:55\,UT on 2008 January 21 \citep{Cummings2008}. It was detected at a refined, ground-based location of $\alpha=09$h09m01.8s and $\delta=+41$d50m21.3s (J2000), with an error region of radius 2.48\arcmin ~with 90\% confidence\footnote{\label{GSFC GRB080121A} \url{swift.gsfc.nasa.gov/results/batgrbcat/GRB080121/web/GRB080121.html}}. We observed the field of this GRB at 6\,GHz with the VLA for an hour (total time including calibration and overhead), starting at 23:04:52.0\,UT on 2020 April 17. Our 3\,GHz observations of the same field started at 18:00:54 UT on 2020 May 22, and lasted about an hour. 

The central noise RMS we measure for this field is $\approx 39\,\mu$Jy at 6\,GHz, and $\approx 110\,\mu$Jy at 3\,GHz (Table \ref{Grandorf_Table}). The relatively high RMS is due to a very bright source outside the BAT error region, which resulted in a limited dynamic range. None of the radio sources identified within the BAT error region passed the criteria described in Section \ref{VLA_Obs}. However, we note that this GRB was investigated further in \citet{Dichiara2020} and \citet{Ricci2021} due to its proximity to possible host galaxies, both of which are within 200 Mpc.

\subsection{GRB\,090417A}
\label{GRB090417A}
GRB\,090417A triggered the \textit{Swift}/BAT at 13:17:23\,UT on 2009 April 17 \citep{Mangano2009}. It was detected at a refined, ground-based location of $\alpha=02$h19m58.3s $\delta=-07$d08m28.9s (J2000) with an uncertainty of 2.8\arcmin and 90\% containment \citep{Baumgartner2009}. We observed the field of GRB\,090417A with the VLA for one hour total time, starting at 20:21:37\,UT on 2020 April 14 at 6\,GHz and at 19:00:42\,UT on 2022 May 23. The central noise RMS we measure for this field is $\approx 9.1\,\mu$Jy at 6\,GHz, and $\approx 16\,\mu$Jy at 3\,GHz (Table \ref{Grandorf_Table}). No radio sources were detected within the BAT error region of this GRB.

\subsection{GRB\,101129A}
\label{GRB101129A}
GRB\,101129A triggered the \textit{Swift}/BAT at 15:39:32\,UT on 2010 November 29\footnote{\label{GSFC GRB101129A} \url{swift.gsfc.nasa.gov/results/batgrbcat/GRB101129A/web/ GRB101129A.html}}. It was detected at a refined, ground-based location of $\alpha=10$h23m41.0s, $\delta=-17$d38m42.0s (J2000), with an uncertainty of 3.1\arcmin ~and 90\% containment. The VLA observed the GRB field for an hour, including calibrations, starting at 01:11:57.0 UTC on April 16th, 2020 for 6 GHz and starting at 02:49:57 UT on 2020 May 25 at 3\,GHz. The central noise RMS we measure for this field is $\approx 5.7\mu$Jy at 6\,GHz and $\approx 21\,\mu$Jy at 3\,GHz. Several radio sources were identified in the images collected for this field, but only one candidate radio counterpart passed the selection criteria described in Section \ref{VLA_Obs} (Table \ref{GRB_Table}).

\subsection{GRB\,111126A}
\label{GRB111126A}
GRB\,111126A triggered the \textit{Swift}/BAT at 18:57:42\,UT on 2011 November 26 \citep{Cummings2011}. It was detected at a refined, ground-based location of $\alpha = 18$h24m07.1s and $\delta=+51$d28m06.1s (J2000) with an uncertainty of 2.5\arcmin \footnote{\label{GSFC GRB111126A} \url{swift.gsfc.nasa.gov/results/batgrbcat/GRB111126A/web/GRB111126A.html}}. We observed the field of this GRB for one hour (total time), starting at 12:12:02 UT on 2020 April 23 at 6\,GHz  and at 08:32:43.0 UT on 2020 May 23 at 3\,GHz. The central noise RMS we measure for this field is $\approx 4.8\,\mu$Jy and $\approx 7.4\,\mu$Jy at 6\,GHz and 3\,GHz, respectively. Several radio sources were identified within the BAT error circle, but only two candidate radio counterparts passed the selection criteria described in Section \ref{VLA_Obs} (Table \ref{GRB_Table}).

\subsection{GRB\,120403A}
\label{GRB120403A}
GRB120403A triggered Swift BAT at 01:05:23 UT on April 3rd, 2012 \citep{Beardmore2012}. It was located at 02h49m49.8s +40d29m21.8s (J2000) with an uncertainty of 2.3 arcminutes and 90\% containment \citep{Sakamoto2012}. The VLA observed GRB120403A for an hour, including calibrations, starting at 00:03:17.0\,UT on April 16th, 2020 for 6 GHz and starting at 11:37:19.0\,UT on May 23rd, 2020 for 3 GHz. The central noise RMS we measure for this field is $\approx 5.2\,\mu$Jy and $\approx 15\,\mu$Jy at 6\,GHz and 3\,GHz, respectively (Table \ref{Grandorf_Table}). No radio sources were detected within the BAT error region of this GRB at either of the observed radio frequencies. 

\subsection{GRB\,140606A}
\label{GRB140606A}
GRB\,140606A triggered the \textit{Swift}/BAT at 10:58:13\,UT on 2014 June 6 \citep{Stroh2014}. It was located at $\alpha=13$h27m11.7s and $\delta=+37$d35m56.5s (J2000), with an uncertainty of 2.4\arcmin and 90\% containment \citep{Cummings2014}. We observed the field of this GRB with the VLA for one hour (total time) at both  6\,GHz and 3\,GHz, starting at 01:52:38\,UT on 2020 April 25 and at 7:44:30\,UT on 2020 May 29, respectively. The detected central RMS values are 5.5 $\mu$Jy for 6 GHz and 9.7 $\mu$Jy for 3 GHz (Table \ref{Grandorf_Table}). Several radio sources were identified within the BAT error circle of this GRB, but none passed the selection criteria described in Section \ref{VLA_Obs}.

\subsection{GRB\,160726A}
\label{GRB160726A}
GRB\,160726A triggered the \textit{Swift}/BAT at 01:34:07.67 UT on 2016 July 26\footnote{\label{GSFC GRB160726A} \url{swift.gsfc.nasa.gov/results/batgrbcat/GRB160726A/web/GRB160726A.html}}. It was located at $\alpha=06$h35m14.3s and $\delta=-06$d37m1.4s (J2000) with an uncertainty of 1.3\arcmin \ (90\% containment). We observed the field of this GRB with the VLA for one hour (total time) in each frequency, starting at 20:20:51.0\,UT on 2020 April 18 at 6\,GHz and at 17:55:04 UT on 2020 May 24 at 6\,GHz. The central noise RMS we measure for this field is $\approx 7.9\,\mu$Jy at 6\,GHz, and $\approx 13.4\,\mu$Jy at 3\,GHz. No radio source passed the selection criteria described in Section \ref{VLA_Obs} (Table \ref{GRB_Table}).

\section{Discussion}
\label{Results}
\subsection{Spectral energy distribution}
Using our observations at 3\,GHz and 6\,GHz, we compute the spectral indices of the three candidate radio counterparts listed in Table \ref{GRB_Table}. We find $\beta_1 = -1.32 \pm 0.15$, $\beta_2 = -0.88 \pm 0.14$, and $\beta_3 = -1.31 \pm 0.17$ for the radio candidates 101129A-Candidate-1, 111126A-Candidate-1, and 111126A-Candidate-2, respectively. While all three of these spectral indices are compatible with optically-thin synchrotron emission, as expected from radio from BNS mergers \citep{Nakar2011}, they are also compatible (within errors) with radio emission from star formation  \citep[for which typical spectral indices are $-1.1 \lesssim \beta \lesssim -0.4$;][]{Seymour2008}. We note that the spectral index of GW170817 was measured to be  $\beta= -0.584 \pm 0.002$ \citep{Makhathini2021}. Therefore, candidate  101129A-Candidate-1 with $\beta_2\approx -0.88$ is the closest to GW170817, though our results all suggest steeper spectral indices. None of the spectral index values we derive for our candidate radio counterparts are suggestive of emission from flat-spectrum AGN \citep[$\beta > -0.6$;][]{Itoh2020}.

\subsection{Contamination from unrelated (persistent or variable) radio sources}
In this Section, we discuss the likelihood that the candidate radio counterparts we have identified in the GRB error regions considered in this work are false positives i.e., radio sources whose origin is unrelated to the GRB itself.

Using the spectral index of each candidate counterpart, we estimate 1.4\,GHx flux densities in the range $\approx 0.4-1$\,mJy. According to \cite{Mooley2013} and \citet{Huynh2005}, the average number of persistent radio sources (of any morphology) with a 1.4\,GHz flux density above 0.1 mJy is $\approx 910$ deg$^{-2}$. Hence, within the BAT error regions for GRB\,101129A and GRB\,111126A we would expect $\lesssim 5 - 8$ unrelated persistent radio sources for observations conducted at $\gtrsim 1.4$\,GHz (having also applied further cuts on their morphology). As discussed in \cite{Mooley2013}, about $\approx 1\%$ of unresolved radio sources above $40\mu$Jy at 1.4\,GHz are variable at the $4\sigma$ level. Hence, we would expect an average of $\lesssim  0.05 - 0.08$ \textit{variable} unrelated radio sources in the error regions of our GRBs \footnote{We note that in \cite{Grandorf2021} the variable source range given should read $\approx 0.026-0.05$ rather than $\approx 0.26-0.5$. Therefore, the average number of unrelated variable radio sources expected for the GRBs in our sample comparable to that found in \citet{Grandorf2021}.}. The Poisson probability of finding one or more unrelated variable sources would then be $\lesssim 5\%-8\%$, which is sufficiently low to motivate further follow-up studies in the radio aimed at establishing the level of time variability of our radio candidates. 

\subsection{AGN or star-formation origin?}
Possible explanations for origin of the radio candidates identified in our search are star-formation in unresolved galaxies and radio emission from AGNs \citep[e.g.][]{Condon1992, Sadler1999, Smol2008, Palliyaguru2016, Padovani2017}. At 1.4\,GHz, star-forming galaxies dominate at lower fluxes (below $\approx 200 \mathrm{\mu Jy}$), whereas AGNs dominate at higher fluxes \citep[1\,mJy and above;][]{Padovani2017,Sadler1999,Smol2017}.

To test whether our radio candidates can be related to star formation in unresolved galaxies, we first consider the constraints arising from the fact that these radio sources have point-like morphologies in our images. Short GRB host galaxies in the cosmological sample have effective radii in the range $0.2\arcsec - 1.2\arcsec$ with a median size of $0.36\arcsec$ \citep{Fong2013b}. For the short GRBs
with known redshifts, the median physical size is $\approx 3.6$\,kpc \citep{Fong2013b}. Hence, if we considered a short-GRB-like host galaxy located at $\approx 200$\,Mpc, its angular radius would be of $\approx 3.7\arcsec$. Our radio candidates have an angular size of $\lesssim 3.5\arcsec$ at 6\,GHz (in C band) due to the requirement we imposed on their morphology and considering the nominal FWHM of the VLA synthesized beam in C configuration. Therefore, if our radio candidates are related to star formation in an unresolved host galaxy, most likely such a galaxy would be located at $\gtrsim 200$\,Mpc.

Radio emission associated with star formation at GHz frequencies is dominated by synchrotron radiation from electrons accelerated by supernovae. The following relation can be used to estimate the star formation rate (SFR) in the galaxy given the measured luminosity at 1.4\,GHz \citep{Murphy2011, Perley2013}:
\begin{equation}
    \Big( \frac{\mathrm{SFR}}{\mathrm{M_{\odot} yr^{-1}}} \Big) = 6.35\times10^{-29} \Big( \frac{L_{1.4 \mathrm{GHz}}}{\mathrm{erg~ s^{-1} Hz^{-1}}} \Big).
\end{equation}
Assuming the candidate radio counterparts listed in Table \ref{GRB_Table} are located at a distance of 200\,Mpc, their estimated flux densities of $\lesssim 0.4 - 1$\,mJy at 1.4\,GHz imply $L_{\mathrm{1.4 GHz}}\lesssim (1.8 - 4.6)\times10^{28}$\,erg\,s$^{-1}$\,Hz$^{-1}$ and, in turn, a SFR rate of $\lesssim (1.2 - 2.9)$ $\mathrm{M_{\odot} yr^{-1}}$. This is compatible with normal galaxies and with cosmological short GRB host galaxies  \citep[$0.2-6$\,M$_{\odot}$\,yr$^{-1}$;][]{Berger2009}. 

On the other hand, if we assume that our candidate radio counterparts are located at the median short GRB redshift of $z =0.72$ \citep{Berger2014}, then the measured fluxes would correspond to radio luminosities in the range $(5.5-14) \times10^{30}$\,erg\,$\mathrm{s^{-1} Hz^{-1}}$ at 1.4 GHz, favoring an AGN origin. 

The AGN scenario can be further tested using AllWISE \citep{AllWISE2019} color information.
%(accessible via the Infrared Science Archive\footnote{\url{https://irsa.ipac.caltech.edu/cgi-bin/Gator/nph-scan?submit=Select&projshort=WISE}}). 
Indeed, AGNs are expected to fall in the wedge defined by the conditions $W2 - W3 > 2.517$ and $W1 - W2 > 0.315\times(W2 - W3) - 0.222$ where $W1$, $W2$, and $W3$ are the instrumental profile-fit photometry magnitudes\footnote{\url{https://wise2.ipac.caltech.edu/docs/release/allwise/expsup/sec2_1a.html\#w1mpro}} in bands one ($3.4\,\mu $m), two ($4.6\,\mu$m), and three \citep[$12\,\mu $m;][]{Wright2010, Mateos2012, Gurkin2014}. The 101129A-Candidate-1 candidate found in the error region of GRB\,101129A has $W1 - W2 = 0.10 \pm .29$ and $W2 - W3 = 3.72 \pm .30$, fulfilling only the first condition of the AGN wedge. Therefore, a SFR origin is more likely. The candidate radio counterparts found in the error region of GRB\,111126A did not have any color information available in the AllWISE Source Catalog. 

%\begin{figure}
%\includegraphics[width=8.5cm]{40GRB111126A-Candidate-1.png}
%\caption{Model light curves derived assuming a two-component, GW170817-like ejecta model and a source located at 40\,Mpc, compared with the data for GRB111126A-Candidate-1. See Section \ref{sec:BNS} for discussion. }
%\label{fig:models_light_40}
%end{figure}

%\begin{figure}
%\includegraphics[width=8.5cm]{200GRB111126A-Candidate-1.png}
%\caption{Same as Figure \ref{fig:models_light_40} but for a source located at 200\,Mpc.}
%\label{fig:models_light_200}
%\end{figure}

\begin{figure}
    \centering
    \includegraphics[width = 8.5cm]{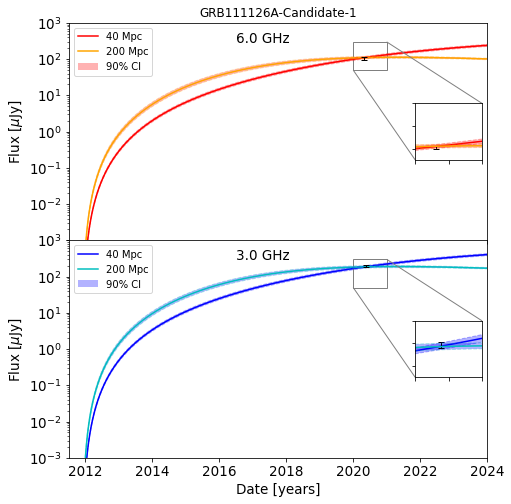}
    \caption{Model light curves derived assuming a two-component, GW170817-like ejecta model and a source located at 40\,Mpc (red and dark blue lines) or 200 Mpc (orange and light blue lines) compared with the data for GRB111126A-Candidate-1 (black bracket). We note the flatter behavior predicted for the light curve best fit at 200\,Mpc compared to the one at 40\,Mpc. Hence, we expect future (after 2022) radio observations to be able to probe more significant radio flux changes if the candidate is at 40\,Mpc. See Section \ref{sec:BNS} for discussion.}
    \label{fig:models_light}
\end{figure}

\begin{figure}
\vbox{\includegraphics[width=8.5cm]{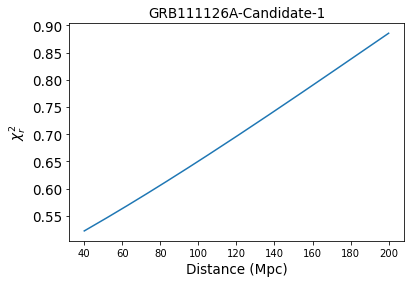}
\includegraphics[width=8.5cm]{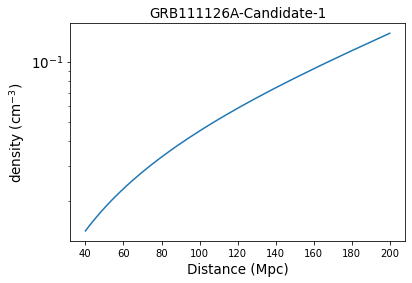}}
\caption{Within the BNS merger ejecta model described in Section \ref{sec:BNS}, we have two free parameters: the distance to the source and the interstellar medium (ISM) number density. Because we only have two observations per candidate radio counterpart,  we fix the distance to a grid of values in between 40\,Mpc and 200\,Mpc and fit for the ISM density. The top panel shows the resulting $\chi2$ values for the fit, and the bottom plot shows the best fit ISM density values.}
\label{fig:models}
\end{figure}

\begin{figure}
\includegraphics[width=8.5cm]{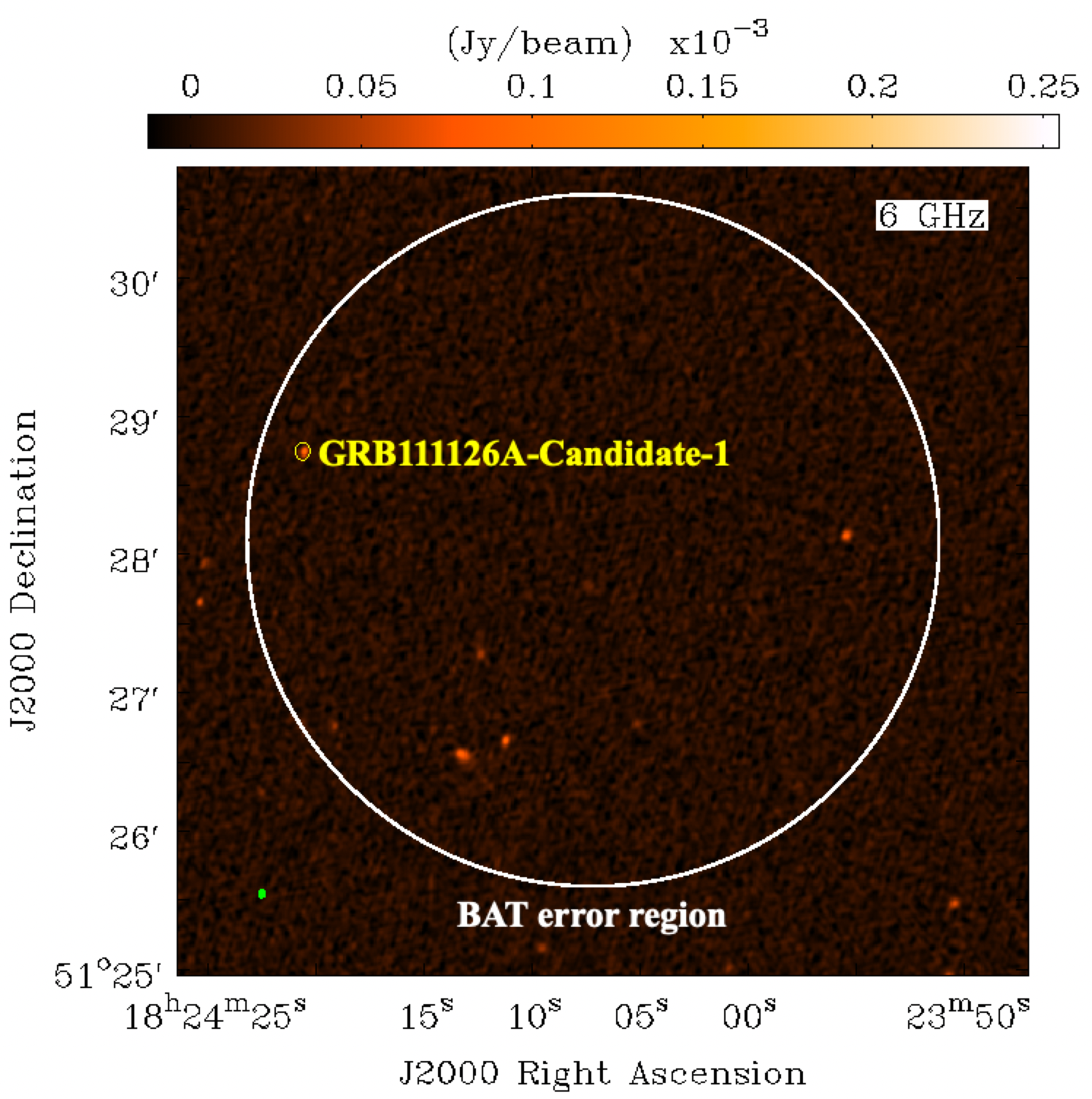}
\caption{VLA field containing the BAT error region for GRB\,11126A (marked with a white circle of radius 2.5\arcmin). The yellow circle is centered on the location of the radio counterpart GRB\,11126A-Candidate-1 and has a radius equal to the FWHM of the nominal VLA synthesized beam at 6\,GHz (3\arcsec). The actual beam is shown as a filled, green ellipse in the bottom-left corner of the image. }
\label{fig:discovery}
\end{figure}

\subsection{BNS Merger Origin}
\label{sec:BNS}
To test the possibility that the radio candidates identified in our analysis are related to radio counterparts of BNS mergers powering the corresponding GRBs, similarly to what done in \citet{Grandorf2021}, we fit $3-6$\,GHz model light curves to our flux density measurements. These light curves are derived assuming a two-component, GW170817-like ejecta model with masses of $0.04$\,M$_{\odot}$ and
$0.01$\,M$_{\odot}$ and respective speeds of $0.1 c$ and $0.3 c$. We set the fractions $\epsilon_e$ and $\epsilon_B$ of energy going into accelerated electrons and magnetic fields, respectively,  equal to their fiducial values of 10\% each. With these model assumptions, we have two free parameters left: the distance to the source and the  interstellar medium (ISM) number density. Because we only have two observations per candidate radio counterpart,  for each of these we fix the distance to a grid of values in between 40\,Mpc and 200\,Mpc and fit for the ISM density. Overall, our fits on the observed radio flux for GRB\,101129A-Candidate-1 and for GRB\,111126A-Candidate-2 return $\chi^2 > 10$. Allowing for $\epsilon_e$ and $\epsilon_B$ to vary in the range $10^{-4}$--$10^{-1}$ returns best fits with ISM densities $\gtrsim 10$\,cm$^{-3}$. For a more realistic range of $0.01$--$0.1$ cm$^{-3}$ \citep[given that $\sim 80$--$95\%$ of short GRBs have densities $<1$\,cm$^{-3}$; ][]{Fong2015}, we get $\chi^2 > 10$. Therefore, we rule out these candidates as promising radio counterparts for late-time radio flares. We obtain better results for GRB\,111126A-Candidate-1, which we show in Figures \ref{fig:models_light}-\ref{fig:models}. All of the ISM density values we determined for GRB\,111126A-Candidate-1 as a function of distance are plausible values within a BNS merger scenario (see Figure \ref{fig:models}). Indeed, short GRBs have been associated with ISM media with densities ranging from $10^{-4}$\,cm$^{-3}$  to $1$\,cm$^{-3}$ \citep{Fong2013a}, and GW170817 had a circum-burst density of $\lesssim 0.03$\,cm$^{-3}$ \citep[e.g.][]{Hallinan2017,Lazzati2018,Makhathini2021}. 

The 6 GHz discovery image of GRB111126A Candidate 1 is shown in Figure \ref{fig:discovery}. If the hypothetical merger behind GRB111126A Candidate 1 is nearby, around 40 Mpc, our best fit model predicts rising light curves at 3--6\,GHz  (Figure \ref{fig:models_light}, red and blue curves) and $7-10 \sigma$ flux variations throughout 2023 and 2024 (compared to our observations in 2020). However, if the merger is farther away, say at 200\,Mpc, the best fit light curves at 3--6\,GHz flatten at late times and we would not expect to see significant flux variations in 2023--2024 compared to our 2020 observations (Figure \ref{fig:models_light}, orange and light blue curves). Hence, re-observing this candidate in 2024 would be a promising strategy for further constraining its nature.

\section{Summary and conclusion}
\label{conclusion}
We used the VLA to observe 7 GRBs without accurate X-ray localizations to identify potential nearby, GW170817-like events.  In the 7 observed fields, we find a total of 3 candidate radio counterparts (one in the error region of GRB\,101129A and two in the error region of GRB\,111126A) passing all our cuts. We have discussed these findings in the context of expectations for false positives, as well as in the context of BNS late-time radio flare models. Overall, one of the radio candidate counterparts found in the error region of GRB\,111126A, GRB 111126A-Candidate-1, appears worthy of further follow-up in the radio. Indeed, detecting radio variability at the level of $\approx 7-10\,\sigma$ appears possible over the next couple of years (at epochs of about $12-13$ years since the GRB). The existence of time variability in the radio would significantly decrease the odds of a false positive origin.

We conclude by stressing that, if 10\% of short GRBs in the known sample of events lacking a redshift measurement is located within 200\,Mpc \citep{Gupte2018}, searches like the one described here and in \citet{Grandorf2021} could quickly become profitable: following up with the VLA a sample of $\gtrsim 23$ short GRBs should result in a $\gtrsim$90\% chance of finding at least one nearby, GW170817-like event. So far with this work and with the observations presented in \citet{Grandorf2021}, we have collected about half of the necessary sample size using only filler-mode observations with the VLA and with a relatively small use of observing resources. With a higher observing priority, we could cover the whole sample at a faster pace. This would also enhance chances for well-timed  follow-up observations aimed at testing expectations for radio flux variability (that could be critical to help discern the nature of any potential radio candidate). Overall, we envision this type of observational programs to continue in the future.

\begin{acknowledgements}
\small A.E. and A.C. gratefully acknowledge support from the National Science Foundation via grant \#1907975. The National Radio Astronomy Observatory is a facility of the National Science Foundation operated under cooperative agreement by Associated Universities, Inc. The authors thank the University of Florida and Columbia University in the City of New York for their generous support. I.B. acknowledges the support of the National Science Foundation under grants PHY-1911796 and PHY-2110060, and the Alfred P. Sloan Foundation. The Columbia Experimental Gravity group acknowledges the support of the support of the National Science Foundation under grant PHY-2012035.

\end{acknowledgements}
\bibliography{bibliography}

\begin{thebibliography}{}
\expandafter\ifx\csname natexlab\endcsname\relax\def\natexlab#1{#1}\fi
\providecommand{\url}[1]{\href{#1}{#1}}
\providecommand{\dodoi}[1]{doi:~\href{http://doi.org/#1}{\nolinkurl{#1}}}
\providecommand{\doeprint}[1]{\href{http://ascl.net/#1}{\nolinkurl{http://ascl.net/#1}}}
\providecommand{\doarXiv}[1]{\href{https://arxiv.org/abs/#1}{\nolinkurl{https://arxiv.org/abs/#1}}}

\bibitem[{{Abbott} {et~al.}(2017{\natexlab{a}}){Abbott}, {Abbott}, {Abbott},
  {Acernese}, {Ackley}, {Adams}, {Adams}, {Addesso}, {Adhikari}, {Adya},
  {Affeldt}, {Afrough}, {Agarwal}, {Agathos}, {Agatsuma}, {Aggarwal}, {Aguiar},
  {Aiello}, {Ain}, {Ajith}, {Allen}, {Allen}, {Allocca}, {Aloy}, {Altin},
  {Amato}, {Ananyeva}, {Anderson}, {Anderson}, {Angelova}, {Antier}, {Appert},
  {Arai}, {Araya}, {Areeda}, {Arnaud}, {Arun}, {Ascenzi}, {Ashton}, {Ast},
  {Aston}, {Astone}, {Atallah}, {Aufmuth}, {Aulbert}, {AultONeal}, {Austin},
  {Avila-Alvarez}, {Babak}, {Bacon}, {Bader}, {Bae}, {Baker}, {Baldaccini},
  {Ballardin}, {Ballmer}, {Banagiri}, {Barayoga}, {Barclay}, {Barish},
  {Barker}, {Barkett}, {Barone}, {Barr}, {Barsotti}, {Barsuglia}, {Barta},
  {Bartlett}, {Bartos}, {Bassiri}, {Basti}, {Batch}, {Bawaj}, {Bayley},
  {Bazzan}, {B{\'e}csy}, {Beer}, {Bejger}, {Belahcene}, {Bell}, {Berger},
  {Bergmann}, {Bero}, {Berry}, {Bersanetti}, {Bertolini}, {Betzwieser},
  {Bhagwat}, {Bhandare}, {Bilenko}, {Billingsley}, {Billman}, {Birch},
  {Birney}, {Birnholtz}, {Biscans}, {Biscoveanu}, {Bisht}, {Bitossi}, {Biwer},
  {Bizouard}, {Blackburn}, {Blackman}, {Blair}, {Blair}, {Blair}, {Bloemen},
  {Bock}, {Bode}, {Boer}, {Bogaert}, {Bohe}, {Bondu}, {Bonilla}, {Bonnand},
  {Boom}, {Bork}, {Boschi}, {Bose}, {Bossie}, {Bouffanais}, {Bozzi},
  {Bradaschia}, {Brady}, {Branchesi}, {Brau}, {Briant}, {Brillet}, {Brinkmann},
  {Brisson}, {Brockill}, {Broida}, {Brooks}, {Brown}, {Brown}, {Brunett},
  {Buchanan}, {Buikema}, {Bulik}, {Bulten}, {Buonanno}, {Buskulic}, {Buy},
  {Byer}, {Cabero}, {Cadonati}, {Cagnoli}, {Cahillane}, {Calder{\'o}n
  Bustillo}, {Callister}, {Calloni}, {Camp}, {Canepa}, {Canizares}, {Cannon},
  {Cao}, {Cao}, {Capano}, {Capocasa}, {Carbognani}, {Caride}, {Carney},
  {Casanueva Diaz}, {Casentini}, {Caudill}, {Cavagli{\`a}}, {Cavalier},
  {Cavalieri}, {Cella}, {Cepeda}, {Cerd{\'a}-Dur{\'a}n}, {Cerretani},
  {Cesarini}, {Chamberlin}, {Chan}, {Chao}, {Charlton}, {Chase},
  {Chassande-Mottin}, {Chatterjee}, {Chatziioannou}, {Cheeseboro}, {Chen},
  {Chen}, {Chen}, {Cheng}, {Chia}, {Chincarini}, {Chiummo}, {Chmiel}, {Cho},
  {Cho}, {Chow}, {Christensen}, {Chu}, {Chua}, {Chua}, {Chung}, {Chung},
  {Ciani}, {Ciolfi}, {Cirelli}, {Cirone}, {Clara}, {Clark}, {Clearwater},
  {Cleva}, {Cocchieri}, {Coccia}, {Cohadon}, {Cohen}, {Colla}, {Collette},
  {Cominsky}, {Constancio}, {Conti}, {Cooper}, {Corban}, {Corbitt},
  {Cordero-Carri{\'o}n}, {Corley}, {Cornish}, {Corsi}, {Cortese}, {Costa},
  {Coughlin}, {Coughlin}, {Coulon}, {Countryman}, {Couvares}, {Covas}, {Cowan},
  {Coward}, {Cowart}, {Coyne}, {Coyne}, {Creighton}, {Creighton}, {Cripe},
  {Crowder}, {Cullen}, {Cumming}, {Cunningham}, {Cuoco}, {Dal Canton},
  {D{\'a}lya}, {Danilishin}, {D'Antonio}, {Danzmann}, {Dasgupta}, {Da Silva
  Costa}, {Dattilo}, {Dave}, {Davier}, {Davis}, {Daw}, {Day}, {De}, {DeBra},
  {Degallaix}, {De Laurentis}, {Del{\'e}glise}, {Del Pozzo}, {Demos}, {Denker},
  {Dent}, {De Pietri}, {Dergachev}, {De Rosa}, {DeRosa}, {De Rossi}, {DeSalvo},
  {de Varona}, {Devenson}, {Dhurandhar}, {D{\'\i}az}, {Di Fiore}, {Di
  Giovanni}, {Di Girolamo}, {Di Lieto}, {Di Pace}, {Di Palma}, {Di Renzo},
  {Doctor}, {Dolique}, {Donovan}, {Dooley}, {Doravari}, {Dorrington},
  {Douglas}, {Dovale {\'A}lvarez}, {Downes}, {Drago}, {Dreissigacker},
  {Driggers}, {Du}, {Ducrot}, {Dupej}, {Dwyer}, {Edo}, {Edwards}, {Effler},
  {Eggenstein}, {Ehrens}, {Eichholz}, {Eikenberry}, {Eisenstein}, {Essick},
  {Estevez}, {Etienne}, {Etzel}, {Evans}, {Evans}, {Factourovich}, {Fafone},
  {Fair}, {Fairhurst}, {Fan}, {Farinon}, {Farr}, {Farr}, {Fauchon-Jones},
  {Favata}, {Fays}, {Fee}, {Fehrmann}, {Feicht}, {Fejer}, {Fernandez-Galiana},
  {Ferrante}, {Ferreira}, {Ferrini}, {Fidecaro}, {Finstad}, {Fiori},
  {Fiorucci}, {Fishbach}, {Fisher}, {Fitz-Axen}, {Flaminio}, {Fletcher},
  {Fong}, {Font}, {Forsyth}, {Forsyth}, {Fournier}, {Frasca}, {Frasconi},
  {Frei}, {Freise}, {Frey}, {Frey}, {Fries}, {Fritschel}, {Frolov}, {Fulda},
  {Fyffe}, {Gabbard}, {Gadre}, {Gaebel}, {Gair}, {Gammaitoni}, {Ganija},
  {Gaonkar}, {Garcia-Quiros}, {Garufi}, {Gateley}, {Gaudio}, {Gaur},
  {Gayathri}, {Gehrels}, {Gemme}, {Genin}, {Gennai}, {George}, {George},
  {Gergely}, {Germain}, {Ghonge}, {Ghosh}, {Ghosh}, {Ghosh}, {Giaime},
  {Giardina}, {Giazotto}, {Gill}, {Glover}, {Goetz}, {Goetz}, {Gomes},
  {Goncharov}, {Gonz{\'a}lez}, {Gonzalez Castro}, {Gopakumar}, {Gorodetsky},
  {Gossan}, {Gosselin}, {Gouaty}, {Grado}, {Graef}, {Granata}, {Grant}, {Gras},
  {Gray}, {Greco}, {Green}, {Gretarsson}, {Groot}, {Grote}, {Grunewald},
  {Gruning}, {Guidi}, {Guo}, {Gupta}, {Gupta}, {Gushwa}, {Gustafson},
  {Gustafson}, {Halim}, {Hall}, {Hall}, {Hamilton}, {Hammond}, {Haney},
  {Hanke}, {Hanks}, {Hanna}, {Hannam}, {Hannuksela}, {Hanson}, {Hardwick},
  {Harms}, {Harry}, {Harry}, {Hart}, {Haster}, {Haughian}, {Healy}, {Heidmann},
  {Heintze}, {Heitmann}, {Hello}, {Hemming}, {Hendry}, {Heng}, {Hennig},
  {Heptonstall}, {Heurs}, {Hild}, {Hinderer}, {Hoak}, {Hofman}, {Holt}, {Holz},
  {Hopkins}, {Horst}, {Hough}, {Houston}, {Howell}, {Hreibi}, {Hu}, {Huerta},
  {Huet}, {Hughey}, {Husa}, {Huttner}, {Huynh-Dinh}, {Indik}, {Inta}, {Intini},
  {Isa}, {Isac}, {Isi}, {Iyer}, {Izumi}, {Jacqmin}, {Jani}, {Jaranowski},
  {Jawahar}, {Jim{\'e}nez-Forteza}, {Johnson}, {Johnson-McDaniel}, {Jones},
  {Jones}, {Jonker}, {Ju}, {Junker}, {Kalaghatgi}, {Kalogera}, {Kamai},
  {Kandhasamy}, {Kang}, {Kanner}, {Kapadia}, {Karki}, {Karvinen}, {Kasprzack},
  {Kastaun}, {Katolik}, {Katsavounidis}, {Katzman}, {Kaufer}, {Kawabe},
  {K{\'e}f{\'e}lian}, {Keitel}, {Kemball}, {Kennedy}, {Kent}, {Key}, {Khalili},
  {Khan}, {Khan}, {Khan}, {Khazanov}, {Kijbunchoo}, {Kim}, {Kim}, {Kim}, {Kim},
  {Kim}, {Kim}, {Kimbrell}, {King}, {King}, {Kinley-Hanlon}, {Kirchhoff},
  {Kissel}, {Kleybolte}, {Klimenko}, {Knowles}, {Koch}, {Koehlenbeck}, {Koley},
  {Kondrashov}, {Kontos}, {Korobko}, {Korth}, {Kowalska}, {Kozak},
  {Kr{\"a}mer}, {Kringel}, {Krishnan}, {Kr{\'o}lak}, {Kuehn}, {Kumar}, {Kumar},
  {Kumar}, {Kuo}, {Kutynia}, {Kwang}, {Lackey}, {Lai}, {Landry}, {Lang},
  {Lange}, {Lantz}, {Lanza}, {Lartaux-Vollard}, {Lasky}, {Laxen}, {Lazzarini},
  {Lazzaro}, {Leaci}, {Leavey}, {Lee}, {Lee}, {Lee}, {Lee}, {Lee}, {Lehmann},
  {Lenon}, {Leonardi}, {Leroy}, {Letendre}, {Levin}, {Li}, {Linker},
  {Littenberg}, {Liu}, {Lo}, {Lockerbie}, {London}, {Lord}, {Lorenzini},
  {Loriette}, {Lormand}, {Losurdo}, {Lough}, {Lousto}, {Lovelace}, {L{\"u}ck},
  {Lumaca}, {Lundgren}, {Lynch}, {Ma}, {Macas}, {Macfoy}, {Machenschalk},
  {MacInnis}, {Macleod}, {Maga{\~n}a Hernandez}, {Maga{\~n}a-Sandoval},
  {Maga{\~n}a Zertuche}, {Magee}, {Majorana}, {Maksimovic}, {Man}, {Mandic},
  {Mangano}, {Mansell}, {Manske}, {Mantovani}, {Marchesoni}, {Marion},
  {M{\'a}rka}, {M{\'a}rka}, {Markakis}, {Markosyan}, {Markowitz}, {Maros},
  {Marquina}, {Martelli}, {Martellini}, {Martin}, {Martin}, {Martynov},
  {Mason}, {Massera}, {Masserot}, {Massinger}, {Masso-Reid}, {Mastrogiovanni},
  {Matas}, {Matichard}, {Matone}, {Mavalvala}, {Mazumder}, {McCarthy},
  {McClelland}, {McCormick}, {McCuller}, {McGuire}, {McIntyre}, {McIver},
  {McManus}, {McNeill}, {McRae}, {McWilliams}, {Meacher}, {Meadors}, {Mehmet},
  {Meidam}, {Mejuto-Villa}, {Melatos}, {Mendell}, {Mercer}, {Merilh},
  {Merzougui}, {Meshkov}, {Messenger}, {Messick}, {Metzdorff}, {Meyers},
  {Miao}, {Michel}, {Middleton}, {Mikhailov}, {Milano}, {Miller}, {Miller},
  {Miller}, {Millhouse}, {Milovich-Goff}, {Minazzoli}, {Minenkov}, {Ming},
  {Mishra}, {Mitra}, {Mitrofanov}, {Mitselmakher}, {Mittleman}, {Moffa},
  {Moggi}, {Mogushi}, {Mohan}, {Mohapatra}, {Montani}, {Moore}, {Moraru},
  {Moreno}, {Morriss}, {Mours}, {Mow-Lowry}, {Mueller}, {Muir}, {Mukherjee},
  {Mukherjee}, {Mukherjee}, {Mukund}, {Mullavey}, {Munch}, {Mu{\~n}iz},
  {Muratore}, {Murray}, {Napier}, {Nardecchia}, {Naticchioni}, {Nayak},
  {Neilson}, {Nelemans}, {Nelson}, {Nery}, {Neunzert}, {Nevin}, {Newport},
  {Newton}, {Ng}, {Nguyen}, {Nichols}, {Nielsen}, {Nissanke}, {Nitz}, {Noack},
  {Nocera}, {Nolting}, {North}, {Nuttall}, {Oberling}, {O'Dea}, {Ogin}, {Oh},
  {Oh}, {Ohme}, {Okada}, {Oliver}, {Oppermann}, {Oram}, {O'Reilly}, {Ormiston},
  {Ortega}, {O'Shaughnessy}, {Ossokine}, {Ottaway}, {Overmier}, {Owen}, {Pace},
  {Page}, {Page}, {Pai}, {Pai}, {Palamos}, {Palashov}, {Palomba}, {Pal-Singh},
  {Pan}, {Pan}, {Pang}, {Pang}, {Pankow}, {Pannarale}, {Pant}, {Paoletti},
  {Paoli}, {Papa}, {Parida}, {Parker}, {Pascucci}, {Pasqualetti},
  {Passaquieti}, {Passuello}, {Patil}, {Patricelli}, {Pearlstone}, {Pedraza},
  {Pedurand}, {Pekowsky}, {Pele}, {Penn}, {Perez}, {Perreca}, {Perri},
  {Pfeiffer}, {Phelps}, {Piccinni}, {Pichot}, {Piergiovanni}, {Pierro},
  {Pillant}, {Pinard}, {Pinto}, {Pirello}, {Pitkin}, {Poe}, {Poggiani},
  {Popolizio}, {Porter}, {Post}, {Powell}, {Prasad}, {Pratt}, {Pratten},
  {Predoi}, {Prestegard}, {Prijatelj}, {Principe}, {Privitera}, {Prodi},
  {Prokhorov}, {Puncken}, {Punturo}, {Puppo}, {P{\"u}rrer}, {Qi}, {Quetschke},
  {Quintero}, {Quitzow-James}, {Raab}, {Rabeling}, {Radkins}, {Raffai}, {Raja},
  {Rajan}, {Rajbhandari}, {Rakhmanov}, {Ramirez}, {Ramos-Buades}, {Rapagnani},
  {Raymond}, {Razzano}, {Read}, {Regimbau}, {Rei}, {Reid}, {Reitze}, {Ren},
  {Reyes}, {Ricci}, {Ricker}, {Rieger}, {Riles}, {Rizzo}, {Robertson}, {Robie},
  {Robinet}, {Rocchi}, {Rolland}, {Rollins}, {Roma}, {Romano}, {Romel},
  {Romie}, {Rosi{\'n}ska}, {Ross}, {Rowan}, {R{\"u}diger}, {Ruggi}, {Rutins},
  {Ryan}, {Sachdev}, {Sadecki}, {Sadeghian}, {Sakellariadou}, {Salconi},
  {Saleem}, {Salemi}, {Samajdar}, {Sammut}, {Sampson}, {Sanchez}, {Sanchez},
  {Sanchis-Gual}, {Sandberg}, {Sanders}, {Sassolas}, {Sathyaprakash},
  {Saulson}, {Sauter}, {Savage}, {Sawadsky}, {Schale}, {Scheel}, {Scheuer},
  {Schmidt}, {Schmidt}, {Schnabel}, {Schofield}, {Sch{\"o}nbeck}, {Schreiber},
  {Schuette}, {Schulte}, {Schutz}, {Schwalbe}, {Scott}, {Scott}, {Seidel},
  {Sellers}, {Sengupta}, {Sentenac}, {Sequino}, {Sergeev}, {Shaddock},
  {Shaffer}, {Shah}, {Shahriar}, {Shaner}, {Shao}, {Shapiro}, {Shawhan},
  {Sheperd}, {Shoemaker}, {Shoemaker}, {Siellez}, {Siemens}, {Sieniawska},
  {Sigg}, {Silva}, {Singer}, {Singh}, {Singhal}, {Sintes}, {Slagmolen},
  {Smith}, {Smith}, {Smith}, {Somala}, {Son}, {Sonnenberg}, {Sorazu},
  {Sorrentino}, {Souradeep}, {Spencer}, {Srivastava}, {Staats}, {Staley},
  {Steinke}, {Steinlechner}, {Steinlechner}, {Steinmeyer}, {Stevenson},
  {Stone}, {Stops}, {Strain}, {Stratta}, {Strigin}, {Strunk}, {Sturani},
  {Stuver}, {Summerscales}, {Sun}, {Sunil}, {Suresh}, {Sutton}, {Swinkels},
  {Szczepa{\'n}czyk}, {Tacca}, {Tait}, {Talbot}, {Talukder}, {Tanner},
  {T{\'a}pai}, {Taracchini}, {Tasson}, {Taylor}, {Taylor}, {Tewari}, {Theeg},
  {Thies}, {Thomas}, {Thomas}, {Thomas}, {Thorne}, {Thorne}, {Thrane},
  {Tiwari}, {Tiwari}, {Tokmakov}, {Toland}, {Tonelli}, {Tornasi},
  {Torres-Forn{\'e}}, {Torrie}, {T{\"o}yr{\"a}}, {Travasso}, {Traylor},
  {Trinastic}, {Tringali}, {Trozzo}, {Tsang}, {Tse}, {Tso}, {Tsukada}, {Tsuna},
  {Tuyenbayev}, {Ueno}, {Ugolini}, {Unnikrishnan}, {Urban}, {Usman},
  {Vahlbruch}, {Vajente}, {Valdes}, {van Bakel}, {van Beuzekom}, {van den
  Brand}, {Van Den Broeck}, {Vander-Hyde}, {van der Schaaf}, {van Heijningen},
  {van Veggel}, {Vardaro}, {Varma}, {Vass}, {Vas{\'u}th}, {Vecchio},
  {Vedovato}, {Veitch}, {Veitch}, {Venkateswara}, {Venugopalan}, {Verkindt},
  {Vetrano}, {Vicer{\'e}}, {Viets}, {Vinciguerra}, {Vine}, {Vinet}, {Vitale},
  {Vo}, {Vocca}, {Vorvick}, {Vyatchanin}, {Wade}, {Wade}, {Wade}, {Walet},
  {Walker}, {Wallace}, {Walsh}, {Wang}, {Wang}, {Wang}, {Wang}, {Wang}, {Ward},
  {Warner}, {Was}, {Watchi}, {Weaver}, {Wei}, {Weinert}, {Weinstein}, {Weiss},
  {Wen}, {Wessel}, {We{\ss}els}, {Westerweck}, {Westphal}, {Wette}, {Whelan},
  {Whitcomb}, {Whiting}, {Whittle}, {Wilken}, {Williams}, {Williams},
  {Williamson}, {Willis}, {Willke}, {Wimmer}, {Winkler}, {Wipf}, {Wittel},
  {Woan}, {Woehler}, {Wofford}, {Wong}, {Worden}, {Wright}, {Wu}, {Wysocki},
  {Xiao}, {Yamamoto}, {Yancey}, {Yang}, {Yap}, {Yazback}, {Yu}, {Yu}, {Yvert},
  {Zadro{\.z}ny}, {Zanolin}, {Zelenova}, {Zendri}, {Zevin}, {Zhang}, {Zhang},
  {Zhang}, {Zhang}, {Zhao}, {Zhou}, {Zhou}, {Zhu}, {Zhu}, {Zimmerman},
  {Zucker}, {Zweizig}, {(LIGO Scientific Collaboration}, {Virgo Collaboration},
  {Burns}, {Veres}, {Kocevski}, {Racusin}, {Goldstein}, {Connaughton},
  {Briggs}, {Blackburn}, {Hamburg}, {Hui}, {von Kienlin}, {McEnery}, {Preece},
  {Wilson-Hodge}, {Bissaldi}, {Cleveland}, {Gibby}, {Giles}, {Kippen},
  {McBreen}, {Meegan}, {Paciesas}, {Poolakkil}, {Roberts}, {Stanbro},
  {Gamma-ray Burst Monitor}, {Savchenko}, {Ferrigno}, {Kuulkers}, {Bazzano},
  {Bozzo}, {Brandt}, {Chenevez}, {Courvoisier}, {Diehl}, {Domingo}, {Hanlon},
  {Jourdain}, {Laurent}, {Lebrun}, {Lutovinov}, {Mereghetti}, {Natalucci},
  {Rodi}, {Roques}, {Sunyaev}, {Ubertini}, \& {(INTEGRAL}}]{Abbott2017a}
{Abbott}, B.~P., {Abbott}, R., {Abbott}, T.~D., {et~al.} 2017{\natexlab{a}},
  \apjl, 848, L13, \dodoi{10.3847/2041-8213/aa920c}

\bibitem[{{Abbott} {et~al.}(2017{\natexlab{b}}){Abbott}, {Abbott}, {Abbott},
  {Acernese}, {Ackley}, {Adams}, {Adams}, {Addesso}, {Adhikari}, {Adya},
  {Affeldt}, {Afrough}, {Agarwal}, {Agathos}, {Agatsuma}, {Aggarwal}, {Aguiar},
  {Aiello}, {Ain}, {Ajith}, {Allen}, {Allen}, {Allocca}, {Altin}, {Amato},
  {Ananyeva}, {Anderson}, {Anderson}, {Angelova}, {Antier}, {Appert}, {Arai},
  {Araya}, {Areeda}, {Arnaud}, {Arun}, {Ascenzi}, {Ashton}, {Ast}, {Aston},
  {Astone}, {Atallah}, {Aufmuth}, {Aulbert}, {AultONeal}, {Austin},
  {Avila-Alvarez}, {Babak}, {Bacon}, {Bader}, {Bae}, {Bailes}, {Baker},
  {Baldaccini}, {Ballardin}, {Ballmer}, {Banagiri}, {Barayoga}, {Barclay},
  {Barish}, {Barker}, {Barkett}, {Barone}, {Barr}, {Barsotti}, {Barsuglia},
  {Barta}, {Barthelmy}, {Bartlett}, {Bartos}, {Bassiri}, {Basti}, {Batch},
  {Bawaj}, {Bayley}, {Bazzan}, {B{\'e}csy}, {Beer}, {Bejger}, {Belahcene},
  {Bell}, {Berger}, {Bergmann}, {Bernuzzi}, {Bero}, {Berry}, {Bersanetti},
  {Bertolini}, {Betzwieser}, {Bhagwat}, {Bhandare}, {Bilenko}, {Billingsley},
  {Billman}, {Birch}, {Birney}, {Birnholtz}, {Biscans}, {Biscoveanu}, {Bisht},
  {Bitossi}, {Biwer}, {Bizouard}, {Blackburn}, {Blackman}, {Blair}, {Blair},
  {Blair}, {Bloemen}, {Bock}, {Bode}, {Boer}, {Bogaert}, {Bohe}, {Bondu},
  {Bonilla}, {Bonnand}, {Boom}, {Bork}, {Boschi}, {Bose}, {Bossie},
  {Bouffanais}, {Bozzi}, {Bradaschia}, {Brady}, {Branchesi}, {Brau}, {Briant},
  {Brillet}, {Brinkmann}, {Brisson}, {Brockill}, {Broida}, {Brooks}, {Brown},
  {Brown}, {Brunett}, {Buchanan}, {Buikema}, {Bulik}, {Bulten}, {Buonanno},
  {Buskulic}, {Buy}, {Byer}, {Cabero}, {Cadonati}, {Cagnoli}, {Cahillane},
  {Calder{\'o}n Bustillo}, {Callister}, {Calloni}, {Camp}, {Canepa},
  {Canizares}, {Cannon}, {Cao}, {Cao}, {Capano}, {Capocasa}, {Carbognani},
  {Caride}, {Carney}, {Carullo}, {Casanueva Diaz}, {Casentini}, {Caudill},
  {Cavagli{\`a}}, {Cavalier}, {Cavalieri}, {Cella}, {Cepeda},
  {Cerd{\'a}-Dur{\'a}n}, {Cerretani}, {Cesarini}, {Chamberlin}, {Chan}, {Chao},
  {Charlton}, {Chase}, {Chassande-Mottin}, {Chatterjee}, {Chatziioannou},
  {Cheeseboro}, {Chen}, {Chen}, {Chen}, {Cheng}, {Chia}, {Chincarini},
  {Chiummo}, {Chmiel}, {Cho}, {Cho}, {Chow}, {Christensen}, {Chu}, {Chua},
  {Chua}, {Chung}, {Chung}, {Ciani}, {Ciolfi}, {Cirelli}, {Cirone}, {Clara},
  {Clark}, {Clearwater}, {Cleva}, {Cocchieri}, {Coccia}, {Cohadon}, {Cohen},
  {Colla}, {Collette}, {Cominsky}, {Constancio}, {Conti}, {Cooper}, {Corban},
  {Corbitt}, {Cordero-Carri{\'o}n}, {Corley}, {Cornish}, {Corsi}, {Cortese},
  {Costa}, {Coughlin}, {Coughlin}, {Coulon}, {Countryman}, {Couvares}, {Covas},
  {Cowan}, {Coward}, {Cowart}, {Coyne}, {Coyne}, {Creighton}, {Creighton},
  {Cripe}, {Crowder}, {Cullen}, {Cumming}, {Cunningham}, {Cuoco}, {Dal Canton},
  {D{\'a}lya}, {Danilishin}, {D'Antonio}, {Danzmann}, {Dasgupta}, {Da Silva
  Costa}, {Dattilo}, {Dave}, {Davier}, {Davis}, {Daw}, {Day}, {De}, {DeBra},
  {Degallaix}, {De Laurentis}, {Del{\'e}glise}, {Del Pozzo}, {Demos}, {Denker},
  {Dent}, {De Pietri}, {Dergachev}, {De Rosa}, {DeRosa}, {De Rossi}, {DeSalvo},
  {de Varona}, {Devenson}, {Dhurandhar}, {D{\'\i}az}, {Dietrich}, {Di Fiore},
  {Di Giovanni}, {Di Girolamo}, {Di Lieto}, {Di Pace}, {Di Palma}, {Di Renzo},
  {Doctor}, {Dolique}, {Donovan}, {Dooley}, {Doravari}, {Dorrington},
  {Douglas}, {Dovale {\'A}lvarez}, {Downes}, {Drago}, {Dreissigacker},
  {Driggers}, {Du}, {Ducrot}, {Dudi}, {Dupej}, {Dwyer}, {Edo}, {Edwards},
  {Effler}, {Eggenstein}, {Ehrens}, {Eichholz}, {Eikenberry}, {Eisenstein},
  {Essick}, {Estevez}, {Etienne}, {Etzel}, {Evans}, {Evans}, {Factourovich},
  {Fafone}, {Fair}, {Fairhurst}, {Fan}, {Farinon}, {Farr}, {Farr},
  {Fauchon-Jones}, {Favata}, {Fays}, {Fee}, {Fehrmann}, {Feicht}, {Fejer},
  {Fernandez-Galiana}, {Ferrante}, {Ferreira}, {Ferrini}, {Fidecaro},
  {Finstad}, {Fiori}, {Fiorucci}, {Fishbach}, {Fisher}, {Fitz-Axen},
  {Flaminio}, {Fletcher}, {Fong}, {Font}, {Forsyth}, {Forsyth}, {Fournier},
  {Frasca}, {Frasconi}, {Frei}, {Freise}, {Frey}, {Frey}, {Fries}, {Fritschel},
  {Frolov}, {Fulda}, {Fyffe}, {Gabbard}, {Gadre}, {Gaebel}, {Gair},
  {Gammaitoni}, {Ganija}, {Gaonkar}, {Garcia-Quiros}, {Garufi}, {Gateley},
  {Gaudio}, {Gaur}, {Gayathri}, {Gehrels}, {Gemme}, {Genin}, {Gennai},
  {George}, {George}, {Gergely}, {Germain}, {Ghonge}, {Ghosh}, {Ghosh},
  {Ghosh}, {Giaime}, {Giardina}, {Giazotto}, {Gill}, {Glover}, {Goetz},
  {Goetz}, {Gomes}, {Goncharov}, {Gonz{\'a}lez}, {Gonzalez Castro},
  {Gopakumar}, {Gorodetsky}, {Gossan}, {Gosselin}, {Gouaty}, {Grado}, {Graef},
  {Granata}, {Grant}, {Gras}, {Gray}, {Greco}, {Green}, {Gretarsson}, {Groot},
  {Grote}, {Grunewald}, {Gruning}, {Guidi}, {Guo}, {Gupta}, {Gupta}, {Gushwa},
  {Gustafson}, {Gustafson}, {Halim}, {Hall}, {Hall}, {Hamilton}, {Hammond},
  {Haney}, {Hanke}, {Hanks}, {Hanna}, {Hannam}, {Hannuksela}, {Hanson},
  {Hardwick}, {Harms}, {Harry}, {Harry}, {Hart}, {Haster}, {Haughian}, {Healy},
  {Heidmann}, {Heintze}, {Heitmann}, {Hello}, {Hemming}, {Hendry}, {Heng},
  {Hennig}, {Heptonstall}, {Heurs}, {Hild}, {Hinderer}, {Ho}, {Hoak}, {Hofman},
  {Holt}, {Holz}, {Hopkins}, {Horst}, {Hough}, {Houston}, {Howell}, {Hreibi},
  {Hu}, {Huerta}, {Huet}, {Hughey}, {Husa}, {Huttner}, {Huynh-Dinh}, {Indik},
  {Inta}, {Intini}, {Isa}, {Isac}, {Isi}, {Iyer}, {Izumi}, {Jacqmin}, {Jani},
  {Jaranowski}, {Jawahar}, {Jim{\'e}nez-Forteza}, {Johnson},
  {Johnson-McDaniel}, {Jones}, {Jones}, {Jonker}, {Ju}, {Junker}, {Kalaghatgi},
  {Kalogera}, {Kamai}, {Kandhasamy}, {Kang}, {Kanner}, {Kapadia}, {Karki},
  {Karvinen}, {Kasprzack}, {Kastaun}, {Katolik}, {Katsavounidis}, {Katzman},
  {Kaufer}, {Kawabe}, {K{\'e}f{\'e}lian}, {Keitel}, {Kemball}, {Kennedy},
  {Kent}, {Key}, {Khalili}, {Khan}, {Khan}, {Khan}, {Khazanov}, {Kijbunchoo},
  {Kim}, {Kim}, {Kim}, {Kim}, {Kim}, {Kim}, {Kimbrell}, {King}, {King},
  {Kinley-Hanlon}, {Kirchhoff}, {Kissel}, {Kleybolte}, {Klimenko}, {Knowles},
  {Koch}, {Koehlenbeck}, {Koley}, {Kondrashov}, {Kontos}, {Korobko}, {Korth},
  {Kowalska}, {Kozak}, {Kr{\"a}mer}, {Kringel}, {Krishnan}, {Kr{\'o}lak},
  {Kuehn}, {Kumar}, {Kumar}, {Kumar}, {Kuo}, {Kutynia}, {Kwang}, {Lackey},
  {Lai}, {Landry}, {Lang}, {Lange}, {Lantz}, {Lanza}, {Larson},
  {Lartaux-Vollard}, {Lasky}, {Laxen}, {Lazzarini}, {Lazzaro}, {Leaci},
  {Leavey}, {Lee}, {Lee}, {Lee}, {Lee}, {Lee}, {Lehmann}, {Lenon}, {Leon},
  {Leonardi}, {Leroy}, {Letendre}, {Levin}, {Li}, {Linker}, {Littenberg},
  {Liu}, {Liu}, {Lo}, {Lockerbie}, {London}, {Lord}, {Lorenzini}, {Loriette},
  {Lormand}, {Losurdo}, {Lough}, {Lousto}, {Lovelace}, {L{\"u}ck}, {Lumaca},
  {Lundgren}, {Lynch}, {Ma}, {Macas}, {Macfoy}, {Machenschalk}, {MacInnis},
  {Macleod}, {Maga{\~n}a Hernandez}, {Maga{\~n}a-Sandoval}, {Maga{\~n}a
  Zertuche}, {Magee}, {Majorana}, {Maksimovic}, {Man}, {Mandic}, {Mangano},
  {Mansell}, {Manske}, {Mantovani}, {Marchesoni}, {Marion}, {M{\'a}rka},
  {M{\'a}rka}, {Markakis}, {Markosyan}, {Markowitz}, {Maros}, {Marquina},
  {Marsh}, {Martelli}, {Martellini}, {Martin}, {Martin}, {Martynov}, {Marx},
  {Mason}, {Massera}, {Masserot}, {Massinger}, {Masso-Reid}, {Mastrogiovanni},
  {Matas}, {Matichard}, {Matone}, {Mavalvala}, {Mazumder}, {McCarthy},
  {McClelland}, {McCormick}, {McCuller}, {McGuire}, {McIntyre}, {McIver},
  {McManus}, {McNeill}, {McRae}, {McWilliams}, {Meacher}, {Meadors}, {Mehmet},
  {Meidam}, {Mejuto-Villa}, {Melatos}, {Mendell}, {Mercer}, {Merilh},
  {Merzougui}, {Meshkov}, {Messenger}, {Messick}, {Metzdorff}, {Meyers},
  {Miao}, {Michel}, {Middleton}, {Mikhailov}, {Milano}, {Miller}, {Miller},
  {Miller}, {Millhouse}, {Milovich-Goff}, {Minazzoli}, {Minenkov}, {Ming},
  {Mishra}, {Mitra}, {Mitrofanov}, {Mitselmakher}, {Mittleman}, {Moffa},
  {Moggi}, {Mogushi}, {Mohan}, {Mohapatra}, {Molina}, {Montani}, {Moore},
  {Moraru}, {Moreno}, {Morisaki}, {Morriss}, {Mours}, {Mow-Lowry}, {Mueller},
  {Muir}, {Mukherjee}, {Mukherjee}, {Mukherjee}, {Mukund}, {Mullavey}, {Munch},
  {Mu{\~n}iz}, {Muratore}, {Murray}, {Nagar}, {Napier}, {Nardecchia},
  {Naticchioni}, {Nayak}, {Neilson}, {Nelemans}, {Nelson}, {Nery}, {Neunzert},
  {Nevin}, {Newport}, {Newton}, {Ng}, {Nguyen}, {Nguyen}, {Nichols}, {Nielsen},
  {Nissanke}, {Nitz}, {Noack}, {Nocera}, {Nolting}, {North}, {Nuttall},
  {Oberling}, {O'Dea}, {Ogin}, {Oh}, {Oh}, {Ohme}, {Okada}, {Oliver},
  {Oppermann}, {Oram}, {O'Reilly}, {Ormiston}, {Ortega}, {O'Shaughnessy},
  {Ossokine}, {Ottaway}, {Overmier}, {Owen}, {Pace}, {Page}, {Page}, {Pai},
  {Pai}, {Palamos}, {Palashov}, {Palomba}, {Pal-Singh}, {Pan}, {Pan}, {Pang},
  {Pang}, {Pankow}, {Pannarale}, {Pant}, {Paoletti}, {Paoli}, {Papa}, {Parida},
  {Parker}, {Pascucci}, {Pasqualetti}, {Passaquieti}, {Passuello}, {Patil},
  {Patricelli}, {Pearlstone}, {Pedraza}, {Pedurand}, {Pekowsky}, {Pele},
  {Penn}, {Perez}, {Perreca}, {Perri}, {Pfeiffer}, {Phelps}, {Piccinni},
  {Pichot}, {Piergiovanni}, {Pierro}, {Pillant}, {Pinard}, {Pinto}, {Pirello},
  {Pitkin}, {Poe}, {Poggiani}, {Popolizio}, {Porter}, {Post}, {Powell},
  {Prasad}, {Pratt}, {Pratten}, {Predoi}, {Prestegard}, {Prijatelj},
  {Principe}, {Privitera}, {Prix}, {Prodi}, {Prokhorov}, {Puncken}, {Punturo},
  {Puppo}, {P{\"u}rrer}, {Qi}, {Quetschke}, {Quintero}, {Quitzow-James},
  {Raab}, {Rabeling}, {Radkins}, {Raffai}, {Raja}, {Rajan}, {Rajbhandari},
  {Rakhmanov}, {Ramirez}, {Ramos-Buades}, {Rapagnani}, {Raymond}, {Razzano},
  {Read}, {Regimbau}, {Rei}, {Reid}, {Reitze}, {Ren}, {Reyes}, {Ricci},
  {Ricker}, {Rieger}, {Riles}, {Rizzo}, {Robertson}, {Robie}, {Robinet},
  {Rocchi}, {Rolland}, {Rollins}, {Roma}, {Romano}, {Romano}, {Romel}, {Romie},
  {Rosi{\'n}ska}, {Ross}, {Rowan}, {R{\"u}diger}, {Ruggi}, {Rutins}, {Ryan},
  {Sachdev}, {Sadecki}, {Sadeghian}, {Sakellariadou}, {Salconi}, {Saleem},
  {Salemi}, {Samajdar}, {Sammut}, {Sampson}, {Sanchez}, {Sanchez},
  {Sanchis-Gual}, {Sandberg}, {Sanders}, {Sassolas}, {Sathyaprakash},
  {Saulson}, {Sauter}, {Savage}, {Sawadsky}, {Schale}, {Scheel}, {Scheuer},
  {Schmidt}, {Schmidt}, {Schnabel}, {Schofield}, {Sch{\"o}nbeck}, {Schreiber},
  {Schuette}, {Schulte}, {Schutz}, {Schwalbe}, {Scott}, {Scott}, {Seidel},
  {Sellers}, {Sengupta}, {Sentenac}, {Sequino}, {Sergeev}, {Shaddock},
  {Shaffer}, {Shah}, {Shahriar}, {Shaner}, {Shao}, {Shapiro}, {Shawhan},
  {Sheperd}, {Shoemaker}, {Shoemaker}, {Siellez}, {Siemens}, {Sieniawska},
  {Sigg}, {Silva}, {Singer}, {Singh}, {Singhal}, {Sintes}, {Slagmolen},
  {Smith}, {Smith}, {Smith}, {Somala}, {Son}, {Sonnenberg}, {Sorazu},
  {Sorrentino}, {Souradeep}, {Spencer}, {Srivastava}, {Staats}, {Staley},
  {Steinke}, {Steinlechner}, {Steinlechner}, {Steinmeyer}, {Stevenson},
  {Stone}, {Stops}, {Strain}, {Stratta}, {Strigin}, {Strunk}, {Sturani},
  {Stuver}, {Summerscales}, {Sun}, {Sunil}, {Suresh}, {Sutton}, {Swinkels},
  {Szczepa{\'n}czyk}, {Tacca}, {Tait}, {Talbot}, {Talukder}, {Tanner},
  {T{\'a}pai}, {Taracchini}, {Tasson}, {Taylor}, {Taylor}, {Tewari}, {Theeg},
  {Thies}, {Thomas}, {Thomas}, {Thomas}, {Thorne}, {Thorne}, {Thrane},
  {Tiwari}, {Tiwari}, {Tokmakov}, {Toland}, {Tonelli}, {Tornasi},
  {Torres-Forn{\'e}}, {Torrie}, {T{\"o}yr{\"a}}, {Travasso}, {Traylor},
  {Trinastic}, {Tringali}, {Trozzo}, {Tsang}, {Tse}, {Tso}, {Tsukada}, {Tsuna},
  {Tuyenbayev}, {Ueno}, {Ugolini}, {Unnikrishnan}, {Urban}, {Usman},
  {Vahlbruch}, {Vajente}, {Valdes}, {Vallisneri}, {van Bakel}, {van Beuzekom},
  {van den Brand}, {Van Den Broeck}, {Vander-Hyde}, {van der Schaaf}, {van
  Heijningen}, {van Veggel}, {Vardaro}, {Varma}, {Vass}, {Vas{\'u}th},
  {Vecchio}, {Vedovato}, {Veitch}, {Veitch}, {Venkateswara}, {Venugopalan},
  {Verkindt}, {Vetrano}, {Vicer{\'e}}, {Viets}, {Vinciguerra}, {Vine}, {Vinet},
  {Vitale}, {Vo}, {Vocca}, {Vorvick}, {Vyatchanin}, {Wade}, {Wade}, {Wade},
  {Walet}, {Walker}, {Wallace}, {Walsh}, {Wang}, {Wang}, {Wang}, {Wang},
  {Wang}, {Ward}, {Warner}, {Was}, {Watchi}, {Weaver}, {Wei}, {Weinert},
  {Weinstein}, {Weiss}, {Wen}, {Wessel}, {We{\ss}els}, {Westerweck},
  {Westphal}, {Wette}, {Whelan}, {Whitcomb}, {Whiting}, {Whittle}, {Wilken},
  {Williams}, {Williams}, {Williamson}, {Willis}, {Willke}, {Wimmer},
  {Winkler}, {Wipf}, {Wittel}, {Woan}, {Woehler}, {Wofford}, {Wong}, {Worden},
  {Wright}, {Wu}, {Wysocki}, {Xiao}, {Yamamoto}, {Yancey}, {Yang}, {Yap},
  {Yazback}, {Yu}, {Yu}, {Yvert}, {Zadro{\.Z}ny}, {Zanolin}, {Zelenova},
  {Zendri}, {Zevin}, {Zhang}, {Zhang}, {Zhang}, {Zhang}, {Zhao}, {Zhou},
  {Zhou}, {Zhu}, {Zhu}, {Zimmerman}, {Zucker}, {Zweizig}, {LIGO Scientific
  Collaboration}, \& {Virgo Collaboration}}]{Abbott2017b}
---. 2017{\natexlab{b}}, \prl, 119, 161101,
  \dodoi{10.1103/PhysRevLett.119.161101}

\bibitem[{{Abbott} {et~al.}(2017{\natexlab{c}}){Abbott}, {Abbott}, {Abbott},
  {Acernese}, {Ackley}, {Adams}, {Adams}, {Addesso}, {Adhikari}, {Adya},
  {Affeldt}, {Afrough}, {Agarwal}, {Agathos}, {Agatsuma}, {Aggarwal}, {Aguiar},
  {Aiello}, {Ain}, {Ajith}, {Allen}, {Allen}, {Allocca}, {Altin}, {Amato},
  {Ananyeva}, {Anderson}, {Anderson}, {Angelova}, {Antier}, {Appert}, {Arai},
  {Araya}, {Areeda}, {Arnaud}, {Arun}, {Ascenzi}, {Ashton}, {Ast}, {Aston},
  {Astone}, {Atallah}, {Aufmuth}, {Aulbert}, {AultONeal}, {Austin},
  {Avila-Alvarez}, {Babak}, {Bacon}, {Bader}, {Bae}, {Baker}, {Baldaccini},
  {Ballardin}, {Ballmer}, {Banagiri}, {Barayoga}, {Barclay}, {Barish},
  {Barker}, {Barkett}, {Barone}, {Barr}, {Barsotti}, {Barsuglia}, {Barta},
  {Barthelmy}, {Bartlett}, {Bartos}, {Bassiri}, {Basti}, {Batch}, {Bawaj},
  {Bayley}, {Bazzan}, {B{\'e}csy}, {Beer}, {Bejger}, {Belahcene}, {Bell},
  {Berger}, {Bergmann}, {Bero}, {Berry}, {Bersanetti}, {Bertolini},
  {Betzwieser}, {Bhagwat}, {Bhandare}, {Bilenko}, {Billingsley}, {Billman},
  {Birch}, {Birney}, {Birnholtz}, {Biscans}, {Biscoveanu}, {Bisht}, {Bitossi},
  {Biwer}, {Bizouard}, {Blackburn}, {Blackman}, {Blair}, {Blair}, {Blair},
  {Bloemen}, {Bock}, {Bode}, {Boer}, {Bogaert}, {Bohe}, {Bondu}, {Bonilla},
  {Bonnand}, {Boom}, {Bork}, {Boschi}, {Bose}, {Bossie}, {Bouffanais}, {Bozzi},
  {Bradaschia}, {Brady}, {Branchesi}, {Brau}, {Briant}, {Brillet}, {Brinkmann},
  {Brisson}, {Brockill}, {Broida}, {Brooks}, {Brown}, {Brown}, {Brunett},
  {Buchanan}, {Buikema}, {Bulik}, {Bulten}, {Buonanno}, {Buskulic}, {Buy},
  {Byer}, {Cabero}, {Cadonati}, {Cagnoli}, {Cahillane}, {Calder{\'o}n
  Bustillo}, {Callister}, {Calloni}, {Camp}, {Canepa}, {Canizares}, {Cannon},
  {Cao}, {Cao}, {Capano}, {Capocasa}, {Carbognani}, {Caride}, {Carney},
  {Casanueva Diaz}, {Casentini}, {Caudill}, {Cavagli{\`a}}, {Cavalier},
  {Cavalieri}, {Cella}, {Cepeda}, {Cerd{\'a}-Dur{\'a}n}, {Cerretani},
  {Cesarini}, {Chamberlin}, {Chan}, {Chao}, {Charlton}, {Chase},
  {Chassande-Mottin}, {Chatterjee}, {Chatziioannou}, {Cheeseboro}, {Chen},
  {Chen}, {Chen}, {Cheng}, {Chia}, {Chincarini}, {Chiummo}, {Chmiel}, {Cho},
  {Cho}, {Chow}, {Christensen}, {Chu}, {Chua}, {Chua}, {Chung}, {Chung},
  {Ciani}, {Ciolfi}, {Cirelli}, {Cirone}, {Clara}, {Clark}, {Clearwater},
  {Cleva}, {Cocchieri}, {Coccia}, {Cohadon}, {Cohen}, {Colla}, {Collette},
  {Cominsky}, {Constancio}, {Conti}, {Cooper}, {Corban}, {Corbitt},
  {Cordero-Carri{\'o}n}, {Corley}, {Cornish}, {Corsi}, {Cortese}, {Costa},
  {Coughlin}, {Coughlin}, {Coulon}, {Countryman}, {Couvares}, {Covas}, {Cowan},
  {Coward}, {Cowart}, {Coyne}, {Coyne}, {Creighton}, {Creighton}, {Cripe},
  {Crowder}, {Cullen}, {Cumming}, {Cunningham}, {Cuoco}, {Dal Canton},
  {D{\'a}lya}, {Danilishin}, {D'Antonio}, {Danzmann}, {Dasgupta}, {Da Silva
  Costa}, {Dattilo}, {Dave}, {Davier}, {Davis}, {Daw}, {Day}, {De}, {DeBra},
  {Degallaix}, {De Laurentis}, {Del{\'e}glise}, {Del Pozzo}, {Demos}, {Denker},
  {Dent}, {De Pietri}, {Dergachev}, {De Rosa}, {DeRosa}, {De Rossi}, {DeSalvo},
  {de Varona}, {Devenson}, {Dhurandhar}, {D{\'\i}az}, {Di Fiore}, {Di
  Giovanni}, {Di Girolamo}, {Di Lieto}, {Di Pace}, {Di Palma}, {Di Renzo},
  {Doctor}, {Dolique}, {Donovan}, {Dooley}, {Doravari}, {Dorrington},
  {Douglas}, {Dovale {\'A}lvarez}, {Downes}, {Drago}, {Dreissigacker},
  {Driggers}, {Du}, {Ducrot}, {Dupej}, {Dwyer}, {Edo}, {Edwards}, {Effler},
  {Ehrens}, {Eichholz}, {Eikenberry}, {Eisenstein}, {Essick}, {Estevez},
  {Etienne}, {Etzel}, {Evans}, {Evans}, {Factourovich}, {Fafone}, {Fair},
  {Fairhurst}, {Fan}, {Farinon}, {Farr}, {Farr}, {Fauchon-Jones}, {Favata},
  {Fays}, {Fee}, {Fehrmann}, {Feicht}, {Fejer}, {Fernandez-Galiana},
  {Ferrante}, {Ferreira}, {Ferrini}, {Fidecaro}, {Finstad}, {Fiori},
  {Fiorucci}, {Fishbach}, {Fisher}, {Fitz-Axen}, {Flaminio}, {Fletcher},
  {Fong}, {Font}, {Forsyth}, {Forsyth}, {Fournier}, {Frasca}, {Frasconi},
  {Frei}, {Freise}, {Frey}, {Frey}, {Fries}, {Fritschel}, {Frolov}, {Fulda},
  {Fyffe}, {Gabbard}, {Gadre}, {Gaebel}, {Gair}, {Gammaitoni}, {Ganija},
  {Gaonkar}, {Garcia-Quiros}, {Garufi}, {Gateley}, {Gaudio}, {Gaur},
  {Gayathri}, {Gehrels}, {Gemme}, {Genin}, {Gennai}, {George}, {George},
  {Gergely}, {Germain}, {Ghonge}, {Ghosh}, {Ghosh}, {Ghosh}, {Giaime},
  {Giardina}, {Giazotto}, {Gill}, {Glover}, {Goetz}, {Goetz}, {Gomes},
  {Goncharov}, {Gonz{\'a}lez}, {Gonzalez Castro}, {Gopakumar}, {Gorodetsky},
  {Gossan}, {Gosselin}, {Gouaty}, {Grado}, {Graef}, {Granata}, {Grant}, {Gras},
  {Gray}, {Greco}, {Green}, {Gretarsson}, {Griswold}, {Groot}, {Grote},
  {Grunewald}, {Gruning}, {Guidi}, {Guo}, {Gupta}, {Gupta}, {Gushwa},
  {Gustafson}, {Gustafson}, {Halim}, {Hall}, {Hall}, {Hamilton}, {Hammond},
  {Haney}, {Hanke}, {Hanks}, {Hanna}, {Hannam}, {Hannuksela}, {Hanson},
  {Hardwick}, {Harms}, {Harry}, {Harry}, {Hart}, {Haster}, {Haughian}, {Healy},
  {Heidmann}, {Heintze}, {Heitmann}, {Hello}, {Hemming}, {Hendry}, {Heng},
  {Hennig}, {Heptonstall}, {Heurs}, {Hild}, {Hinderer}, {Hoak}, {Hofman},
  {Holt}, {Holz}, {Hopkins}, {Horst}, {Hough}, {Houston}, {Howell}, {Hreibi},
  {Hu}, {Huerta}, {Huet}, {Hughey}, {Husa}, {Huttner}, {Huynh-Dinh}, {Indik},
  {Inta}, {Intini}, {Isa}, {Isac}, {Isi}, {Iyer}, {Izumi}, {Jacqmin}, {Jani},
  {Jaranowski}, {Jawahar}, {Jim{\'e}nez-Forteza}, {Johnson}, {Jones}, {Jones},
  {Jonker}, {Ju}, {Junker}, {Kalaghatgi}, {Kalogera}, {Kamai}, {Kandhasamy},
  {Kang}, {Kanner}, {Kapadia}, {Karki}, {Karvinen}, {Kasprzack}, {Katolik},
  {Katsavounidis}, {Katzman}, {Kaufer}, {Kawabe}, {K{\'e}f{\'e}lian}, {Keitel},
  {Kemball}, {Kennedy}, {Kent}, {Key}, {Khalili}, {Khan}, {Khan}, {Khan},
  {Khazanov}, {Kijbunchoo}, {Kim}, {Kim}, {Kim}, {Kim}, {Kim}, {Kim},
  {Kimbrell}, {King}, {King}, {Kinley-Hanlon}, {Kirchhoff}, {Kissel},
  {Kleybolte}, {Klimenko}, {Knowles}, {Koch}, {Koehlenbeck}, {Koley},
  {Kondrashov}, {Kontos}, {Korobko}, {Korth}, {Kowalska}, {Kozak},
  {Kr{\"a}mer}, {Kringel}, {Krishnan}, {Kr{\'o}lak}, {Kuehn}, {Kumar}, {Kumar},
  {Kumar}, {Kuo}, {Kutynia}, {Kwang}, {Lackey}, {Lai}, {Landry}, {Lang},
  {Lange}, {Lantz}, {Lanza}, {Larson}, {Lartaux-Vollard}, {Lasky}, {Laxen},
  {Lazzarini}, {Lazzaro}, {Leaci}, {Leavey}, {Lee}, {Lee}, {Lee}, {Lee}, {Lee},
  {Lehmann}, {Lenon}, {Leonardi}, {Leroy}, {Letendre}, {Levin}, {Li}, {Linker},
  {Littenberg}, {Liu}, {Lo}, {Lockerbie}, {London}, {Lord}, {Lorenzini},
  {Loriette}, {Lormand}, {Losurdo}, {Lough}, {Lousto}, {Lovelace}, {L{\"u}ck},
  {Lumaca}, {Lundgren}, {Lynch}, {Ma}, {Macas}, {Macfoy}, {Machenschalk},
  {MacInnis}, {Macleod}, {Maga{\~n}a Hernandez}, {Maga{\~n}a-Sandoval},
  {Maga{\~n}a Zertuche}, {Magee}, {Majorana}, {Maksimovic}, {Man}, {Mandic},
  {Mangano}, {Mansell}, {Manske}, {Mantovani}, {Marchesoni}, {Marion},
  {M{\'a}rka}, {M{\'a}rka}, {Markakis}, {Markosyan}, {Markowitz}, {Maros},
  {Marquina}, {Marsh}, {Martelli}, {Martellini}, {Martin}, {Martin},
  {Martynov}, {Mason}, {Massera}, {Masserot}, {Massinger}, {Masso-Reid},
  {Mastrogiovanni}, {Matas}, {Matichard}, {Matone}, {Mavalvala}, {Mazumder},
  {McCarthy}, {McClelland}, {McCormick}, {McCuller}, {McGuire}, {McIntyre},
  {McIver}, {McManus}, {McNeill}, {McRae}, {McWilliams}, {Meacher}, {Meadors},
  {Mehmet}, {Meidam}, {Mejuto-Villa}, {Melatos}, {Mendell}, {Mercer}, {Merilh},
  {Merzougui}, {Meshkov}, {Messenger}, {Messick}, {Metzdorff}, {Meyers},
  {Miao}, {Michel}, {Middleton}, {Mikhailov}, {Milano}, {Miller}, {Miller},
  {Miller}, {Millhouse}, {Milovich-Goff}, {Minazzoli}, {Minenkov}, {Ming},
  {Mishra}, {Mitra}, {Mitrofanov}, {Mitselmakher}, {Mittleman}, {Moffa},
  {Moggi}, {Mogushi}, {Mohan}, {Mohapatra}, {Montani}, {Moore}, {Moraru},
  {Moreno}, {Morriss}, {Mours}, {Mow-Lowry}, {Mueller}, {Muir}, {Mukherjee},
  {Mukherjee}, {Mukherjee}, {Mukund}, {Mullavey}, {Munch}, {Mu{\~n}iz},
  {Muratore}, {Murray}, {Napier}, {Nardecchia}, {Naticchioni}, {Nayak},
  {Neilson}, {Nelemans}, {Nelson}, {Nery}, {Neunzert}, {Nevin}, {Newport},
  {Newton}, {Ng}, {Nguyen}, {Nguyen}, {Nichols}, {Nielsen}, {Nissanke}, {Nitz},
  {Noack}, {Nocera}, {Nolting}, {North}, {Nuttall}, {Oberling}, {O'Dea},
  {Ogin}, {Oh}, {Oh}, {Ohme}, {Okada}, {Oliver}, {Oppermann}, {Oram},
  {O'Reilly}, {Ormiston}, {Ortega}, {O'Shaughnessy}, {Ossokine}, {Ottaway},
  {Overmier}, {Owen}, {Pace}, {Page}, {Page}, {Pai}, {Pai}, {Palamos},
  {Palashov}, {Palomba}, {Pal-Singh}, {Pan}, {Pan}, {Pang}, {Pang}, {Pankow},
  {Pannarale}, {Pant}, {Paoletti}, {Paoli}, {Papa}, {Parida}, {Parker},
  {Pascucci}, {Pasqualetti}, {Passaquieti}, {Passuello}, {Patil}, {Patricelli},
  {Pearlstone}, {Pedraza}, {Pedurand}, {Pekowsky}, {Pele}, {Penn}, {Perez},
  {Perreca}, {Perri}, {Pfeiffer}, {Phelps}, {Piccinni}, {Pichot},
  {Piergiovanni}, {Pierro}, {Pillant}, {Pinard}, {Pinto}, {Pirello}, {Pitkin},
  {Poe}, {Poggiani}, {Popolizio}, {Porter}, {Post}, {Powell}, {Prasad},
  {Pratt}, {Pratten}, {Predoi}, {Prestegard}, {Price}, {Prijatelj}, {Principe},
  {Privitera}, {Prodi}, {Prokhorov}, {Puncken}, {Punturo}, {Puppo},
  {P{\"u}rrer}, {Qi}, {Quetschke}, {Quintero}, {Quitzow-James}, {Raab},
  {Rabeling}, {Radkins}, {Raffai}, {Raja}, {Rajan}, {Rajbhandari}, {Rakhmanov},
  {Ramirez}, {Ramos-Buades}, {Rapagnani}, {Raymond}, {Razzano}, {Read},
  {Regimbau}, {Rei}, {Reid}, {Reitze}, {Ren}, {Reyes}, {Ricci}, {Ricker},
  {Rieger}, {Riles}, {Rizzo}, {Robertson}, {Robie}, {Robinet}, {Rocchi},
  {Rolland}, {Rollins}, {Roma}, {Romano}, {Romel}, {Romie}, {Rosi{\'n}ska},
  {Ross}, {Rowan}, {R{\"u}diger}, {Ruggi}, {Rutins}, {Ryan}, {Sachdev},
  {Sadecki}, {Sadeghian}, {Sakellariadou}, {Salconi}, {Saleem}, {Salemi},
  {Samajdar}, {Sammut}, {Sampson}, {Sanchez}, {Sanchez}, {Sanchis-Gual},
  {Sandberg}, {Sanders}, {Sassolas}, {Sathyaprakash}, {Saulson}, {Sauter},
  {Savage}, {Sawadsky}, {Schale}, {Scheel}, {Scheuer}, {Schmidt}, {Schmidt},
  {Schnabel}, {Schofield}, {Sch{\"o}nbeck}, {Schreiber}, {Schuette}, {Schulte},
  {Schutz}, {Schwalbe}, {Scott}, {Scott}, {Seidel}, {Sellers}, {Sengupta},
  {Sentenac}, {Sequino}, {Sergeev}, {Shaddock}, {Shaffer}, {Shah}, {Shahriar},
  {Shaner}, {Shao}, {Shapiro}, {Shawhan}, {Sheperd}, {Shoemaker}, {Shoemaker},
  {Siellez}, {Siemens}, {Sieniawska}, {Sigg}, {Silva}, {Singer}, {Singh},
  {Singhal}, {Sintes}, {Slagmolen}, {Smith}, {Smith}, {Smith}, {Somala}, {Son},
  {Sonnenberg}, {Sorazu}, {Sorrentino}, {Souradeep}, {Spencer}, {Srivastava},
  {Staats}, {Staley}, {Steinke}, {Steinlechner}, {Steinlechner}, {Steinmeyer},
  {Stevenson}, {Stone}, {Stops}, {Strain}, {Stratta}, {Strigin}, {Strunk},
  {Sturani}, {Stuver}, {Summerscales}, {Sun}, {Sunil}, {Suresh}, {Sutton},
  {Swinkels}, {Szczepa{\'n}czyk}, {Tacca}, {Tait}, {Talbot}, {Talukder},
  {Tanner}, {T{\'a}pai}, {Taracchini}, {Tasson}, {Taylor}, {Taylor}, {Tewari},
  {Theeg}, {Thies}, {Thomas}, {Thomas}, {Thomas}, {Thorne}, {Thorne}, {Thrane},
  {Tiwari}, {Tiwari}, {Tokmakov}, {Toland}, {Tonelli}, {Tornasi},
  {Torres-Forn{\'e}}, {Torrie}, {T{\"o}yr{\"a}}, {Travasso}, {Traylor},
  {Trinastic}, {Tringali}, {Trozzo}, {Tsang}, {Tse}, {Tso}, {Tsukada}, {Tsuna},
  {Tuyenbayev}, {Ueno}, {Ugolini}, {Unnikrishnan}, {Urban}, {Usman},
  {Vahlbruch}, {Vajente}, {Valdes}, {van Bakel}, {van Beuzekom}, {van den
  Brand}, {Van Den Broeck}, {Vander-Hyde}, {van der Schaaf}, {van Heijningen},
  {van Veggel}, {Vardaro}, {Varma}, {Vass}, {Vas{\'u}th}, {Vecchio},
  {Vedovato}, {Veitch}, {Veitch}, {Venkateswara}, {Venugopalan}, {Verkindt},
  {Vetrano}, {Vicer{\'e}}, {Viets}, {Vinciguerra}, {Vine}, {Vinet}, {Vitale},
  {Vo}, {Vocca}, {Vorvick}, {Vyatchanin}, {Wade}, {Wade}, {Wade}, {Walet},
  {Walker}, {Wallace}, {Walsh}, {Wang}, {Wang}, {Wang}, {Wang}, {Wang}, {Ward},
  {Warner}, {Was}, {Watchi}, {Weaver}, {Wei}, {Weinert}, {Weinstein}, {Weiss},
  {Wen}, {Wessel}, {Wessels}, {Westerweck}, {Westphal}, {Wette}, {Whelan},
  {Whitcomb}, {Whiting}, {Whittle}, {Wilken}, {Williams}, {Williams},
  {Williamson}, {Willis}, {Willke}, {Wimmer}, {Winkler}, {Wipf}, {Wittel},
  {Woan}, {Woehler}, {Wofford}, {Wong}, {Worden}, {Wright}, {Wu}, {Wysocki},
  {Xiao}, {Yamamoto}, {Yancey}, {Yang}, {Yap}, {Yazback}, {Yu}, {Yu}, {Yvert},
  {Zadro{\.z}ny}, {Zanolin}, {Zelenova}, {Zendri}, {Zevin}, {Zhang}, {Zhang},
  {Zhang}, {Zhang}, {Zhao}, {Zhou}, {Zhou}, {Zhu}, {Zhu}, {Zimmerman},
  {Zucker}, {Zweizig}, {LIGO Scientific Collaboration}, {Virgo Collaboration},
  {Wilson-Hodge}, {Bissaldi}, {Blackburn}, {Briggs}, {Burns}, {Cleveland},
  {Connaughton}, {Gibby}, {Giles}, {Goldstein}, {Hamburg}, {Jenke}, {Hui},
  {Kippen}, {Kocevski}, {McBreen}, {Meegan}, {Paciesas}, {Poolakkil}, {Preece},
  {Racusin}, {Roberts}, {Stanbro}, {Veres}, {von Kienlin}, {GBM}, {Savchenko},
  {Ferrigno}, {Kuulkers}, {Bazzano}, {Bozzo}, {Brandt}, {Chenevez},
  {Courvoisier}, {Diehl}, {Domingo}, {Hanlon}, {Jourdain}, {Laurent}, {Lebrun},
  {Lutovinov}, {Martin-Carrillo}, {Mereghetti}, {Natalucci}, {Rodi}, {Roques},
  {Sunyaev}, {Ubertini}, {INTEGRAL}, {Aartsen}, {Ackermann}, {Adams},
  {Aguilar}, {Ahlers}, {Ahrens}, {Samarai}, {Altmann}, {Andeen}, {Anderson},
  {Ansseau}, {Anton}, {Arg{\"u}elles}, {Auffenberg}, {Axani}, {Bagherpour},
  {Bai}, {Barron}, {Barwick}, {Baum}, {Bay}, {Beatty}, {Becker Tjus},
  {Bernardini}, {Besson}, {Binder}, {Bindig}, {Blaufuss}, {Blot}, {Bohm},
  {B{\"o}rner}, {Bos}, {Bose}, {B{\"o}ser}, {Botner}, {Bourbeau}, {Bourbeau},
  {Bradascio}, {Braun}, {Brayeur}, {Brenzke}, {Bretz}, {Bron},
  {Brostean-Kaiser}, {Burgman}, {Carver}, {Casey}, {Casier}, {Cheung},
  {Chirkin}, {Christov}, {Clark}, {Classen}, {Coenders}, {Collin}, {Conrad},
  {Cowen}, {Cross}, {Day}, {de Andr{\'e}}, {De Clercq}, {DeLaunay},
  {Dembinski}, {De Ridder}, {Desiati}, {de Vries}, {de Wasseige}, {de With},
  {DeYoung}, {D{\'\i}az-V{\'e}lez}, {di Lorenzo}, {Dujmovic}, {Dumm},
  {Dunkman}, {Dvorak}, {Eberhardt}, {Ehrhardt}, {Eichmann}, {Eller}, {Evenson},
  {Fahey}, {Fazely}, {Felde}, {Filimonov}, {Finley}, {Flis}, {Franckowiak},
  {Friedman}, {Fuchs}, {Gaisser}, {Gallagher}, {Gerhardt}, {Ghorbani}, {Giang},
  {Glauch}, {Gl{\"u}senkamp}, {Goldschmidt}, {Gonzalez}, {Grant}, {Griffith},
  {Haack}, {Hallgren}, {Halzen}, {Hanson}, {Hebecker}, {Heereman}, {Helbing},
  {Hellauer}, {Hickford}, {Hignight}, {Hill}, {Hoffman}, {Hoffmann},
  {Hokanson-Fasig}, {Hoshina}, {Huang}, {Huber}, {Hultqvist}, {H{\"u}nnefeld},
  {In}, {Ishihara}, {Jacobi}, {Japaridze}, {Jeong}, {Jero}, {Jones},
  {Kalaczynski}, {Kang}, {Kappes}, {Karg}, {Karle}, {Kauer}, {Keivani},
  {Kelley}, {Kheirandish}, {Kim}, {Kim}, {Kintscher}, {Kiryluk}, {Kittler},
  {Klein}, {Kohnen}, {Koirala}, {Kolanoski}, {K{\"o}pke}, {Kopper}, {Kopper},
  {Koschinsky}, {Koskinen}, {Kowalski}, {Krings}, {Kroll}, {Kr{\"u}ckl},
  {Kunnen}, {Kunwar}, {Kurahashi}, {Kuwabara}, {Kyriacou}, {Labare},
  {Lanfranchi}, {Larson}, {Lauber}, {Lesiak-Bzdak}, {Leuermann}, {Liu}, {Lu},
  {L{\"u}nemann}, {Luszczak}, {Madsen}, {Maggi}, {Mahn}, {Mancina}, {Maruyama},
  {Mase}, {Maunu}, {McNally}, {Meagher}, {Medici}, {Meier}, {Menne}, {Merino},
  {Meures}, {Miarecki}, {Micallef}, {Moment{\'e}}, {Montaruli}, {Moore},
  {Moulai}, {Nahnhauer}, {Nakarmi}, {Naumann}, {Neer}, {Niederhausen},
  {Nowicki}, {Nygren}, {Obertacke Pollmann}, {Olivas}, {O'Murchadha},
  {Palczewski}, {Pandya}, {Pankova}, {Peiffer}, {Pepper}, {P{\'e}rez de los
  Heros}, {Pieloth}, {Pinat}, {Price}, {Przybylski}, {Raab}, {R{\"a}del},
  {Rameez}, {Rawlins}, {Rea}, {Reimann}, {Relethford}, {Relich}, {Resconi},
  {Rhode}, {Richman}, {Robertson}, {Rongen}, {Rott}, {Ruhe}, {Ryckbosch},
  {Rysewyk}, {S{\"a}lzer}, {Sanchez Herrera}, {Sandrock}, {Sandroos},
  {Santander}, {Sarkar}, {Sarkar}, {Satalecka}, {Schlunder}, {Schmidt},
  {Schneider}, {Schoenen}, {Sch{\"o}neberg}, {Schumacher}, {Seckel},
  {Seunarine}, {Soedingrekso}, {Soldin}, {Song}, {Spiczak}, {Spiering},
  {Stachurska}, {Stamatikos}, {Stanev}, {Stasik}, {Stettner}, {Steuer},
  {Stezelberger}, {Stokstad}, {St{\"o}ssl}, {Strotjohann}, {Stuttard},
  {Sullivan}, {Sutherland}, {Taboada}, {Tatar}, {Tenholt}, {Ter-Antonyan},
  {Terliuk}, {Te{\v{s}}i{\'c}}, {Tilav}, {Toale}, {Tobin}, {Toscano}, {Tosi},
  {Tselengidou}, {Tung}, {Turcati}, {Turley}, {Ty}, {Unger}, {Usner},
  {Vandenbroucke}, {Van Driessche}, {van Eijndhoven}, {Vanheule}, {van Santen},
  {Vehring}, {Vogel}, {Vraeghe}, {Walck}, {Wallace}, {Wallraff}, {Wandler},
  {Wandkowsky}, {Waza}, {Weaver}, {Weiss}, {Wendt}, {Werthebach}, {Whelan},
  {Wiebe}, {Wiebusch}, {Wille}, {Williams}, {Wills}, {Wolf}, {Wood}, {Woolsey},
  {Woschnagg}, {Xu}, {Xu}, {Xu}, {Yanez}, {Yodh}, {Yoshida}, {Yuan}, {Zoll},
  {IceCube Collaboration}, {Balasubramanian}, {Mate}, {Bhalerao},
  {Bhattacharya}, {Vibhute}, {Dewangan}, {Rao}, {Vadawale}, {AstroSat Cadmium
  Zinc Telluride Imager Team}, {Svinkin}, {Hurley}, {Aptekar}, {Frederiks},
  {Golenetskii}, {Kozlova}, {Lysenko}, {Oleynik}, {Tsvetkova}, {Ulanov},
  {Cline}, {IPN Collaboration}, {Li}, {Xiong}, {Zhang}, {Lu}, {Song}, {Cao},
  {Chang}, {Chen}, {Chen}, {Chen}, {Chen}, {Chen}, {Chen}, {Cui}, {Cui},
  {Deng}, {Dong}, {Du}, {Fu}, {Gao}, {Gao}, {Gao}, {Ge}, {Gu}, {Guan}, {Guo},
  {Han}, {Hu}, {Huang}, {Huo}, {Jia}, {Jiang}, {Jiang}, {Jin}, {Jin}, {Li},
  {Li}, {Li}, {Li}, {Li}, {Li}, {Li}, {Li}, {Li}, {Li}, {Li}, {Liang}, {Liao},
  {Liu}, {Liu}, {Liu}, {Liu}, {Liu}, {Liu}, {Liu}, {Lu}, {Lu}, {Luo}, {Ma},
  {Meng}, {Nang}, {Nie}, {Ou}, {Qu}, {Sai}, {Sun}, {Tan}, {Tao}, {Tao}, {Tuo},
  {Wang}, {Wang}, {Wang}, {Wang}, {Wang}, {Wen}, {Wu}, {Wu}, {Xiao}, {Xu},
  {Xu}, {Yan}, {Yang}, {Yang}, {Yang}, {Zhang}, {Zhang}, {Zhang}, {Zhang},
  {Zhang}, {Zhang}, {Zhang}, {Zhang}, {Zhang}, {Zhang}, {Zhang}, {Zhang},
  {Zhang}, {Zhang}, {Zhang}, {Zhang}, {Zhang}, {Zhang}, {Zhao}, {Zhao}, {Zhao},
  {Zheng}, {Zhu}, {Zhu}, {Zou}, {Insight-HXMT Collaboration}, {Albert},
  {Andr{\'e}}, {Anghinolfi}, {Ardid}, {Aubert}, {Aublin}, {Avgitas}, {Baret},
  {Barrios-Mart{\'\i}}, {Basa}, {Belhorma}, {Bertin}, {Biagi}, {Bormuth},
  {Bourret}, {Bouwhuis}, {Br{\^a}nza{\c{s}}}, {Bruijn}, {Brunner}, {Busto},
  {Capone}, {Caramete}, {Carr}, {Celli}, {Cherkaoui El Moursli}, {Chiarusi},
  {Circella}, {Coelho}, {Coleiro}, {Coniglione}, {Costantini}, {Coyle},
  {Creusot}, {D{\'\i}az}, {Deschamps}, {De Bonis}, {Distefano}, {Di Palma},
  {Domi}, {Donzaud}, {Dornic}, {Drouhin}, {Eberl}, {El Bojaddaini}, {El
  Khayati}, {Els{\"a}sser}, {Enzenh{\"o}fer}, {Ettahiri}, {Fassi}, {Felis},
  {Fusco}, {Gay}, {Giordano}, {Glotin}, {Gr{\'e}goire}, {Ruiz}, {Graf},
  {Hallmann}, {van Haren}, {Heijboer}, {Hello}, {Hern{\'a}ndez-Rey},
  {H{\"o}ssl}, {Hofest{\"a}dt}, {Hugon}, {Illuminati}, {James}, {de Jong},
  {Jongen}, {Kadler}, {Kalekin}, {Katz}, {Kiessling}, {Kouchner}, {Kreter},
  {Kreykenbohm}, {Kulikovskiy}, {Lachaud}, {Lahmann}, {Lef{\`e}vre}, {Leonora},
  {Lotze}, {Loucatos}, {Marcelin}, {Margiotta}, {Marinelli},
  {Mart{\'\i}nez-Mora}, {Mele}, {Melis}, {Michael}, {Migliozzi}, {Moussa},
  {Navas}, {Nezri}, {Organokov}, {P{\u{a}}v{\u{a}}la{\c{s}}}, {Pellegrino},
  {Perrina}, {Piattelli}, {Popa}, {Pradier}, {Quinn}, {Racca}, {Riccobene},
  {S{\'a}nchez-Losa}, {Salda{\~n}a}, {Salvadori}, {Samtleben}, {Sanguineti},
  {Sapienza}, {Sieger}, {Spurio}, {Stolarczyk}, {Taiuti}, {Tayalati},
  {Trovato}, {Turpin}, {T{\"o}nnis}, {Vallage}, {Van Elewyck}, {Versari},
  {Vivolo}, {Vizzoca}, {Wilms}, {Zornoza}, {Z{\'u}{\~n}iga}, {ANTARES
  Collaboration}, {Beardmore}, {Breeveld}, {Burrows}, {Cenko}, {Cusumano},
  {D'A{\`\i}}, {de Pasquale}, {Emery}, {Evans}, {Giommi}, {Gronwall}, {Kennea},
  {Krimm}, {Kuin}, {Lien}, {Marshall}, {Melandri}, {Nousek}, {Oates},
  {Osborne}, {Pagani}, {Page}, {Palmer}, {Perri}, {Siegel}, {Sbarufatti},
  {Tagliaferri}, {Tohuvavohu}, {Swift Collaboration}, {Tavani}, {Verrecchia},
  {Bulgarelli}, {Evangelista}, {Pacciani}, {Feroci}, {Pittori}, {Giuliani},
  {Del Monte}, {Donnarumma}, {Argan}, {Trois}, {Ursi}, {Cardillo}, {Piano},
  {Longo}, {Lucarelli}, {Munar-Adrover}, {Fuschino}, {Labanti}, {Marisaldi},
  {Minervini}, {Fioretti}, {Parmiggiani}, {Gianotti}, {Trifoglio}, {Di Persio},
  {Antonelli}, {Barbiellini}, {Caraveo}, {Cattaneo}, {Costa}, {Colafrancesco},
  {D'Amico}, {Ferrari}, {Morselli}, {Paoletti}, {Picozza}, {Pilia}, {Rappoldi},
  {Soffitta}, {Vercellone}, {AGILE Team}, {Foley}, {Coulter}, {Kilpatrick},
  {Drout}, {Piro}, {Shappee}, {Siebert}, {Simon}, {Ulloa}, {Kasen}, {Madore},
  {Murguia-Berthier}, {Pan}, {Prochaska}, {Ramirez-Ruiz}, {Rest},
  {Rojas-Bravo}, {1M2H Team}, {Berger}, {Soares-Santos}, {Annis}, {Alexander},
  {Allam}, {Balbinot}, {Blanchard}, {Brout}, {Butler}, {Chornock}, {Cook},
  {Cowperthwaite}, {Diehl}, {Drlica-Wagner}, {Drout}, {Durret}, {Eftekhari},
  {Finley}, {Fong}, {Frieman}, {Fryer}, {Garc{\'\i}a-Bellido}, {Gruendl},
  {Hartley}, {Herner}, {Kessler}, {Lin}, {Lopes}, {Louren{\c{c}}o}, {Margutti},
  {Marshall}, {Matheson}, {Medina}, {Metzger}, {Mu{\~n}oz}, {Muir}, {Nicholl},
  {Nugent}, {Palmese}, {Paz-Chinch{\'o}n}, {Quataert}, {Sako}, {Sauseda},
  {Schlegel}, {Scolnic}, {Secco}, {Smith}, {Sobreira}, {Villar}, {Vivas},
  {Wester}, {Williams}, {Yanny}, {Zenteno}, {Zhang}, {Abbott}, {Banerji},
  {Bechtol}, {Benoit-L{\'e}vy}, {Bertin}, {Brooks}, {Buckley-Geer}, {Burke},
  {Capozzi}, {Carnero Rosell}, {Carrasco Kind}, {Castander}, {Crocce}, {Cunha},
  {D'Andrea}, {da Costa}, {Davis}, {DePoy}, {Desai}, {Dietrich}, {Eifler},
  {Fernandez}, {Flaugher}, {Fosalba}, {Gaztanaga}, {Gerdes}, {Giannantonio},
  {Goldstein}, {Gruen}, {Gschwend}, {Gutierrez}, {Honscheid}, {James},
  {Jeltema}, {Johnson}, {Johnson}, {Kent}, {Krause}, {Kron}, {Kuehn}, {Lahav},
  {Lima}, {Maia}, {March}, {Martini}, {McMahon}, {Menanteau}, {Miller},
  {Miquel}, {Mohr}, {Nichol}, {Ogando}, {Plazas}, {Romer}, {Roodman}, {Rykoff},
  {Sanchez}, {Scarpine}, {Schindler}, {Schubnell}, {Sevilla-Noarbe}, {Sheldon},
  {Smith}, {Smith}, {Stebbins}, {Suchyta}, {Swanson}, {Tarle}, {Thomas},
  {Troxel}, {Tucker}, {Vikram}, {Walker}, {Wechsler}, {Weller}, {Carlin},
  {Gill}, {Li}, {Marriner}, {Neilsen}, {Dark Energy Camera GW-EM
  Collaboration}, {DES Collaboration}, {Haislip}, {Kouprianov}, {Reichart},
  {Sand}, {Tartaglia}, {Valenti}, {Yang}, {DLT40 Collaboration}, {Benetti},
  {Brocato}, {Campana}, {Cappellaro}, {Covino}, {D'Avanzo}, {D'Elia}, {Getman},
  {Ghirlanda}, {Ghisellini}, {Limatola}, {Nicastro}, {Palazzi}, {Pian},
  {Piranomonte}, {Possenti}, {Rossi}, {Salafia}, {Tomasella}, {Amati},
  {Antonelli}, {Bernardini}, {Bufano}, {Capaccioli}, {Casella}, {Dadina}, {De
  Cesare}, {Di Paola}, {Giuffrida}, {Giunta}, {Israel}, {Lisi}, {Maiorano},
  {Mapelli}, {Masetti}, {Pescalli}, {Pulone}, {Salvaterra}, {Schipani},
  {Spera}, {Stamerra}, {Stella}, {Testa}, {Turatto}, {Vergani}, {Aresu},
  {Bachetti}, {Buffa}, {Burgay}, {Buttu}, {Caria}, {Carretti}, {Casasola},
  {Castangia}, {Carboni}, {Casu}, {Concu}, {Corongiu}, {Deiana}, {Egron},
  {Fara}, {Gaudiomonte}, {Gusai}, {Ladu}, {Loru}, {Leurini}, {Marongiu},
  {Melis}, {Melis}, {Migoni}, {Milia}, {Navarrini}, {Orlati}, {Ortu}, {Palmas},
  {Pellizzoni}, {Perrodin}, {Pisanu}, {Poppi}, {Righini}, {Saba}, {Serra},
  {Serrau}, {Stagni}, {Surcis}, {Vacca}, {Vargiu}, {Hunt}, {Jin}, {Klose},
  {Kouveliotou}, {Mazzali}, {M{\o}ller}, {Nava}, {Piran}, {Selsing}, {Vergani},
  {Wiersema}, {Toma}, {Higgins}, {Mundell}, {di Serego Alighieri}, {G{\'o}tz},
  {Gao}, {Gomboc}, {Kaper}, {Kobayashi}, {Kopac}, {Mao}, {Starling}, {Steele},
  {van der Horst}, {GRAWITA: GRAvitational Wave Inaf TeAm}, {Acero}, {Atwood},
  {Baldini}, {Barbiellini}, {Bastieri}, {Berenji}, {Bellazzini}, {Bissaldi},
  {Blandford}, {Bloom}, {Bonino}, {Bottacini}, {Bregeon}, {Buehler}, {Buson},
  {Cameron}, {Caputo}, {Caraveo}, {Cavazzuti}, {Chekhtman}, {Cheung}, {Chiang},
  {Ciprini}, {Cohen-Tanugi}, {Cominsky}, {Costantin}, {Cuoco}, {D'Ammando}, {de
  Palma}, {Digel}, {Di Lalla}, {Di Mauro}, {Di Venere}, {Dubois}, {Fegan},
  {Focke}, {Franckowiak}, {Fukazawa}, {Funk}, {Fusco}, {Gargano}, {Gasparrini},
  {Giglietto}, {Giordano}, {Giroletti}, {Glanzman}, {Green}, {Grondin},
  {Guillemot}, {Guiriec}, {Harding}, {Horan}, {J{\'o}hannesson}, {Kamae},
  {Kensei}, {Kuss}, {La Mura}, {Latronico}, {Lemoine-Goumard}, {Longo},
  {Loparco}, {Lovellette}, {Lubrano}, {Magill}, {Maldera}, {Manfreda},
  {Mazziotta}, {McEnery}, {Meyer}, {Michelson}, {Mirabal}, {Monzani},
  {Moretti}, {Morselli}, {Moskalenko}, {Negro}, {Nuss}, {Ojha}, {Omodei},
  {Orienti}, {Orlando}, {Palatiello}, {Paliya}, {Paneque}, {Pesce-Rollins},
  {Piron}, {Porter}, {Principe}, {Rain{\`o}}, {Rando}, {Razzano}, {Razzaque},
  {Reimer}, {Reimer}, {Reposeur}, {Rochester}, {Saz Parkinson}, {Sgr{\`o}},
  {Siskind}, {Spada}, {Spandre}, {Suson}, {Takahashi}, {Tanaka}, {Thayer},
  {Thayer}, {Thompson}, {Tibaldo}, {Torres}, {Torresi}, {Troja}, {Venters},
  {Vianello}, {Zaharijas}, {Fermi Large Area Telescope Collaboration},
  {Allison}, {Bannister}, {Dobie}, {Kaplan}, {Lenc}, {Lynch}, {Murphy},
  {Sadler}, {Australia Telescope Compact Array}, {Hotan}, {James}, {Oslowski},
  {Raja}, {Shannon}, {Whiting}, {Australian SKA Pathfinder}, {Arcavi},
  {Howell}, {McCully}, {Hosseinzadeh}, {Hiramatsu}, {Poznanski}, {Barnes},
  {Zaltzman}, {Vasylyev}, {Maoz}, {Las Cumbres Observatory Group}, {Cooke},
  {Bailes}, {Wolf}, {Deller}, {Lidman}, {Wang}, {Gendre}, {Andreoni}, {Ackley},
  {Pritchard}, {Bessell}, {Chang}, {M{\"o}ller}, {Onken}, {Scalzo},
  {Ridden-Harper}, {Sharp}, {Tucker}, {Farrell}, {Elmer}, {Johnston},
  {Venkatraman Krishnan}, {Keane}, {Green}, {Jameson}, {Hu}, {Ma}, {Sun}, {Wu},
  {Wang}, {Shang}, {Hu}, {Ashley}, {Yuan}, {Li}, {Tao}, {Zhu}, {Zhang},
  {Suntzeff}, {Zhou}, {Yang}, {Orange}, {Morris}, {Cucchiara}, {Giblin},
  {Klotz}, {Staff}, {Thierry}, {Schmidt}, {OzGrav}, {(Deeper}, {Wider},
  {program}, {AST3}, {CAASTRO Collaborations}, {Tanvir}, {Levan}, {Cano}, {de
  Ugarte-Postigo}, {Gonz{\'a}lez-Fern{\'a}ndez}, {Greiner}, {Hjorth}, {Irwin},
  {Kr{\"u}hler}, {Mandel}, {Milvang-Jensen}, {O'Brien}, {Rol}, {Rosetti},
  {Rosswog}, {Rowlinson}, {Steeghs}, {Th{\"o}ne}, {Ulaczyk}, {Watson}, {Bruun},
  {Cutter}, {Figuera Jaimes}, {Fujii}, {Fruchter}, {Gompertz}, {Jakobsson},
  {Hodosan}, {J{\`e}rgensen}, {Kangas}, {Kann}, {Rabus}, {Schr{\o}der},
  {Stanway}, {Wijers}, {VINROUGE Collaboration}, {Lipunov}, {Gorbovskoy},
  {Kornilov}, {Tyurina}, {Balanutsa}, {Kuznetsov}, {Vlasenko}, {Podesta},
  {Lopez}, {Podesta}, {Levato}, {Saffe}, {Mallamaci}, {Budnev}, {Gress},
  {Kuvshinov}, {Gorbunov}, {Vladimirov}, {Zimnukhov}, {Gabovich}, {Yurkov},
  {Sergienko}, {Rebolo}, {Serra-Ricart}, {Tlatov}, {Ishmuhametova}, {MASTER
  Collaboration}, {Abe}, {Aoki}, {Aoki}, {Asakura}, {Baar}, {Barway}, {Bond},
  {Doi}, {Finet}, {Fujiyoshi}, {Furusawa}, {Honda}, {Itoh}, {Kanda},
  {Kawabata}, {Kawabata}, {Kim}, {Koshida}, {Kuroda}, {Lee}, {Liu},
  {Matsubayashi}, {Miyazaki}, {Morihana}, {Morokuma}, {Motohara}, {Murata},
  {Nagai}, {Nagashima}, {Nagayama}, {Nakaoka}, {Nakata}, {Ohsawa}, {Ohshima},
  {Ohta}, {Okita}, {Saito}, {Saito}, {Sako}, {Sekiguchi}, {Sumi}, {Tajitsu},
  {Takahashi}, {Takayama}, {Tamura}, {Tanaka}, {Tanaka}, {Terai}, {Tominaga},
  {Tristram}, {Uemura}, {Utsumi}, {Yamaguchi}, {Yasuda}, {Yoshida}, {Zenko},
  {J-GEM}, {Adams}, {Anupama}, {Bally}, {Barway}, {Bellm}, {Blagorodnova},
  {Cannella}, {Chandra}, {Chatterjee}, {Clarke}, {Cobb}, {Cook}, {Copperwheat},
  {De}, {Emery}, {Feindt}, {Foster}, {Fox}, {Frail}, {Fremling}, {Frohmaier},
  {Garcia}, {Ghosh}, {Giacintucci}, {Goobar}, {Gottlieb}, {Grefenstette},
  {Hallinan}, {Harrison}, {Heida}, {Helou}, {Ho}, {Horesh}, {Hotokezaka}, {Ip},
  {Itoh}, {Jacobs}, {Jencson}, {Kasen}, {Kasliwal}, {Kassim}, {Kim}, {Kiran},
  {Kuin}, {Kulkarni}, {Kupfer}, {Lau}, {Madsen}, {Mazzali}, {Miller},
  {Miyasaka}, {Mooley}, {Myers}, {Nakar}, {Ngeow}, {Nugent}, {Ofek},
  {Palliyaguru}, {Pavana}, {Perley}, {Peters}, {Pike}, {Piran}, {Qi}, {Quimby},
  {Rana}, {Rosswog}, {Rusu}, {Sadler}, {Van Sistine}, {Sollerman}, {Xu}, {Yan},
  {Yatsu}, {Yu}, {Zhang}, {Zhao}, {GROWTH}, {JAGWAR}, {Caltech-NRAO},
  {TTU-NRAO}, {NuSTAR Collaborations}, {Chambers}, {Huber}, {Schultz},
  {Bulger}, {Flewelling}, {Magnier}, {Lowe}, {Wainscoat}, {Waters}, {Willman},
  {Pan-STARRS}, {Ebisawa}, {Hanyu}, {Harita}, {Hashimoto}, {Hidaka}, {Hori},
  {Ishikawa}, {Isobe}, {Iwakiri}, {Kawai}, {Kawai}, {Kawamuro}, {Kawase},
  {Kitaoka}, {Makishima}, {Matsuoka}, {Mihara}, {Morita}, {Morita}, {Nakahira},
  {Nakajima}, {Nakamura}, {Negoro}, {Oda}, {Sakamaki}, {Sasaki}, {Serino},
  {Shidatsu}, {Shimomukai}, {Sugawara}, {Sugita}, {Sugizaki}, {Tachibana},
  {Takao}, {Tanimoto}, {Tomida}, {Tsuboi}, {Tsunemi}, {Ueda}, {Ueno}, {Yamada},
  {Yamaoka}, {Yamauchi}, {Yatabe}, {Yoneyama}, {Yoshii}, {MAXI Team}, {Coward},
  {Crisp}, {Macpherson}, {Andreoni}, {Laugier}, {Noysena}, {Klotz}, {Gendre},
  {Thierry}, {Turpin}, {Consortium}, {Im}, {Choi}, {Kim}, {Yoon}, {Lim}, {Lee},
  {Lee}, {Kim}, {Ko}, {Joe}, {Kwon}, {Kim}, {Lim}, {Choi}, {KU Collaboration},
  {Fynbo}, {Malesani}, {Xu}, {Optical Telescope}, {Smartt}, {Jerkstrand},
  {Kankare}, {Sim}, {Fraser}, {Inserra}, {Maguire}, {Leloudas}, {Magee},
  {Shingles}, {Smith}, {Young}, {Kotak}, {Gal-Yam}, {Lyman}, {Homan},
  {Agliozzo}, {Anderson}, {Angus}, {Ashall}, {Barbarino}, {Bauer}, {Berton},
  {Botticella}, {Bulla}, {Cannizzaro}, {Cartier}, {Cikota}, {Clark}, {De Cia},
  {Della Valle}, {Dennefeld}, {Dessart}, {Dimitriadis}, {Elias-Rosa}, {Firth},
  {Fl{\"o}rs}, {Frohmaier}, {Galbany}, {Gonz{\'a}lez-Gait{\'a}n}, {Gromadzki},
  {Guti{\'e}rrez}, {Hamanowicz}, {Harmanen}, {Heintz}, {Hernandez}, {Hodgkin},
  {Hook}, {Izzo}, {James}, {Jonker}, {Kerzendorf}, {Kostrzewa-Rutkowska},
  {Kromer}, {Kuncarayakti}, {Lawrence}, {Manulis}, {Mattila}, {McBrien},
  {M{\"u}ller}, {Nordin}, {O'Neill}, {Onori}, {Palmerio}, {Pastorello},
  {Patat}, {Pignata}, {Podsiadlowski}, {Razza}, {Reynolds}, {Roy}, {Ruiter},
  {Rybicki}, {Salmon}, {Pumo}, {Prentice}, {Seitenzahl}, {Smith}, {Sollerman},
  {Sullivan}, {Szegedi}, {Taddia}, {Taubenberger}, {Terreran}, {Van Soelen},
  {Vos}, {Walton}, {Wright}, {Wyrzykowski}, {Yaron}, {pre=''(''>ePESSTO},
  {Chen}, {Kr{\"u}hler}, {Schady}, {Wiseman}, {Greiner}, {Rau}, {Schweyer},
  {Klose}, {Nicuesa Guelbenzu}, {GROND}, {Palliyaguru}, {Tech University},
  {Shara}, {Williams}, {Vaisanen}, {Potter}, {Romero Colmenero}, {Crawford},
  {Buckley}, {Mao}, {SALT Group}, {D{\'\i}az}, {Macri}, {Garc{\'\i}a Lambas},
  {Mendes de Oliveira}, {Nilo Castell{\'o}n}, {Ribeiro}, {S{\'a}nchez},
  {Schoenell}, {Abramo}, {Akras}, {Alcaniz}, {Artola}, {Beroiz}, {Bonoli},
  {Cabral}, {Camuccio}, {Chavushyan}, {Coelho}, {Colazo}, {Costa-Duarte},
  {Cuevas Larenas}, {Dom{\'\i}nguez Romero}, {Dultzin}, {Fern{\'a}ndez},
  {Garc{\'\i}a}, {Girardini}, {Gon{\c{c}}alves}, {Gon{\c{c}}alves}, {Gurovich},
  {Jim{\'e}nez-Teja}, {Kanaan}, {Lares}, {Lopes de Oliveira}, {L{\'o}pez-Cruz},
  {Melia}, {Molino}, {Padilla}, {Pe{\~n}uela}, {Placco}, {Qui{\~n}ones},
  {Ram{\'\i}rez Rivera}, {Renzi}, {Riguccini}, {R{\'\i}os-L{\'o}pez},
  {Rodriguez}, {Sampedro}, {Schneiter}, {Sodr{\'e}}, {Starck}, {Torres-Flores},
  {Tornatore}, {Zadro{\.z}ny}, {Castillo}, {TOROS: Transient Robotic
  Observatory of South Collaboration}, {Castro-Tirado}, {Tello}, {Hu}, {Zhang},
  {Cunniffe}, {Castell{\'o}n}, {Hiriart}, {Caballero-Garc{\'\i}a},
  {Jel{\'\i}nek}, {Kub{\'a}nek}, {P{\'e}rez del Pulgar}, {Park}, {Jeong},
  {Castro Cer{\'o}n}, {Pandey}, {Yock}, {Querel}, {Fan}, {Wang}, {BOOTES
  Collaboration}, {Beardsley}, {Brown}, {Crosse}, {Emrich}, {Franzen},
  {Gaensler}, {Horsley}, {Johnston-Hollitt}, {Kenney}, {Morales}, {Pallot},
  {Sokolowski}, {Steele}, {Tingay}, {Trott}, {Walker}, {Wayth}, {Williams},
  {Wu}, {Murchison Widefield Array}, {Yoshida}, {Sakamoto}, {Kawakubo},
  {Yamaoka}, {Takahashi}, {Asaoka}, {Ozawa}, {Torii}, {Shimizu}, {Tamura},
  {Ishizaki}, {Cherry}, {Ricciarini}, {Penacchioni}, {Marrocchesi}, {CALET
  Collaboration}, {Pozanenko}, {Volnova}, {Mazaeva}, {Minaev}, {Krugov},
  {Kusakin}, {Reva}, {Moskvitin}, {Rumyantsev}, {Inasaridze}, {Klunko},
  {Tungalag}, {Schmalz}, {Burhonov}, {IKI-GW Follow-up Collaboration},
  {Abdalla}, {Abramowski}, {Aharonian}, {Ait Benkhali}, {Ang{\"u}ner},
  {Arakawa}, {Arrieta}, {Aubert}, {Backes}, {Balzer}, {Barnard}, {Becherini},
  {Becker Tjus}, {Berge}, {Bernhard}, {Bernl{\"o}hr}, {Blackwell},
  {B{\"o}ttcher}, {Boisson}, {Bolmont}, {Bonnefoy}, {Bordas}, {Bregeon},
  {Brun}, {Brun}, {Bryan}, {B{\"u}chele}, {Bulik}, {Capasso}, {Caroff},
  {Carosi}, {Casanova}, {Cerruti}, {Chakraborty}, {Chaves}, {Chen},
  {Chevalier}, {Colafrancesco}, {Condon}, {Conrad}, {Davids}, {Decock}, {Deil},
  {Devin}, {deWilt}, {Dirson}, {Djannati-Ata{\"\i}}, {Donath}, {O'C. Drury},
  {Dutson}, {Dyks}, {Edwards}, {Egberts}, {Emery}, {Ernenwein}, {Eschbach},
  {Farnier}, {Fegan}, {Fernandes}, {Fiasson}, {Fontaine}, {Funk},
  {F{\"u}ssling}, {Gabici}, {Gallant}, {Garrigoux}, {Gat{\'e}}, {Giavitto},
  {Giebels}, {Glawion}, {Glicenstein}, {Gottschall}, {Grondin}, {Hahn},
  {Haupt}, {Hawkes}, {Heinzelmann}, {Henri}, {Hermann}, {Hinton}, {Hofmann},
  {Hoischen}, {Holch}, {Holler}, {Horns}, {Ivascenko}, {Iwasaki},
  {Jacholkowska}, {Jamrozy}, {Jankowsky}, {Jankowsky}, {Jingo}, {Jouvin},
  {Jung-Richardt}, {Kastendieck}, {Katarzy{\'n}ski}, {Katsuragawa},
  {Kerszberg}, {Khangulyan}, {Kh{\'e}lifi}, {King}, {Klepser}, {Klochkov},
  {Klu{\'z}niak}, {Komin}, {Kosack}, {Krakau}, {Kraus}, {Kr{\"u}ger}, {Laffon},
  {Lamanna}, {Lau}, {Lees}, {Lefaucheur}, {Lemi{\`e}re}, {Lemoine-Goumard},
  {Lenain}, {Leser}, {Lohse}, {Lorentz}, {Liu}, {Lypova}, {Malyshev},
  {Marandon}, {Marcowith}, {Mariaud}, {Marx}, {Maurin}, {Maxted}, {Mayer},
  {Meintjes}, {Meyer}, {Mitchell}, {Moderski}, {Mohamed}, {Mohrmann},
  {Mor{\r{a}}}, {Moulin}, {Murach}, {Nakashima}, {de Naurois}, {Ndiyavala},
  {Niederwanger}, {Niemiec}, {Oakes}, {O'Brien}, {Odaka}, {Ohm}, {Ostrowski},
  {Oya}, {Padovani}, {Panter}, {Parsons}, {Pekeur}, {Pelletier}, {Perennes},
  {Petrucci}, {Peyaud}, {Piel}, {Pita}, {Poireau}, {Poon}, {Prokhorov},
  {Prokoph}, {P{\"u}hlhofer}, {Punch}, {Quirrenbach}, {Raab}, {Rauth},
  {Reimer}, {Reimer}, {Renaud}, {de los Reyes}, {Rieger}, {Rinchiuso},
  {Romoli}, {Rowell}, {Rudak}, {Rulten}, {Sahakian}, {Saito}, {Sanchez},
  {Santangelo}, {Sasaki}, {Schlickeiser}, {Sch{\"u}ssler}, {Schulz},
  {Schwanke}, {Schwemmer}, {Seglar-Arroyo}, {Settimo}, {Seyffert}, {Shafi},
  {Shilon}, {Shiningayamwe}, {Simoni}, {Sol}, {Spanier}, {Spir-Jacob},
  {Stawarz}, {Steenkamp}, {Stegmann}, {Steppa}, {Sushch}, {Takahashi},
  {Tavernet}, {Tavernier}, {Taylor}, {Terrier}, {Tibaldo}, {Tiziani},
  {Tluczykont}, {Trichard}, {Tsirou}, {Tsuji}, {Tuffs}, {Uchiyama}, {van der
  Walt}, {van Eldik}, {van Rensburg}, {van Soelen}, {Vasileiadis}, {Veh},
  {Venter}, {Viana}, {Vincent}, {Vink}, {Voisin}, {V{\"o}lk}, {Vuillaume},
  {Wadiasingh}, {Wagner}, {Wagner}, {Wagner}, {White}, {Wierzcholska},
  {Willmann}, {W{\"o}rnlein}, {Wouters}, {Yang}, {Zaborov}, {Zacharias},
  {Zanin}, {Zdziarski}, {Zech}, {Zefi}, {Ziegler}, {Zorn}, {{\.Z}ywucka},
  {H.~E.~S.~S. Collaboration}, {Fender}, {Broderick}, {Rowlinson}, {Wijers},
  {Stewart}, {ter Veen}, {Shulevski}, {LOFAR Collaboration}, {Kavic},
  {Simonetti}, {League}, {Tsai}, {Obenberger}, {Nathaniel}, {Taylor}, {Dowell},
  {Liebling}, {Estes}, {Lippert}, {Sharma}, {Vincent}, {Farella}, {Wavelength
  Array}, {Abeysekara}, {Albert}, {Alfaro}, {Alvarez}, {Arceo},
  {Arteaga-Vel{\'a}zquez}, {Avila Rojas}, {Ayala Solares}, {Barber}, {Becerra
  Gonzalez}, {Becerril}, {Belmont-Moreno}, {BenZvi}, {Berley}, {Bernal},
  {Braun}, {Brisbois}, {Caballero-Mora}, {Capistr{\'a}n}, {Carrami{\~n}ana},
  {Casanova}, {Castillo}, {Cotti}, {Cotzomi}, {Couti{\~n}o de Le{\'o}n}, {De
  Le{\'o}n}, {De la Fuente}, {Diaz Hernandez}, {Dichiara}, {Dingus},
  {DuVernois}, {D{\'\i}az-V{\'e}lez}, {Ellsworth}, {Engel},
  {Enr{\'\i}quez-Rivera}, {Fiorino}, {Fleischhack}, {Fraija},
  {Garc{\'\i}a-Gonz{\'a}lez}, {Garfias}, {Gerhardt}, {Gonz{\~o}lez Mu{\~n}oz},
  {Gonz{\'a}lez}, {Goodman}, {Hampel-Arias}, {Harding}, {Hernandez},
  {Hernandez-Almada}, {Hona}, {H{\"u}ntemeyer}, {Iriarte}, {Jardin-Blicq},
  {Joshi}, {Kaufmann}, {Kieda}, {Lara}, {Lauer}, {Lennarz}, {Le{\'o}n Vargas},
  {Linnemann}, {Longinotti}, {Raya}, {Luna-Garc{\'\i}a}, {L{\'o}pez-Coto},
  {Malone}, {Marinelli}, {Martinez}, {Martinez-Castellanos},
  {Mart{\'\i}nez-Castro}, {Mart{\'\i}nez-Huerta}, {Matthews},
  {Miranda-Romagnoli}, {Moreno}, {Mostaf{\'a}}, {Nellen}, {Newbold}, {Nisa},
  {Noriega-Papaqui}, {Pelayo}, {Pretz}, {P{\'e}rez-P{\'e}rez}, {Ren}, {Rho},
  {Rivi{\`e}re}, {Rosa-Gonz{\'a}lez}, {Rosenberg}, {Ruiz-Velasco}, {Salazar},
  {Salesa Greus}, {Sandoval}, {Schneider}, {Schoorlemmer}, {Sinnis}, {Smith},
  {Springer}, {Surajbali}, {Tibolla}, {Tollefson}, {Torres}, {Ukwatta},
  {Weisgarber}, {Westerhoff}, {Wisher}, {Wood}, {Yapici}, {Yodh}, {Younk},
  {Zhou}, {{\'A}lvarez}, {HAWC Collaboration}, {Aab}, {Abreu}, {Aglietta},
  {Albuquerque}, {Albury}, {Allekotte}, {Almela}, {Alvarez Castillo},
  {Alvarez-Mu{\~n}iz}, {Anastasi}, {Anchordoqui}, {Andrada}, {Andringa},
  {Aramo}, {Arsene}, {Asorey}, {Assis}, {Avila}, {Badescu}, {Balaceanu},
  {Barbato}, {Barreira Luz}, {Becker}, {Bellido}, {Berat}, {Bertaina},
  {Bertou}, {Biermann}, {Biteau}, {Blaess}, {Blanco}, {Blazek}, {Bleve},
  {Boh{\'a}{\v{c}}ov{\'a}}, {Bonifazi}, {Borodai}, {Botti}, {Brack}, {Brancus},
  {Bretz}, {Bridgeman}, {Briechle}, {Buchholz}, {Bueno}, {Buitink}, {Buscemi},
  {Caballero-Mora}, {Caccianiga}, {Cancio}, {Canfora}, {Caruso}, {Castellina},
  {Catalani}, {Cataldi}, {Cazon}, {Chavez}, {Chinellato}, {Chudoba}, {Clay},
  {Cobos Cerutti}, {Colalillo}, {Coleman}, {Collica}, {Coluccia},
  {Concei{\c{c}}{\~a}o}, {Consolati}, {Contreras}, {Cooper}, {Coutu},
  {Covault}, {Cronin}, {D'Amico}, {Daniel}, {Dasso}, {Daumiller}, {Dawson},
  {Day}, {de Almeida}, {de Jong}, {De Mauro}, {de Mello Neto}, {De Mitri}, {de
  Oliveira}, {de Souza}, {Debatin}, {Deligny}, {D{\'\i}az Castro}, {Diogo},
  {Dobrigkeit}, {D'Olivo}, {Dorosti}, {Dos Anjos}, {Dova}, {Dundovic}, {Ebr},
  {Engel}, {Erdmann}, {Erfani}, {Escobar}, {Espadanal}, {Etchegoyen}, {Falcke},
  {Farmer}, {Farrar}, {Fauth}, {Fazzini}, {Feldbusch}, {Fenu}, {Fick},
  {Figueira}, {Filip{\v{c}}i{\v{c}}}, {Freire}, {Fujii}, {Fuster},
  {Ga{\"\i}or}, {Garc{\'\i}a}, {Gat{\'e}}, {Gemmeke}, {Gherghel-Lascu}, {Ghia},
  {Giaccari}, {Giammarchi}, {Giller}, {G{\l}as}, {Glaser}, {Golup}, {G{\'o}mez
  Berisso}, {G{\'o}mez Vitale}, {Gonz{\'a}lez}, {Gorgi}, {Gottowik}, {Grillo},
  {Grubb}, {Guarino}, {Guedes}, {Halliday}, {Hampel}, {Hansen}, {Harari},
  {Harrison}, {Harvey}, {Haungs}, {Hebbeker}, {Heck}, {Heimann}, {Herve},
  {Hill}, {Hojvat}, {Holt}, {Homola}, {H{\"o}randel}, {Horvath},
  {Hrabovsk{\'y}}, {Huege}, {Hulsman}, {Insolia}, {Isar}, {Jandt}, {Johnsen},
  {Josebachuili}, {Jurysek}, {K{\"a}{\"a}p{\"a}}, {Kampert}, {Keilhauer},
  {Kemmerich}, {Kemp}, {Kieckhafer}, {Klages}, {Kleifges}, {Kleinfeller},
  {Krause}, {Krohm}, {Kuempel}, {Kukec Mezek}, {Kunka}, {Kuotb Awad}, {Lago},
  {LaHurd}, {Lang}, {Lauscher}, {Legumina}, {Leigui de Oliveira},
  {Letessier-Selvon}, {Lhenry-Yvon}, {Link}, {Lo Presti}, {Lopes}, {L{\'o}pez},
  {L{\'o}pez Casado}, {Lorek}, {Luce}, {Lucero}, {Malacari}, {Mallamaci},
  {Mandat}, {Mantsch}, {Mariazzi}, {Maris}, {Marsella}, {Martello}, {Martinez},
  {Mart{\'\i}nez Bravo}, {Mas{\'\i}as Meza}, {Mathes}, {Mathys}, {Matthews},
  {Matthiae}, {Mayotte}, {Mazur}, {Medina}, {Medina-Tanco}, {Melo},
  {Menshikov}, {Merenda}, {Michal}, {Micheletti}, {Middendorf}, {Miramonti},
  {Mitrica}, {Mockler}, {Mollerach}, {Montanet}, {Morello}, {Morlino},
  {M{\"u}ller}, {M{\"u}ller}, {Muller}, {M{\"u}ller}, {Mussa}, {Naranjo},
  {Nguyen}, {Niculescu-Oglinzanu}, {Niechciol}, {Niemietz}, {Niggemann},
  {Nitz}, {Nosek}, {Novotny}, {No{\v{z}}ka}, {N{\'u}{\~n}ez}, {Oikonomou},
  {Olinto}, {Palatka}, {Pallotta}, {Papenbreer}, {Parente}, {Parra}, {Paul},
  {Pech}, {Pedreira}, {P{\c{e}}kala}, {Pe{\~n}a-Rodriguez}, {Pereira},
  {Perlin}, {Perrone}, {Peters}, {Petrera}, {Phuntsok}, {Pierog}, {Pimenta},
  {Pirronello}, {Platino}, {Plum}, {Poh}, {Porowski}, {Prado}, {Privitera},
  {Prouza}, {Quel}, {Querchfeld}, {Quinn}, {Ramos-Pollan}, {Rautenberg},
  {Ravignani}, {Ridky}, {Riehn}, {Risse}, {Ristori}, {Rizi}, {Rodrigues de
  Carvalho}, {Rodriguez Fernandez}, {Rodriguez Rojo}, {Roncoroni}, {Roth},
  {Roulet}, {Rovero}, {Ruehl}, {Saffi}, {Saftoiu}, {Salamida}, {Salazar},
  {Saleh}, {Salina}, {S{\'a}nchez}, {Sanchez-Lucas}, {Santos}, {Santos},
  {Sarazin}, {Sarmento}, {Sarmiento-Cano}, {Sato}, {Schauer}, {Scherini},
  {Schieler}, {Schimp}, {Schmidt}, {Scholten}, {Schov{\'a}nek}, {Schr{\"o}der},
  {Schr{\"o}der}, {Schulz}, {Schumacher}, {Sciutto}, {Segreto}, {Shadkam},
  {Shellard}, {Sigl}, {Silli}, {{\v{S}}m{\'\i}da}, {Snow}, {Sommers},
  {Sonntag}, {Soriano}, {Squartini}, {Stanca}, {Stani{\v{c}}}, {Stasielak},
  {Stassi}, {Stolpovskiy}, {Strafella}, {Streich}, {Suarez},
  {Suarez-Dur{\'a}n}, {Sudholz}, {Suomij{\"a}rvi}, {Supanitsky},
  {{\v{S}}up{\'\i}k}, {Swain}, {Szadkowski}, {Taboada}, {Taborda},
  {Timmermans}, {Todero Peixoto}, {Tomankova}, {Tom{\'e}}, {Torralba Elipe},
  {Travnicek}, {Trini}, {Tueros}, {Ulrich}, {Unger}, {Urban}, {Vald{\'e}s
  Galicia}, {Vali{\~n}o}, {Valore}, {van Aar}, {van Bodegom}, {van den Berg},
  {van Vliet}, {Varela}, {Vargas C{\'a}rdenas}, {V{\'a}zquez}, {Veberi{\v{c}}},
  {Ventura}, {Vergara Quispe}, {Verzi}, {Vicha}, {Villase{\~n}or}, {Vorobiov},
  {Wahlberg}, {Wainberg}, {Walz}, {Watson}, {Weber}, {Weindl}, {Wiede{\'n}ski},
  {Wiencke}, {Wilczy{\'n}ski}, {Wirtz}, {Wittkowski}, {Wundheiler}, {Yang},
  {Yushkov}, {Zas}, {Zavrtanik}, {Zavrtanik}, {Zepeda}, {Zimmermann},
  {Ziolkowski}, {Zong}, {Zuccarello}, {Pierre Auger Collaboration}, {Kim},
  {Schulze}, {Bauer}, {Corral-Santana}, {de Gregorio-Monsalvo},
  {Gonz{\'a}lez-L{\'o}pez}, {Hartmann}, {Ishwara-Chandra}, {Mart{\'\i}n},
  {Mehner}, {Misra}, {Micha{\l}owski}, {Resmi}, {ALMA Collaboration}, {Paragi},
  {Agudo}, {An}, {Beswick}, {Casadio}, {Frey}, {Jonker}, {Kettenis}, {Marcote},
  {Moldon}, {Szomoru}, {van Langevelde}, {Yang}, {Euro VLBI Team}, {Cwiek},
  {Cwiok}, {Czyrkowski}, {Dabrowski}, {Kasprowicz}, {Mankiewicz}, {Nawrocki},
  {Opiela}, {Piotrowski}, {Wrochna}, {Zaremba}, {{\.Z}arnecki}, {Pi of the Sky
  Collaboration}, {Haggard}, {Nynka}, {Ruan}, {Chandra Team at McGill
  University}, {Bland}, {Booler}, {Devillepoix}, {de Gois}, {Hancock}, {Howie},
  {Paxman}, {Sansom}, {Towner}, {Desert Fireball Network}, {Tonry}, {Coughlin},
  {Stubbs}, {Denneau}, {Heinze}, {Stalder}, {Weiland}, {ATLAS}, {Eatough},
  {Kramer}, {Kraus}, {Time Resolution Universe Survey}, {Troja}, {Piro},
  {Becerra Gonz{\'a}lez}, {Butler}, {Fox}, {Khandrika}, {Kutyrev}, {Lee},
  {Ricci}, {Ryan}, {S{\'a}nchez-Ram{\'\i}rez}, {Veilleux}, {Watson},
  {Wieringa}, {Burgess}, {van Eerten}, {Fontes}, {Fryer}, {Korobkin},
  {Wollaeger}, {RIMAS}, {RATIR}, {Camilo}, {Foley}, {Goedhart}, {Makhathini},
  {Oozeer}, {Smirnov}, {Fender}, {Woudt}, \& {South
  Africa/MeerKAT}}]{Abbott2017c}
---. 2017{\natexlab{c}}, \apjl, 848, L12, \dodoi{10.3847/2041-8213/aa91c9}

\bibitem[{{Alexander} {et~al.}(2017){Alexander}, {Berger}, {Fong}, {Williams},
  {Guidorzi}, {Margutti}, {Metzger}, {Annis}, {Blanchard}, {Brout}, {Brown},
  {Chen}, {Chornock}, {Cowperthwaite}, {Drout}, {Eftekhari}, {Frieman}, {Holz},
  {Nicholl}, {Rest}, {Sako}, {Soares-Santos}, \& {Villar}}]{Alexander2017}
{Alexander}, K.~D., {Berger}, E., {Fong}, W., {et~al.} 2017, \apjl, 848, L21,
  \dodoi{10.3847/2041-8213/aa905d}

\bibitem[{{Arcavi} {et~al.}(2017){Arcavi}, {Hosseinzadeh}, {Howell}, {McCully},
  {Poznanski}, {Kasen}, {Barnes}, {Zaltzman}, {Vasylyev}, {Maoz}, \&
  {Valenti}}]{Arcavi2017}
{Arcavi}, I., {Hosseinzadeh}, G., {Howell}, D.~A., {et~al.} 2017, \nat, 551,
  64, \dodoi{10.1038/nature24291}

\bibitem[{{Balasubramanian} {et~al.}(2021){Balasubramanian}, {Corsi}, {Mooley},
  {Brightman}, {Hallinan}, {Hotokezaka}, {Kaplan}, {Lazzati}, \&
  {Murphy}}]{Balasubramanian2021}
{Balasubramanian}, A., {Corsi}, A., {Mooley}, K.~P., {et~al.} 2021, \apjl, 914,
  L20, \dodoi{10.3847/2041-8213/abfd38}

\bibitem[{{Balasubramanian} {et~al.}(2022){Balasubramanian}, {Corsi}, {Mooley},
  {Hotokezaka}, {Kaplan}, {Frail}, {Hallinan}, {Lazzati}, \&
  {Murphy}}]{Balasubramanian2022}
---. 2022, \apj, 938, 12, \dodoi{10.3847/1538-4357/ac9133}

\bibitem[{{Bartos} {et~al.}(2019){Bartos}, {Lee}, {Corsi}, {M{\'a}rka}, \&
  {M{\'a}rka}}]{Bartos2019}
{Bartos}, I., {Lee}, K.~H., {Corsi}, A., {M{\'a}rka}, Z., \& {M{\'a}rka}, S.
  2019, \mnras, 485, 4150, \dodoi{10.1093/mnras/stz719}

\bibitem[{{Baumgartner} {et~al.}(2009){Baumgartner}, {Barthelmy}, {Cummings},
  {Fenimore}, {Gehrels}, {Krimm}, {Mangano}, {Markwardt}, {Palmer}, {Sakamoto},
  {Sato}, {Stamatikos}, {Tueller}, \& {Ukwatta}}]{Baumgartner2009}
{Baumgartner}, W.~H., {Barthelmy}, S.~D., {Cummings}, J.~R., {et~al.} 2009,
  Gamma-Ray Coordinate Network, 9138

\bibitem[{{Beardmore} {et~al.}(2012){Beardmore}, {Barthelmy}, {Burrows},
  {Cummings}, {D'Elia}, {Markwardt}, {Marshall}, {Page}, {Sakamoto},
  {Sbarufatti}, {Swenson}, {et~al.}}]{Beardmore2012}
{Beardmore}, A.~P., {Barthelmy}, S.~D., {Burrows}, D.~N., {et~al.} 2012,
  Gamma-Ray Coordinate Network, 13191

\bibitem[{{Becker} {et~al.}(1994){Becker}, {White}, \& {Helfand}}]{Becker1994}
{Becker}, R.~H., {White}, R.~L., \& {Helfand}, D.~J. 1994, in Astronomical
  Society of the Pacific Conference Series, Vol.~61, Astronomical Data Analysis
  Software and Systems III, ed. D.~R. {Crabtree}, R.~J. {Hanisch}, \&
  J.~{Barnes}, 165

\bibitem[{{Bennett} {et~al.}(2014){Bennett}, {Larson}, {Weiland}, \&
  {Hinshaw}}]{Bennett2014}
{Bennett}, C.~L., {Larson}, D., {Weiland}, J.~L., \& {Hinshaw}, G. 2014, \apj,
  794, 135, \dodoi{10.1088/0004-637X/794/2/135}

\bibitem[{{Berger}(2009)}]{Berger2009}
{Berger}, E. 2009, \apj, 690, 231, \dodoi{10.1088/0004-637X/690/1/231}

\bibitem[{{Berger}(2014)}]{Berger2014}
---. 2014, \araa, 52, 43, \dodoi{10.1146/annurev-astro-081913-035926}

\bibitem[{{Bruni} {et~al.}(2021){Bruni}, {O'Connor}, {Matsumoto}, {Troja},
  {Piran}, {Piro}, \& {Ricci}}]{Bruni2021}
{Bruni}, G., {O'Connor}, B., {Matsumoto}, T., {et~al.} 2021, \mnras, 505, L41,
  \dodoi{10.1093/mnrasl/slab046}

\bibitem[{{Chornock} {et~al.}(2017){Chornock}, {Berger}, {Kasen},
  {Cowperthwaite}, {Nicholl}, {Villar}, {Alexander}, {Blanchard}, {Eftekhari},
  {Fong}, {Margutti}, {Williams}, {Annis}, {Brout}, {Brown}, {Chen}, {Drout},
  {Farr}, {Foley}, {Frieman}, {Fryer}, {Herner}, {Holz}, {Kessler}, {Matheson},
  {Metzger}, {Quataert}, {Rest}, {Sako}, {Scolnic}, {Smith}, \&
  {Soares-Santos}}]{Chornock2017}
{Chornock}, R., {Berger}, E., {Kasen}, D., {et~al.} 2017, \apjl, 848, L19,
  \dodoi{10.3847/2041-8213/aa905c}

\bibitem[{{Condon}(1992)}]{Condon1992}
{Condon}, J.~J. 1992, \araa, 30, 575,
  \dodoi{10.1146/annurev.aa.30.090192.003043}

\bibitem[{{Coulter} {et~al.}(2017){Coulter}, {Foley}, {Kilpatrick}, {Drout},
  {Piro}, {Shappee}, {Siebert}, {Simon}, {Ulloa}, {Kasen}, {Madore},
  {Murguia-Berthier}, {Pan}, {Prochaska}, {Ramirez-Ruiz}, {Rest}, \&
  {Rojas-Bravo}}]{Coulter2017}
{Coulter}, D.~A., {Foley}, R.~J., {Kilpatrick}, C.~D., {et~al.} 2017, Science,
  358, 1556, \dodoi{10.1126/science.aap9811}

\bibitem[{{Cowperthwaite} {et~al.}(2017){Cowperthwaite}, {Berger}, {Villar},
  {Metzger}, {Nicholl}, {Chornock}, {Blanchard}, {Fong}, {Margutti},
  {Soares-Santos}, {Alexander}, {Allam}, {Annis}, {Brout}, {Brown}, {Butler},
  {Chen}, {Diehl}, {Doctor}, {Drout}, {Eftekhari}, {Farr}, {Finley}, {Foley},
  {Frieman}, {Fryer}, {Garc{\'\i}a-Bellido}, {Gill}, {Guillochon}, {Herner},
  {Holz}, {Kasen}, {Kessler}, {Marriner}, {Matheson}, {Neilsen}, {Quataert},
  {Palmese}, {Rest}, {Sako}, {Scolnic}, {Smith}, {Tucker}, {Williams},
  {Balbinot}, {Carlin}, {Cook}, {Durret}, {Li}, {Lopes}, {Louren{\c{c}}o},
  {Marshall}, {Medina}, {Muir}, {Mu{\~n}oz}, {Sauseda}, {Schlegel}, {Secco},
  {Vivas}, {Wester}, {Zenteno}, {Zhang}, {Abbott}, {Banerji}, {Bechtol},
  {Benoit-L{\'e}vy}, {Bertin}, {Buckley-Geer}, {Burke}, {Capozzi}, {Carnero
  Rosell}, {Carrasco Kind}, {Castander}, {Crocce}, {Cunha}, {D'Andrea}, {da
  Costa}, {Davis}, {DePoy}, {Desai}, {Dietrich}, {Drlica-Wagner}, {Eifler},
  {Evrard}, {Fernandez}, {Flaugher}, {Fosalba}, {Gaztanaga}, {Gerdes},
  {Giannantonio}, {Goldstein}, {Gruen}, {Gruendl}, {Gutierrez}, {Honscheid},
  {Jain}, {James}, {Jeltema}, {Johnson}, {Johnson}, {Kent}, {Krause}, {Kron},
  {Kuehn}, {Nuropatkin}, {Lahav}, {Lima}, {Lin}, {Maia}, {March}, {Martini},
  {McMahon}, {Menanteau}, {Miller}, {Miquel}, {Mohr}, {Neilsen}, {Nichol},
  {Ogando}, {Plazas}, {Roe}, {Romer}, {Roodman}, {Rykoff}, {Sanchez},
  {Scarpine}, {Schindler}, {Schubnell}, {Sevilla-Noarbe}, {Smith}, {Smith},
  {Sobreira}, {Suchyta}, {Swanson}, {Tarle}, {Thomas}, {Thomas}, {Troxel},
  {Vikram}, {Walker}, {Wechsler}, {Weller}, {Yanny}, \&
  {Zuntz}}]{Cowperthwaite2017}
{Cowperthwaite}, P.~S., {Berger}, E., {Villar}, V.~A., {et~al.} 2017, \apjl,
  848, L17, \dodoi{10.3847/2041-8213/aa8fc7}

\bibitem[{{Cummings} {et~al.}(2011){Cummings}, {Palmer}, \&
  {others}}]{Cummings2011}
{Cummings}, J.~R., {Palmer}, D.~M., \& {others}. 2011, Gamma-Ray Coordinate
  Network, 12599

\bibitem[{{Cummings} {et~al.}(2008){Cummings}, {Palmer},
  {et~al.}}]{Cummings2008}
{Cummings}, J.~R., {Palmer}, D.~M., {et~al.} 2008, Gamma-Ray Coordinate
  Network, 7209

\bibitem[{{Cummings} {et~al.}(2014){Cummings}, {Barthelmy}, {Baumgartner},
  {Gehrels}, {Krimm}, {Lien}, {Markwardt}, {Palmer}, {Sakamoto}, {Stamatikos},
  {Stroh}, {Tueller}, \& {Ukwatta}}]{Cummings2014}
{Cummings}, J.~R., {Barthelmy}, S.~D., {Baumgartner}, W.~H., {et~al.} 2014,
  Gamma-Ray Coordinate Network, 16354

\bibitem[{{Dichiara} {et~al.}(2020){Dichiara}, {Troja}, {O'Connor}, {Marshall},
  {Beniamini}, {Cannizzo}, {Lien}, \& {Sakamoto}}]{Dichiara2020}
{Dichiara}, S., {Troja}, E., {O'Connor}, B., {et~al.} 2020, \mnras, 492, 5011,
  \dodoi{10.1093/mnras/staa124}

\bibitem[{{Drout} {et~al.}(2017){Drout}, {Piro}, {Shappee}, {Kilpatrick},
  {Simon}, {Contreras}, {Coulter}, {Foley}, {Siebert}, {Morrell}, {Boutsia},
  {Di Mille}, {Holoien}, {Kasen}, {Kollmeier}, {Madore}, {Monson},
  {Murguia-Berthier}, {Pan}, {Prochaska}, {Ramirez-Ruiz}, {Rest}, {Adams},
  {Alatalo}, {Ba{\~n}ados}, {Baughman}, {Beers}, {Bernstein}, {Bitsakis},
  {Campillay}, {Hansen}, {Higgs}, {Ji}, {Maravelias}, {Marshall}, {Moni Bidin},
  {Prieto}, {Rasmussen}, {Rojas-Bravo}, {Strom}, {Ulloa},
  {Vargas-Gonz{\'a}lez}, {Wan}, \& {Whitten}}]{Dorut2017}
{Drout}, M.~R., {Piro}, A.~L., {Shappee}, B.~J., {et~al.} 2017, Science, 358,
  1570, \dodoi{10.1126/science.aaq0049}

\bibitem[{{Evans} {et~al.}(2010){Evans}, {Primini}, {Glotfelty}, {Anderson},
  {Bonaventura}, {Chen}, {Davis}, {Doe}, {Evans}, {Fabbiano}, {Galle}, {Gibbs},
  {Grier}, {Hain}, {Hall}, {Harbo}, {He}, {Houck}, {Karovska}, {Kashyap},
  {Lauer}, {McCollough}, {McDowell}, {Miller}, {Mitschang}, {Morgan},
  {Mossman}, {Nichols}, {Nowak}, {Plummer}, {Refsdal}, {Rots}, {Siemiginowska},
  {Sundheim}, {Tibbetts}, {Van Stone}, {Winkelman}, \& {Zografou}}]{Evans2010}
{Evans}, I.~N., {Primini}, F.~A., {Glotfelty}, K.~J., {et~al.} 2010, \apjs,
  189, 37, \dodoi{10.1088/0067-0049/189/1/37}

\bibitem[{{Fong} \& {Berger}(2013)}]{Fong2013b}
{Fong}, W., \& {Berger}, E. 2013, \apj, 776, 18,
  \dodoi{10.1088/0004-637X/776/1/18}

\bibitem[{{Fong} {et~al.}(2015){Fong}, {Berger}, {Margutti}, \&
  {Zauderer}}]{Fong2015}
{Fong}, W., {Berger}, E., {Margutti}, R., \& {Zauderer}, B.~A. 2015, \apj, 815,
  102, \dodoi{10.1088/0004-637X/815/2/102}

\bibitem[{{Fong} {et~al.}(2013){Fong}, {Berger}, {Chornock}, {Margutti},
  {Levan}, {Tanvir}, {Tunnicliffe}, {Czekala}, {Fox}, {Perley}, {Cenko},
  {Zauderer}, {Laskar}, {Persson}, {Monson}, {Kelson}, {Birk}, {Murphy},
  {Servillat}, \& {Anglada}}]{Fong2013a}
{Fong}, W., {Berger}, E., {Chornock}, R., {et~al.} 2013, \apj, 769, 56,
  \dodoi{10.1088/0004-637X/769/1/56}

\bibitem[{{Ghirlanda} {et~al.}(2019){Ghirlanda}, {Salafia}, {Paragi},
  {Giroletti}, {Yang}, {Marcote}, {Blanchard}, {Agudo}, {An}, {Bernardini},
  {Beswick}, {Branchesi}, {Campana}, {Casadio}, {Chassande-Mottin}, {Colpi},
  {Covino}, {D'Avanzo}, {D'Elia}, {Frey}, {Gawronski}, {Ghisellini}, {Gurvits},
  {Jonker}, {van Langevelde}, {Melandri}, {Moldon}, {Nava}, {Perego},
  {Perez-Torres}, {Reynolds}, {Salvaterra}, {Tagliaferri}, {Venturi},
  {Vergani}, \& {Zhang}}]{Ghirlanda2019}
{Ghirlanda}, G., {Salafia}, O.~S., {Paragi}, Z., {et~al.} 2019, Science, 363,
  968, \dodoi{10.1126/science.aau8815}

\bibitem[{{Ghosh} {et~al.}(2022){Ghosh}, {Vaishnava}, {Resmi}, {Misra}, {Arun},
  {Omar}, \& {Chakradhari}}]{Ghosh2022}
{Ghosh}, A., {Vaishnava}, C.~S., {Resmi}, L., {et~al.} 2022, arXiv e-prints,
  arXiv:2207.10001.
\newblock \doarXiv{2207.10001}

\bibitem[{{Grandorf} {et~al.}(2021){Grandorf}, {McCarty}, {Rajkumar}, {Harbin},
  {Lee}, {Corsi}, {Bartos}, {M{\'a}rka}, {Balasubramanian}, \&
  {M{\'a}rka}}]{Grandorf2021}
{Grandorf}, C., {McCarty}, J., {Rajkumar}, P., {et~al.} 2021, \apj, 908, 63,
  \dodoi{10.3847/1538-4357/abd315}

\bibitem[{{Granot} {et~al.}(2018){Granot}, {Gill}, {Guetta}, \& {De
  Colle}}]{Granot2018}
{Granot}, J., {Gill}, R., {Guetta}, D., \& {De Colle}, F. 2018, \mnras, 481,
  1597, \dodoi{10.1093/mnras/sty2308}

\bibitem[{{Gupte} \& {Bartos}(2018)}]{Gupte2018}
{Gupte}, N., \& {Bartos}, I. 2018, arXiv e-prints, arXiv:1808.06238.
\newblock \doarXiv{1808.06238}

\bibitem[{{G{\"u}rkan} {et~al.}(2014){G{\"u}rkan}, {Hardcastle}, \&
  {Jarvis}}]{Gurkin2014}
{G{\"u}rkan}, G., {Hardcastle}, M.~J., \& {Jarvis}, M.~J. 2014, \mnras, 438,
  1149, \dodoi{10.1093/mnras/stt2264}

\bibitem[{{Haggard} {et~al.}(2017){Haggard}, {Nynka}, {Ruan}, {Kalogera},
  {Cenko}, {Evans}, \& {Kennea}}]{Haggard2017}
{Haggard}, D., {Nynka}, M., {Ruan}, J.~J., {et~al.} 2017, \apjl, 848, L25,
  \dodoi{10.3847/2041-8213/aa8ede}

\bibitem[{{Hajela} {et~al.}(2022){Hajela}, {Margutti}, {Bright}, {Alexander},
  {Metzger}, {Nedora}, {Kathirgamaraju}, {Margalit}, {Radice}, {Guidorzi},
  {Berger}, {MacFadyen}, {Giannios}, {Chornock}, {Heywood}, {Sironi},
  {Gottlieb}, {Coppejans}, {Laskar}, {Cendes}, {Duran}, {Eftekhari}, {Fong},
  {McDowell}, {Nicholl}, {Xie}, {Zrake}, {Bernuzzi}, {Broekgaarden},
  {Kilpatrick}, {Terreran}, {Villar}, {Blanchard}, {Gomez}, {Hosseinzadeh},
  {Matthews}, \& {Rastinejad}}]{Hajela2022}
{Hajela}, A., {Margutti}, R., {Bright}, J.~S., {et~al.} 2022, \apjl, 927, L17,
  \dodoi{10.3847/2041-8213/ac504a}

\bibitem[{{Hales} {et~al.}(2012{\natexlab{a}}){Hales}, {Murphy}, {Curran},
  {Middelberg}, {Gaensler}, \& {Norris}}]{Hales2012a}
{Hales}, C.~A., {Murphy}, T., {Curran}, J.~R., {et~al.} 2012{\natexlab{a}},
  {BLOBCAT: Software to Catalog Blobs}.
\newblock \doeprint{1208.009}

\bibitem[{{Hales} {et~al.}(2012{\natexlab{b}}){Hales}, {Murphy}, {Curran},
  {Middelberg}, {Gaensler}, \& {Norris}}]{Hales2012b}
---. 2012{\natexlab{b}}, \mnras, 425, 979,
  \dodoi{10.1111/j.1365-2966.2012.21373.x}

\bibitem[{{Hallinan} {et~al.}(2017){Hallinan}, {Corsi}, {Mooley}, {Hotokezaka},
  {Nakar}, {Kasliwal}, {Kaplan}, {Frail}, {Myers}, {Murphy}, {De}, {Dobie},
  {Allison}, {Bannister}, {Bhalerao}, {Chandra}, {Clarke}, {Giacintucci}, {Ho},
  {Horesh}, {Kassim}, {Kulkarni}, {Lenc}, {Lockman}, {Lynch}, {Nichols},
  {Nissanke}, {Palliyaguru}, {Peters}, {Piran}, {Rana}, {Sadler}, \&
  {Singer}}]{Hallinan2017}
{Hallinan}, G., {Corsi}, A., {Mooley}, K.~P., {et~al.} 2017, Science, 358,
  1579, \dodoi{10.1126/science.aap9855}

\bibitem[{{Horesh} {et~al.}(2016){Horesh}, {Hotokezaka}, {Piran}, {Nakar}, \&
  {Hancock}}]{Horesh2016}
{Horesh}, A., {Hotokezaka}, K., {Piran}, T., {Nakar}, E., \& {Hancock}, P.
  2016, \apjl, 819, L22, \dodoi{10.3847/2041-8205/819/2/L22}

\bibitem[{{Hotokezaka} {et~al.}(2018){Hotokezaka}, {Kiuchi}, {Shibata},
  {Nakar}, \& {Piran}}]{Hotokezaka2018}
{Hotokezaka}, K., {Kiuchi}, K., {Shibata}, M., {Nakar}, E., \& {Piran}, T.
  2018, \apj, 867, 95, \dodoi{10.3847/1538-4357/aadf92}

\bibitem[{{Hotokezaka} \& {Piran}(2015)}]{Hotokezaka2015}
{Hotokezaka}, K., \& {Piran}, T. 2015, \mnras, 450, 1430,
  \dodoi{10.1093/mnras/stv620}

\bibitem[{{Huynh} {et~al.}(2005){Huynh}, {Jackson}, {Norris}, \&
  {Prandoni}}]{Huynh2005}
{Huynh}, M.~T., {Jackson}, C.~A., {Norris}, R.~P., \& {Prandoni}, I. 2005, \aj,
  130, 1373, \dodoi{10.1086/432873}

\bibitem[{{Itoh} {et~al.}(2020){Itoh}, {Utsumi}, {Inoue}, {Ohta}, {Doi},
  {Morokuma}, {Kawabata}, \& {Tanaka}}]{Itoh2020}
{Itoh}, R., {Utsumi}, Y., {Inoue}, Y., {et~al.} 2020, \apjl, 1, 12,
  \dodoi{10.3847/1538-4357/abab07}

\bibitem[{{Kasen} {et~al.}(2017){Kasen}, {Metzger}, {Barnes}, {Quataert}, \&
  {Ramirez-Ruiz}}]{Kasen2017}
{Kasen}, D., {Metzger}, B., {Barnes}, J., {Quataert}, E., \& {Ramirez-Ruiz}, E.
  2017, \nat, 551, 80, \dodoi{10.1038/nature24453}

\bibitem[{{Kasliwal} {et~al.}(2017){Kasliwal}, {Nakar}, {Singer}, {Kaplan},
  {Cook}, {Van Sistine}, {Lau}, {Fremling}, {Gottlieb}, {Jencson}, {Adams},
  {Feindt}, {Hotokezaka}, {Ghosh}, {Perley}, {Yu}, {Piran}, {Allison},
  {Anupama}, {Balasubramanian}, {Bannister}, {Bally}, {Barnes}, {Barway},
  {Bellm}, {Bhalerao}, {Bhattacharya}, {Blagorodnova}, {Bloom}, {Brady},
  {Cannella}, {Chatterjee}, {Cenko}, {Cobb}, {Copperwheat}, {Corsi}, {De},
  {Dobie}, {Emery}, {Evans}, {Fox}, {Frail}, {Frohmaier}, {Goobar}, {Hallinan},
  {Harrison}, {Helou}, {Hinderer}, {Ho}, {Horesh}, {Ip}, {Itoh}, {Kasen},
  {Kim}, {Kuin}, {Kupfer}, {Lynch}, {Madsen}, {Mazzali}, {Miller}, {Mooley},
  {Murphy}, {Ngeow}, {Nichols}, {Nissanke}, {Nugent}, {Ofek}, {Qi}, {Quimby},
  {Rosswog}, {Rusu}, {Sadler}, {Schmidt}, {Sollerman}, {Steele}, {Williamson},
  {Xu}, {Yan}, {Yatsu}, {Zhang}, \& {Zhao}}]{Kasliwal2017}
{Kasliwal}, M.~M., {Nakar}, E., {Singer}, L.~P., {et~al.} 2017, Science, 358,
  1559, \dodoi{10.1126/science.aap9455}

\bibitem[{{Kathirgamaraju} {et~al.}(2019){Kathirgamaraju}, {Giannios}, \&
  {Beniamini}}]{Kathirgamaraju2019}
{Kathirgamaraju}, A., {Giannios}, D., \& {Beniamini}, P. 2019, \mnras, 487,
  3914, \dodoi{10.1093/mnras/stz1564}

\bibitem[{{Kilpatrick} {et~al.}(2017){Kilpatrick}, {Foley}, {Kasen},
  {Murguia-Berthier}, {Ramirez-Ruiz}, {Coulter}, {Drout}, {Piro}, {Shappee},
  {Boutsia}, {Contreras}, {Di Mille}, {Madore}, {Morrell}, {Pan}, {Prochaska},
  {Rest}, {Rojas-Bravo}, {Siebert}, {Simon}, \& {Ulloa}}]{Kilpatrick2017}
{Kilpatrick}, C.~D., {Foley}, R.~J., {Kasen}, D., {et~al.} 2017, Science, 358,
  1583, \dodoi{10.1126/science.aaq0073}

\bibitem[{{Kilpatrick} {et~al.}(2021){Kilpatrick}, {Fong}, {Blanchard}, {Leja},
  {Nugent}, {Palmese}, {Paterson}, {Starkenburg}, {Alexander}, {Berger},
  {Chornock}, {Hajela}, \& {Margutti}}]{Kilpatrick2021}
{Kilpatrick}, C.~D., {Fong}, W.-f., {Blanchard}, P.~K., {et~al.} 2021, arXiv
  e-prints, arXiv:2109.06211.
\newblock \doarXiv{2109.06211}

\bibitem[{{Lacy} {et~al.}(2020){Lacy}, {Baum}, {Chandler}, {Chatterjee},
  {Clarke}, {Deustua}, {English}, {Farnes}, {Gaensler}, {Gugliucci},
  {Hallinan}, {Kent}, {Kimball}, {Law}, {Lazio}, {Marvil}, {Mao}, {Medlin},
  {Mooley}, {Murphy}, {Myers}, {Osten}, {Richards}, {Rosolowsky}, {Rudnick},
  {Schinzel}, {Sivakoff}, {Sjouwerman}, {Taylor}, {White}, {Wrobel},
  {Andernach}, {Beasley}, {Berger}, {Bhatnager}, {Birkinshaw}, {Bower},
  {Brandt}, {Brown}, {Burke-Spolaor}, {Butler}, {Comerford}, {Demorest}, {Fu},
  {Giacintucci}, {Golap}, {G{\"u}th}, {Hales}, {Hiriart}, {Hodge}, {Horesh},
  {Ivezi{\'c}}, {Jarvis}, {Kamble}, {Kassim}, {Liu}, {Loinard}, {Lyons},
  {Masters}, {Mezcua}, {Moellenbrock}, {Mroczkowski}, {Nyland}, {O'Dea},
  {O'Sullivan}, {Peters}, {Radford}, {Rao}, {Robnett}, {Salcido}, {Shen},
  {Sobotka}, {Witz}, {Vaccari}, {van Weeren}, {Vargas}, {Williams}, \&
  {Yoon}}]{Lacy2020}
{Lacy}, M., {Baum}, S.~A., {Chandler}, C.~J., {et~al.} 2020, \pasp, 132,
  035001, \dodoi{10.1088/1538-3873/ab63eb}

\bibitem[{{Lazzati} {et~al.}(2017){Lazzati}, {Deich}, {Morsony}, \&
  {Workman}}]{Lazzati2017}
{Lazzati}, D., {Deich}, A., {Morsony}, B.~J., \& {Workman}, J.~C. 2017, \mnras,
  471, 1652, \dodoi{10.1093/mnras/stx1683}

\bibitem[{{Lazzati} {et~al.}(2018){Lazzati}, {Perna}, {Morsony},
  {Lopez-Camara}, {Cantiello}, {Ciolfi}, {Giacomazzo}, \&
  {Workman}}]{Lazzati2018}
{Lazzati}, D., {Perna}, R., {Morsony}, B.~J., {et~al.} 2018, \prl, 120, 241103,
  \dodoi{10.1103/PhysRevLett.120.241103}

\bibitem[{{Makhathini} {et~al.}(2021){Makhathini}, {Mooley}, {Brightman},
  {Hotokezaka}, {Nayana}, {Intema}, {Dobie}, {Lenc}, {Perley}, {Fremling},
  {Mold{\`o}n}, {Lazzati}, {Kaplan}, {Balasubramanian}, {Brown}, {Carbone},
  {Chandra}, {Corsi}, {Camilo}, {Deller}, {Frail}, {Murphy}, {Murphy}, {Nakar},
  {Smirnov}, {Beswick}, {Fender}, {Hallinan}, {Heywood}, {Kasliwal}, {Lee},
  {Lu}, {Rana}, {Perkins}, {White}, {J{\'o}zsa}, {Hugo}, \&
  {Kamphuis}}]{Makhathini2021}
{Makhathini}, S., {Mooley}, K.~P., {Brightman}, M., {et~al.} 2021, \apj, 922,
  154, \dodoi{10.3847/1538-4357/ac1ffc}

\bibitem[{{Mangano} {et~al.}(2009){Mangano}, {Barthelmy}, {Baumgartner},
  {Burrows}, {Evans}, {Gehrels}, {Guidorzi}, {Holland}, {Hoversten}, {Kennea},
  {Krimm}, {Kuin}, {Markwardt}, {Marshall}, {Osborne}, {Page}, {Palmer},
  {Perri}, {Romano}, {Rowlinson}, {Sbarufatti}, {Stroh}, {Ukwatta}, {Vetere},
  {Ziaeepour}, {et~al.}}]{Mangano2009}
{Mangano}, V., {Barthelmy}, S.~D., {Baumgartner}, W.~H., {et~al.} 2009,
  Gamma-Ray Coordinate Network, 9133

\bibitem[{{Margutti} {et~al.}(2017){Margutti}, {Berger}, {Fong}, {Guidorzi},
  {Alexander}, {Metzger}, {Blanchard}, {Cowperthwaite}, {Chornock},
  {Eftekhari}, {Nicholl}, {Villar}, {Williams}, {Annis}, {Brown}, {Chen},
  {Doctor}, {Frieman}, {Holz}, {Sako}, \& {Soares-Santos}}]{Margutti2017}
{Margutti}, R., {Berger}, E., {Fong}, W., {et~al.} 2017, \apjl, 848, L20,
  \dodoi{10.3847/2041-8213/aa9057}

\bibitem[{{Mateos} {et~al.}(2012){Mateos}, {Alonso-Herrero}, {Carrera},
  {Blain}, {Watson}, {Barcons}, {Braito}, {Severgnini}, {Donley}, \&
  {Stern}}]{Mateos2012}
{Mateos}, S., {Alonso-Herrero}, A., {Carrera}, F.~J., {et~al.} 2012, \mnras,
  426, 3271, \dodoi{10.1111/j.1365-2966.2012.21843.x}

\bibitem[{{Matsumoto} \& {Piran}(2020)}]{Matsumoto2020}
{Matsumoto}, T., \& {Piran}, T. 2020, \mnras, 492, 4283,
  \dodoi{10.1093/mnras/staa050}

\bibitem[{{Metzger}(2019)}]{Metzger2019}
{Metzger}, B.~D. 2019, Living Reviews in Relativity, 23, 1,
  \dodoi{10.1007/s41114-019-0024-0}

\bibitem[{{Metzger} \& {Fern{\'a}ndez}(2021)}]{Metzger2021}
{Metzger}, B.~D., \& {Fern{\'a}ndez}, R. 2021, \apjl, 916, L3,
  \dodoi{10.3847/2041-8213/ac1169}

\bibitem[{{Mooley} {et~al.}(2013){Mooley}, {Frail}, {Ofek}, {Miller},
  {Kulkarni}, \& {Horesh}}]{Mooley2013}
{Mooley}, K.~P., {Frail}, D.~A., {Ofek}, E.~O., {et~al.} 2013, \apj, 768, 165,
  \dodoi{10.1088/0004-637X/768/2/165}

\bibitem[{{Mooley} {et~al.}(2018{\natexlab{a}}){Mooley}, {Deller}, {Gottlieb},
  {Nakar}, {Hallinan}, {Bourke}, {Frail}, {Horesh}, {Corsi}, \&
  {Hotokezaka}}]{Mooley2018a}
{Mooley}, K.~P., {Deller}, A.~T., {Gottlieb}, O., {et~al.} 2018{\natexlab{a}},
  \nat, 561, 355, \dodoi{10.1038/s41586-018-0486-3}

\bibitem[{{Mooley} {et~al.}(2018{\natexlab{b}}){Mooley}, {Frail}, {Dobie},
  {Lenc}, {Corsi}, {De}, {Nayana}, {Makhathini}, {Heywood}, {Murphy}, {Kaplan},
  {Chandra}, {Smirnov}, {Nakar}, {Hallinan}, {Camilo}, {Fender}, {Goedhart},
  {Groot}, {Kasliwal}, {Kulkarni}, \& {Woudt}}]{Mooley2018b}
{Mooley}, K.~P., {Frail}, D.~A., {Dobie}, D., {et~al.} 2018{\natexlab{b}},
  \apjl, 868, L11, \dodoi{10.3847/2041-8213/aaeda7}

\bibitem[{{Murphy} {et~al.}(2011){Murphy}, {Condon}, {Schinnerer}, {Kennicutt},
  {Calzetti}, {Armus}, {Helou}, {Turner}, {Aniano}, {Beir{\~a}o}, {Bolatto},
  {Brandl}, {Croxall}, {Dale}, {Donovan Meyer}, {Draine}, {Engelbracht},
  {Hunt}, {Hao}, {Koda}, {Roussel}, {Skibba}, \& {Smith}}]{Murphy2011}
{Murphy}, E.~J., {Condon}, J.~J., {Schinnerer}, E., {et~al.} 2011, \apj, 737,
  67, \dodoi{10.1088/0004-637X/737/2/67}

\bibitem[{{Nakar}(2007)}]{Nakar2007}
{Nakar}, E. 2007, \physrep, 442, 166, \dodoi{10.1016/j.physrep.2007.02.005}

\bibitem[{{Nakar} \& {Piran}(2011)}]{Nakar2011}
{Nakar}, E., \& {Piran}, T. 2011, \nat, 478, 82, \dodoi{10.1038/nature10365}

\bibitem[{{Nakar} \& {Piran}(2018)}]{Nakar2018}
---. 2018, \mnras, 478, 407, \dodoi{10.1093/mnras/sty952}

\bibitem[{{NASA/IPAC Extragalactic Database}(2019)}]{NED2019}
{NASA/IPAC Extragalactic Database}. 2019,  IPAC, \dodoi{10.26132/NED1}

\bibitem[{{Nedora} {et~al.}(2021){Nedora}, {Radice}, {Bernuzzi}, {Perego},
  {Daszuta}, {Endrizzi}, {Prakash}, \& {Schianchi}}]{Nedora2021}
{Nedora}, V., {Radice}, D., {Bernuzzi}, S., {et~al.} 2021, \mnras, 506, 5908,
  \dodoi{10.1093/mnras/stab2004}

\bibitem[{{O'Connor} \& {Troja}(2022)}]{Oconnor2022}
{O'Connor}, B., \& {Troja}, E. 2022, GRB Coordinates Network, 32065, 1

\bibitem[{{Padovani} {et~al.}(2017){Padovani}, {Alexander}, {Assef}, {De
  Marco}, {Giommi}, {Hickox}, {Richards}, {Smol{\v{c}}i{\'c}},
  {Hatziminaoglou}, {Mainieri}, \& {Salvato}}]{Padovani2017}
{Padovani}, P., {Alexander}, D.~M., {Assef}, R.~J., {et~al.} 2017, \aapr, 25,
  2, \dodoi{10.1007/s00159-017-0102-9}

\bibitem[{{Palliyaguru} {et~al.}(2016){Palliyaguru}, {Corsi}, {Kasliwal},
  {Cenko}, {Frail}, {Perley}, {Mishra}, {Singer}, {Gal-Yam}, {Nugent}, \&
  {Surace}}]{Palliyaguru2016}
{Palliyaguru}, N.~T., {Corsi}, A., {Kasliwal}, M.~M., {et~al.} 2016, \apjl,
  829, L28, \dodoi{10.3847/2041-8205/829/2/L28}

\bibitem[{{Perley} \& {Perley}(2013)}]{Perley2013}
{Perley}, D.~A., \& {Perley}, R.~A. 2013, \apj, 778, 172,
  \dodoi{10.1088/0004-637X/778/2/172}

\bibitem[{{Pian} {et~al.}(2017){Pian}, {D'Avanzo}, {Benetti}, {Branchesi},
  {Brocato}, {Campana}, {Cappellaro}, {Covino}, {D'Elia}, {Fynbo}, {Getman},
  {Ghirlanda}, {Ghisellini}, {Grado}, {Greco}, {Hjorth}, {Kouveliotou},
  {Levan}, {Limatola}, {Malesani}, {Mazzali}, {Melandri}, {M{\o}ller},
  {Nicastro}, {Palazzi}, {Piranomonte}, {Rossi}, {Salafia}, {Selsing},
  {Stratta}, {Tanaka}, {Tanvir}, {Tomasella}, {Watson}, {Yang}, {Amati},
  {Antonelli}, {Ascenzi}, {Bernardini}, {Bo{\"e}r}, {Bufano}, {Bulgarelli},
  {Capaccioli}, {Casella}, {Castro-Tirado}, {Chassande-Mottin}, {Ciolfi},
  {Copperwheat}, {Dadina}, {De Cesare}, {di Paola}, {Fan}, {Gendre},
  {Giuffrida}, {Giunta}, {Hunt}, {Israel}, {Jin}, {Kasliwal}, {Klose}, {Lisi},
  {Longo}, {Maiorano}, {Mapelli}, {Masetti}, {Nava}, {Patricelli}, {Perley},
  {Pescalli}, {Piran}, {Possenti}, {Pulone}, {Razzano}, {Salvaterra},
  {Schipani}, {Spera}, {Stamerra}, {Stella}, {Tagliaferri}, {Testa}, {Troja},
  {Turatto}, {Vergani}, \& {Vergani}}]{Pian2017}
{Pian}, E., {D'Avanzo}, P., {Benetti}, S., {et~al.} 2017, \nat, 551, 67,
  \dodoi{10.1038/nature24298}

\bibitem[{{Radice} {et~al.}(2020){Radice}, {Bernuzzi}, \&
  {Perego}}]{Radice2020}
{Radice}, D., {Bernuzzi}, S., \& {Perego}, A. 2020, Annual Review of Nuclear
  and Particle Science, 70, 95, \dodoi{10.1146/annurev-nucl-013120-114541}

\bibitem[{{Radice} {et~al.}(2018){Radice}, {Perego}, {Hotokezaka}, {Fromm},
  {Bernuzzi}, \& {Roberts}}]{Radice2018}
{Radice}, D., {Perego}, A., {Hotokezaka}, K., {et~al.} 2018, \apj, 869, 130,
  \dodoi{10.3847/1538-4357/aaf054}

\bibitem[{{Rastinejad} {et~al.}(2022){Rastinejad}, {Gompertz}, {Levan}, {Fong},
  {Nicholl}, {Lamb}, {Malesani}, {Nugent}, {Oates}, {Tanvir}, {de Ugarte
  Postigo}, {Kilpatrick}, {Moore}, {Metzger}, {Ravasio}, {Rossi}, {Schroeder},
  {Jencson}, {Sand}, {Smith}, {Ag{\"u}{\'\i} Fern{\'a}ndez}, {Berger},
  {Blanchard}, {Chornock}, {Cobb}, {De Pasquale}, {Fynbo}, {Izzo}, {Kann},
  {Laskar}, {Marini}, {Paterson}, {Rouco Escorial}, {Sears}, \&
  {Th{\"o}ne}}]{Rastinejad2022}
{Rastinejad}, J.~C., {Gompertz}, B.~P., {Levan}, A.~J., {et~al.} 2022, arXiv
  e-prints, arXiv:2204.10864.
\newblock \doarXiv{2204.10864}

\bibitem[{{Ricci} {et~al.}(2021){Ricci}, {Troja}, {Bruni}, {Matsumoto}, {Piro},
  {O'Connor}, {Piran}, {Navaieelavasani}, {Corsi}, {Giacomazzo}, \&
  {Wieringa}}]{Ricci2021}
{Ricci}, R., {Troja}, E., {Bruni}, G., {et~al.} 2021, \mnras, 500, 1708,
  \dodoi{10.1093/mnras/staa3241}

\bibitem[{{Sadler} {et~al.}(1999){Sadler}, {McIntyre}, {Jackson}, \&
  {Cannon}}]{Sadler1999}
{Sadler}, E.~M., {McIntyre}, V.~J., {Jackson}, C.~A., \& {Cannon}, R.~D. 1999,
  \pasa, 16, 247, \dodoi{10.1071/AS99247}

\bibitem[{{Sakamoto} {et~al.}(2012){Sakamoto}, {Barthelmy}, {Baumgartner},
  {Beardmore}, {Cummings}, {Fenimore}, {Gehrels}, {Krimm}, {Markwardt},
  {Palmer}, {Sato}, {Stamatikos}, {Tueller}, \& {Ukwatta}}]{Sakamoto2012}
{Sakamoto}, T., {Barthelmy}, S.~D., {Baumgartner}, W.~H., {et~al.} 2012,
  Gamma-Ray Coordinate Network, 13195

\bibitem[{{Schroeder} {et~al.}(2020){Schroeder}, {Margalit}, {Fong}, {Metzger},
  {Williams}, {Paterson}, {Alexander}, {Laskar}, {Goyal}, \&
  {Berger}}]{Schroeder2020}
{Schroeder}, G., {Margalit}, B., {Fong}, W.-f., {et~al.} 2020, \apj, 902, 82,
  \dodoi{10.3847/1538-4357/abb407}

\bibitem[{{Seymour} {et~al.}(2008){Seymour}, {Dwelly}, {Moss}, {McHardy},
  {Zoghbi}, {Rieke}, {Page}, {Hopkins}, \& {Loaring}}]{Seymour2008}
{Seymour}, N., {Dwelly}, T., {Moss}, D., {et~al.} 2008, \mnras, 386, 1695,
  \dodoi{10.1111/j.1365-2966.2008.13166.x}

\bibitem[{{Shappee} {et~al.}(2017){Shappee}, {Simon}, {Drout}, {Piro},
  {Morrell}, {Prieto}, {Kasen}, {Holoien}, {Kollmeier}, {Kelson}, {Coulter},
  {Foley}, {Kilpatrick}, {Siebert}, {Madore}, {Murguia-Berthier}, {Pan},
  {Prochaska}, {Ramirez-Ruiz}, {Rest}, {Adams}, {Alatalo}, {Ba{\~n}ados},
  {Baughman}, {Bernstein}, {Bitsakis}, {Boutsia}, {Bravo}, {Di Mille}, {Higgs},
  {Ji}, {Maravelias}, {Marshall}, {Placco}, {Prieto}, \& {Wan}}]{Shappee2017}
{Shappee}, B.~J., {Simon}, J.~D., {Drout}, M.~R., {et~al.} 2017, Science, 358,
  1574, \dodoi{10.1126/science.aaq0186}

\bibitem[{{Smartt} {et~al.}(2017){Smartt}, {Chen}, {Jerkstrand}, {Coughlin},
  {Kankare}, {Sim}, {Fraser}, {Inserra}, {Maguire}, {Chambers}, {Huber},
  {Kr{\"u}hler}, {Leloudas}, {Magee}, {Shingles}, {Smith}, {Young}, {Tonry},
  {Kotak}, {Gal-Yam}, {Lyman}, {Homan}, {Agliozzo}, {Anderson}, {Angus},
  {Ashall}, {Barbarino}, {Bauer}, {Berton}, {Botticella}, {Bulla}, {Bulger},
  {Cannizzaro}, {Cano}, {Cartier}, {Cikota}, {Clark}, {De Cia}, {Della Valle},
  {Denneau}, {Dennefeld}, {Dessart}, {Dimitriadis}, {Elias-Rosa}, {Firth},
  {Flewelling}, {Fl{\"o}rs}, {Franckowiak}, {Frohmaier}, {Galbany},
  {Gonz{\'a}lez-Gait{\'a}n}, {Greiner}, {Gromadzki}, {Guelbenzu},
  {Guti{\'e}rrez}, {Hamanowicz}, {Hanlon}, {Harmanen}, {Heintz}, {Heinze},
  {Hernandez}, {Hodgkin}, {Hook}, {Izzo}, {James}, {Jonker}, {Kerzendorf},
  {Klose}, {Kostrzewa-Rutkowska}, {Kowalski}, {Kromer}, {Kuncarayakti},
  {Lawrence}, {Lowe}, {Magnier}, {Manulis}, {Martin-Carrillo}, {Mattila},
  {McBrien}, {M{\"u}ller}, {Nordin}, {O'Neill}, {Onori}, {Palmerio},
  {Pastorello}, {Patat}, {Pignata}, {Podsiadlowski}, {Pumo}, {Prentice}, {Rau},
  {Razza}, {Rest}, {Reynolds}, {Roy}, {Ruiter}, {Rybicki}, {Salmon}, {Schady},
  {Schultz}, {Schweyer}, {Seitenzahl}, {Smith}, {Sollerman}, {Stalder},
  {Stubbs}, {Sullivan}, {Szegedi}, {Taddia}, {Taubenberger}, {Terreran}, {van
  Soelen}, {Vos}, {Wainscoat}, {Walton}, {Waters}, {Weiland}, {Willman},
  {Wiseman}, {Wright}, {Wyrzykowski}, \& {Yaron}}]{Smartt2017}
{Smartt}, S.~J., {Chen}, T.~W., {Jerkstrand}, A., {et~al.} 2017, \nat, 551, 75,
  \dodoi{10.1038/nature24303}

\bibitem[{{Smol{\v{c}}i{\'c}} {et~al.}(2017){Smol{\v{c}}i{\'c}}, {Delvecchio},
  {Zamorani}, {Baran}, {Novak}, {Delhaize}, {Schinnerer}, {Berta}, {Bondi},
  {Ciliegi}, {Capak}, {Civano}, {Karim}, {Le Fevre}, {Ilbert}, {Laigle},
  {Marchesi}, {McCracken}, {Tasca}, {Salvato}, \& {Vardoulaki}}]{Smol2017}
{Smol{\v{c}}i{\'c}}, V., {Delvecchio}, I., {Zamorani}, G., {et~al.} 2017, \aap,
  602, A2, \dodoi{10.1051/0004-6361/201630223}

\bibitem[{{Smolčić} {et~al.}(2008){Smolčić}, {Schinnerer}, {Scodeggio},
  {Franzetti}, {Aussel}, {Bondi}, {Brusa}, {Carilli}, {Capak}, \&
  {Charlot}}]{Smol2008}
{Smolčić}, V., {Schinnerer}, E., {Scodeggio}, M., {et~al.} 2008, \apjs, 177

\bibitem[{{Stroh} {et~al.}(2014){Stroh}, {Beardmore}, {Cummings}, {Evans},
  {Gehrels}, {Kennea}, {Marshall}, {Romano}, {Starling}, {et~al.}}]{Stroh2014}
{Stroh}, M.~C., {Beardmore}, A.~P., {Cummings}, J.~R., {et~al.} 2014, Gamma-Ray
  Coordinate Network, 16353

\bibitem[{{Tanvir} {et~al.}(2017){Tanvir}, {Levan},
  {Gonz{\'a}lez-Fern{\'a}ndez}, {Korobkin}, {Mandel}, {Rosswog}, {Hjorth},
  {D'Avanzo}, {Fruchter}, {Fryer}, {Kangas}, {Milvang-Jensen}, {Rosetti},
  {Steeghs}, {Wollaeger}, {Cano}, {Copperwheat}, {Covino}, {D'Elia}, {de Ugarte
  Postigo}, {Evans}, {Even}, {Fairhurst}, {Figuera Jaimes}, {Fontes}, {Fujii},
  {Fynbo}, {Gompertz}, {Greiner}, {Hodosan}, {Irwin}, {Jakobsson},
  {J{\o}rgensen}, {Kann}, {Lyman}, {Malesani}, {McMahon}, {Melandri},
  {O'Brien}, {Osborne}, {Palazzi}, {Perley}, {Pian}, {Piranomonte}, {Rabus},
  {Rol}, {Rowlinson}, {Schulze}, {Sutton}, {Th{\"o}ne}, {Ulaczyk}, {Watson},
  {Wiersema}, \& {Wijers}}]{Tanvir2017}
{Tanvir}, N.~R., {Levan}, A.~J., {Gonz{\'a}lez-Fern{\'a}ndez}, C., {et~al.}
  2017, \apjl, 848, L27, \dodoi{10.3847/2041-8213/aa90b6}

\bibitem[{{Troja} {et~al.}(2017){Troja}, {Piro}, {van Eerten}, {Wollaeger},
  {Im}, {Fox}, {Butler}, {Cenko}, {Sakamoto}, {Fryer}, {Ricci}, {Lien}, {Ryan},
  {Korobkin}, {Lee}, {Burgess}, {Lee}, {Watson}, {Choi}, {Covino}, {D'Avanzo},
  {Fontes}, {Gonz{\'a}lez}, {Khandrika}, {Kim}, {Kim}, {Lee}, {Lee}, {Kutyrev},
  {Lim}, {S{\'a}nchez-Ram{\'\i}rez}, {Veilleux}, {Wieringa}, \&
  {Yoon}}]{Troja2017}
{Troja}, E., {Piro}, L., {van Eerten}, H., {et~al.} 2017, \nat, 551, 71,
  \dodoi{10.1038/nature24290}

\bibitem[{{Troja} {et~al.}(2021){Troja}, {O'Connor}, {Ryan}, {Piro}, {Ricci},
  {Zhang}, {Piran}, {Bruni}, {Cenko}, \& {van Eerten}}]{Troja2021}
{Troja}, E., {O'Connor}, B., {Ryan}, G., {et~al.} 2021, arXiv e-prints,
  arXiv:2104.13378.
\newblock \doarXiv{2104.13378}

\bibitem[{{Troja} {et~al.}(2022){Troja}, {O'Connor}, {Ryan}, {Piro}, {Ricci},
  {Zhang}, {Piran}, {Bruni}, {Cenko}, \& {van Eerten}}]{Troja2022}
---. 2022, \mnras, 510, 1902, \dodoi{10.1093/mnras/stab3533}

\bibitem[{{Valenti} {et~al.}(2017){Valenti}, {Sand}, {Yang}, {Cappellaro},
  {Tartaglia}, {Corsi}, {Jha}, {Reichart}, {Haislip}, \&
  {Kouprianov}}]{Valenti2017}
{Valenti}, S., {Sand}, D.~J., {Yang}, S., {et~al.} 2017, \apjl, 848, L24,
  \dodoi{10.3847/2041-8213/aa8edf}

\bibitem[{{Wright} {et~al.}(2010){Wright}, {Eisenhardt}, {Mainzer}, {Ressler},
  {Cutri}, {Jarrett}, {Kirkpatrick}, {Padgett}, {McMillan}, {Skrutskie},
  {Stanford}, {Cohen}, {Walker}, {Mather}, {Leisawitz}, {Gautier}, {McLean},
  {Benford}, {Lonsdale}, {Blain}, {Mendez}, {Irace}, {Duval}, {Liu}, {Royer},
  {Heinrichsen}, {Howard}, {Shannon}, {Kendall}, {Walsh}, {Larsen}, {Cardon},
  {Schick}, {Schwalm}, {Abid}, {Fabinsky}, {Naes}, \& {Tsai}}]{Wright2010}
{Wright}, E.~L., {Eisenhardt}, P. R.~M., {Mainzer}, A.~K., {et~al.} 2010, \aj,
  140, 1868, \dodoi{10.1088/0004-6256/140/6/1868}

\bibitem[{{Wright} {et~al.}(2019){Wright}, {Eisenhardt}, {Mainzer}, {Ressler},
  {Cutri}, {Jarrett}, {Kirkpatrick}, {Padgett}, {McMillan}, {Skrutskie},
  {Stanford}, {Cohen}, {Walker}, {Mather}, {Leisawitz}, {Gautier}, {McLean},
  {Benford}, {Lonsdale}, {Blain}, {Mendez}, {Irace}, {Duval}, {Liu}, {Royer},
  {Heinrichsen}, {Howard}, {Shannon}, {Kendall}, {Walsh}, {Larsen}, {Cardon},
  {Schick}, {Schwalm}, {Abid}, {Fabinsky}, {Naes}, \& {Tsai}}]{AllWISE2019}
---. 2019,  IPAC, \dodoi{10.26131/IRSA1}

\bibitem[{{Xu} {et~al.}(2014){Xu}, {Niu}, {Yang}, {Esamdin}, \& {Ma}}]{Xu2014}
{Xu}, D., {Niu}, H.-B., {Yang}, T.-Z., {Esamdin}, A., \& {Ma}, L. 2014,
  Gamma-Ray Coordinate Network, 16359

\bibitem[{Yurkov {et~al.}(2012)}]{Yurkov2012}
Yurkov, V., {et~al.} 2012, Gamma-Ray Coordinate Network, 13197

\end{thebibliography}

\end{document}